\documentclass[traditabstract]{aa}

\usepackage{graphicx}
\usepackage{txfonts}
\usepackage{url}
\usepackage{aalongtable}

\begin{document}

\title{
The first IRAM/PdBI polarimetric millimeter survey of \\active galactic nuclei
}

\subtitle{II. Activity and properties of individual sources\thanks{This study is based on observations carried out with the IRAM Plateau de Bure Interferometer. IRAM is supported by INSU/CNRS (France), MPG (Germany), and IGN (Spain).}}

\titlerunning{Polarimetric Millimeter Survey of AGN. II.}

\author{S. Trippe\inst{1} \and R. Neri\inst{2} \and M. Krips\inst{2} \and A. Castro-Carrizo\inst{2} \and M. Bremer\inst{2} \and V. Pi\'etu\inst{2} \and J.M. Winters\inst{2}}

\institute{
Seoul National University, Department of Physics and Astronomy, 599 Gwanak-ro, Gwanak-gu, Seoul 151-742, South Korea \\
e-mail: \url{trippe@astro.snu.ac.kr} \and
Institut de Radioastronomie Millim\'etrique (IRAM), 300 rue de la Piscine, F-38406 Saint Martin d'H\`eres, France \\
}

\date{Received 2 December 2011; accepted 15 February 2012}

\abstract{
We present an analysis of the linear polarization of six active galactic nuclei -- 0415+379 (3C~111), 0507+179, 0528+134 (OG+134), 0954+658, 1418+546 (OQ+530), and 1637+574 (OS+562). Our targets were monitored from 2007 to 2011 in the observatory-frame frequency range 80--253~GHz, corresponding to a rest-frame frequency range 88--705~GHz. We find average degrees of polarization $m_L\approx2-7$\%; this indicates that the polarization signals are effectively averaged out by the emitter geometries. From a comparison of the fluctuation rates in flux and degree of polarization we conclude that the spatial scales relevant for polarized emission are of the same order of, but probably not smaller than, the spatial scales relevant for the emission of the total flux. We see indication for fairly strong shocks and/or complex, variable emission region geometries in our sources, with compression factors $\lesssim$0.9 and/or changes in viewing angles by $\gtrsim$10$^{\circ}$. An analysis of correlations between source fluxes and polarization parameter points out special cases: the presence of (at least) two distinct emission regions with different levels of polarization (for 0415+379) as well as emission from a single, predominant component (for 0507+179 and 1418+546).
Regarding the evolution of flux and polarization, we find good agreement between observations and the signal predicted by ``oblique shock in jet'' scenarios in one source (1418+546).
We attempt to derive rotation measures for all sources, leading to actual measurements for two AGN and upper limits for three sources. We derive values of ${\rm RM}=(-39\pm1_{\rm stat}\pm13_{\rm sys})\times10^3$~rad\,m$^{-2}$ and ${\rm RM=(42\pm1_{\rm stat}\pm11_{\rm sys})\times10^4}$~rad\,m$^{-2}$ for 1418+546 and 1637+574, respectively; \rm these are the highest values reported to date for AGN.  These values indicate magnetic field strengths of the order $\sim$10$^{-4}$~G. For 0415+379, 0507+179, and 0954+658 we derive upper limits ${\rm|RM|}<1.7\times10^4$~rad\,m$^{-2}$. From the relation $|{\rm RM}|\propto\nu^a$ we find $a=1.9\pm0.3$ for 1418+546, \rm in good agreement with $a=2$ as expected for a spherical or conical outflow.
}


\maketitle

\section{Introduction}

Active galactic nuclei (AGN) have been studied extensively in the wavelength range from cm-radio to $\gamma$ radiation in the last decades (see, e.g., Kembhavi \& Narlikar \cite{kembhavi1999}, or Krolik \cite{krolik1999}, and references therein for a review). There is overwhelming observational evidence that the main source of their emission is accretion onto supermassive black holes (SMBH) with masses $M_{\bullet}\simeq10^{6...9}M_{\odot}$ (e.g., Ferrarese \& Ford \cite{ferrarese2005}, and references therein). Important information on the physics of active galacic nuclei is encoded in their linear polarization $m_L$. Degrees and angles of polarization provide details on synchrotron emission, the geometry of emission regions, strength and orientation of magnetic fields, and (via Faraday rotation and/or depolarization) on particle densities and matter distributions of the surrounding or outflowing matter (see, e.g., Saikia \& Salter \cite{saikia1988}, and references therein).

In Trippe et al. (\cite{trippe2010}), hereafter Paper~I, we presented the results of our polarimetric AGN survey with the IRAM Plateau de Bure Interferometer (PdBI; e.g., Winters \& Neri \cite{winters2011})\footnote{http://www.iram.fr/IRAMFR/GILDAS/doc/pdf/pdbi-intro.pdf}. This survey observed 86 sources in the (observatory frame) frequency range 80--267~GHz. Sources were not resolved spatially, all properties observed were source-integrated ones. Paper~I provided a statistical analysis of the entire survey sample. The main results of this analysis were

\begin{itemize}

\item  For 73 out of the 86 AGN observed linear polarization was detected. The median degree of polarization was $\approx$4\%, the highest value ever measured was $\approx$19\%. These numbers are far below the value $\gtrsim$60\% expected from ``ideal'' synchrotron emission from optically thin sources, pointing toward a substantial averaging of polarization due to limited spatial resolution and non-uniform B-field geometries, and/or a substantial contribution by optically thick emission from galactic cores.

\item  On average, BL~Lacertae objects show a higher degree of polarization ($m_L\simeq7$\%) than Seyfert galaxies or QSOs ($m_L\simeq3$\% and $m_L\simeq5$\%, respectively). This ``polarization level sequence'' BL~Lac $\rightarrow$ QSOs $\rightarrow$ Seyferts can be understood approximately in the context of the viewing angle unification scheme of AGN: a more direct view into a more active nucleus permits observation of more concentrated emission. This reduces the impact of averaging the of polarization through source geometry.

\item  We did not find evidence for correlations of polarization with rest-frame frequency or redshift $z$. The absence of a $m_L-z$ correlation indicates that there has been no noticeable change of polarization properties since $z\simeq2.4$.

\item  There is a trend toward higher degrees of polarization in sources with flat spectral mm/radio indices that indicate partially optically thick, core-dominated emission. This is somewhat unexpected because, in general, optically thin AGN outflows are more polarized than optically thick cores. A possible explanation is -- again -- provided by source geometry: because the jet emission is distributed over much larger spatial areas than the nuclear emission, averaging of polarization through non-uniform magnetic field geometries could be more efficient in the outflows than in the cores, resulting in observation of the highest polarization levels in the most core-dominated sources.

\end{itemize}

In this article, Paper~II, we focus on the linear polarization properties of AGN that have been observed multiple times and resolved in time and frequency. We selected sources from the sample analyzed in Paper~I (1) for which significant polarization was detected at least 20 times, or (2) that had been observed at least 25 times and polarization had been detected at least 15 times. When selecting targets, these limits proved to be the minimum numbers necessary for reliable statistical statements. Observations were obtained between February 2007 and January 2011 at frequencies 82--253~GHz. Eventually, this left us with a sample of six targets: 0415+379, 0507+179, 0528+134, 0954+658, 1418+546, and 1637+574. We present the basic physical properties of our target sources in Table~\ref{tab_journal}.

\begin{table*}
\caption{Physical properties and observed frequency ranges for our six target AGN. Source types, redshifts, and kpc-morphology are taken from the NED and the MOJAVE database (Lister et al. \cite{lister2009}). Black hole masses $M_{\bullet}$ are taken from the corresponding references. We give the ranges of observed frequencies $\nu$ as well as the corresponding rest-frame frequencies $\nu_0$. Millimeter spectral indices $\alpha$ are taken from Paper~I; they are defined via $S_{\nu}\propto\nu^{-\alpha}$.}
\label{tab_journal}
\centering
\begin{tabular}{l l c c l c c c c}
\hline\hline
Object & Type$^{\mathrm{a}}$ & Redshift & $M_{\bullet} [10^8M_{\odot}]$ & kpc-morphology & $\alpha$ & $\nu$ [GHz] & $\nu_0$ [GHz] &  \\
\hline
0415+379 (3C~111) & Sy~1   & 0.049 & 36$^{\mathrm{b}}$ & 2-sided jet  & 0.9$\pm$0.1 & 84--227 & ~88--238 &  \\
0507+179          & QSO    & 0.416 &  2$^{\mathrm{c}}$ & 1-sided jet  & 0.3$\pm$0.1 & 84--231 & 120--326 &  \\
0528+134 (OG+134) & HPQ    & 2.060 & 10$^{\mathrm{d}}$ & 2-sided jet  & 0.9$\pm$0.1 & 84--231 & 258--705 &  \\
0954+658          & BL~Lac & 0.367 &  3$^{\mathrm{e}}$ & 1-sided jet  & 0.3$\pm$0.1 & 82--235 & 112--321 &  \\
1418+546 (OQ+530) & BL~Lac & 0.153 &  9$^{\mathrm{f}}$ & 1-sided halo & 0.2$\pm$0.1 & 82--253 & ~95--291 &  \\
1637+574 (OS+562) & Sy~1   & 0.751 & 17$^{\mathrm{g}}$ & 1-sided halo & 0.1$\pm$0.3 & 80--115 & 141--201 &  \\
\hline
\end{tabular}

\begin{list}{}{}
\item[$^{\mathrm{a}}$] ``Sy'' = Seyfert galaxy, ``HPQ'' = high polarization quasar, ``BL~Lac'' = BL~Lacertae object
\item[$^{\mathrm{b}}$] Marchesini et al. (\cite{marchesini2004})
\item[$^{\mathrm{c}}$] D'Elia et al. (\cite{elia2003})
\item[$^{\mathrm{d}}$] Ghisellini et al. (\cite{ghisellini2009})
\item[$^{\mathrm{e}}$] Fan \& Cao (\cite{fan2004})
\item[$^{\mathrm{f}}$] Falomo et al. (\cite{falomo2003})
\item[$^{\mathrm{g}}$] Liu et al. (\cite{liu2006})
\end{list}

\end{table*}

\section{Observations and data processing}

\subsection{Polarization}

In January 2007 (January 2008 for the 2mm-band), the antennas of the PdBI were equipped with dual linear polarization Cassegrain focus receivers. Since then, it is possible to observe both orthogonal polarizations -- ``horizontal'' (H) and ``vertical'' (V) with respect to the antenna frame -- simultaneously. Observations can be carried out (non-simultaneously) in three atmospheric windows located around wavelengths of 1.3~mm, 2~mm, and 3~mm.   Each of these bands covers a continuous range of frequencies; their ranges are
201--267~GHz for the 1.3-mm band,
129--174~GHz for the 2-mm band,
and 80--116~GHz for the 3-mm band.
Within a given band, any frequency is available for observations. Receivers covering a fourth wavelength band centered at 0.8~mm began operations at the end of 2010 but did not provide polarization data as of January 2011.

At the time of the observations presented here, the PdBI was not yet equipped for observations of all Stokes parameters. We collected linear polarization data on point sources via Earth rotation polarimetry, i.e. by monitoring the fluxes in the H and V channels as functions of parallactic angle $\psi$. For assessing the polarization of a source, we calculated the parameter

\begin{equation}
q(\psi) = \frac{V-H}{V+H}(\psi) = \frac{Q}{I}\cos(2\psi) + \frac{U}{I}\sin(2\psi)
\label{eq_q}
\end{equation}

\noindent
from the fluxes $H(\psi)$ and $V(\psi)$. The second equality means that $q(\psi)$ provides full information on linear polarization (see, e.g., Sault, Hamaker \& Bregman \cite{sault1996}; Thompson, Moran \& Swenson \cite{thompson2001}; but also Beltr\'an et al. \cite{beltran2004}) if a sufficient range of $\psi$ is observed.

Owing to the nature of polarized light and because the PdBI antenna receivers are located in the Cassegrain foci, observing a polarized target results in $q(\psi)$ being a cosinusoidal signal with a period of $180^{\circ}$. The functional form of $q(\psi)$ is thus

\begin{equation}
q(\psi) \equiv m_L\cos\left[2(\psi-\chi)\right]  .
\label{eq_polformula}
\end{equation}

\noindent
Here $m_L$ is the fraction of linear polarization (ranging from 0 to 1; in the following, we will express $m_L$ in units of \%) and $\chi$ is the polarization angle (ranging from 0$^{\circ}$ to 180$^{\circ}$). A more detailed discussion of the methodology is provided in Paper~I. Our observational results are summarized in Table~\ref{tab_obsjournal}.

\subsection{Flux densities}

Source fluxes are initially recorded as antenna temperatures. Antenna temperatures are converted into physical flux densities (quoted in units of Jansky throughout this paper) by using empirical antenna efficiencies as conversion factors. Those factors are functions of frequencies and are located in the range from $\approx$22~Jy\,K$^{-1}$ (for the 3-mm band) to $\approx$37~Jy\,K$^{-1}$ (for the 1.3-mm band). To estimate the systematic uncertainties of our dataset, we analyzed observations of the radio continuum star MWC~349A (e.g. Tafoya et al. \cite{tafoya2004}) that were obtained and calibrated like the AGN measurements. Owing to its well-known flux properties MWC~349 serves as a ``standard candle'' for the PdBI. From the systematic scatter of the MWC~349 lightcurves we conclude that the \emph{systematic} relative uncertainties of our AGN observations are $\approx$10\% for the 2-mm and 3-mm bands and $\approx$16\% for the 1.3-mm band. For more details regarding AGN flux monitoring with the PdBI, please see Trippe et al. (\cite{trippe2011}).

We note that the systematic accuracy of our flux data is limited by the automatic monitoring and calibration. When calibrating observations individually and interactively while using MWC~349 as reference, systematic errors can be reduced to $\lesssim$5\%; Krips et al. (\emph{in prep.}) are going to discuss the recent developments regarding the PdBI flux calibration and the resulting improved performance.

\section{Analysis}

\subsection{Variability of polarization and flux}

\begin{figure*}
\centering
\includegraphics[height=8.9cm,angle=-90]{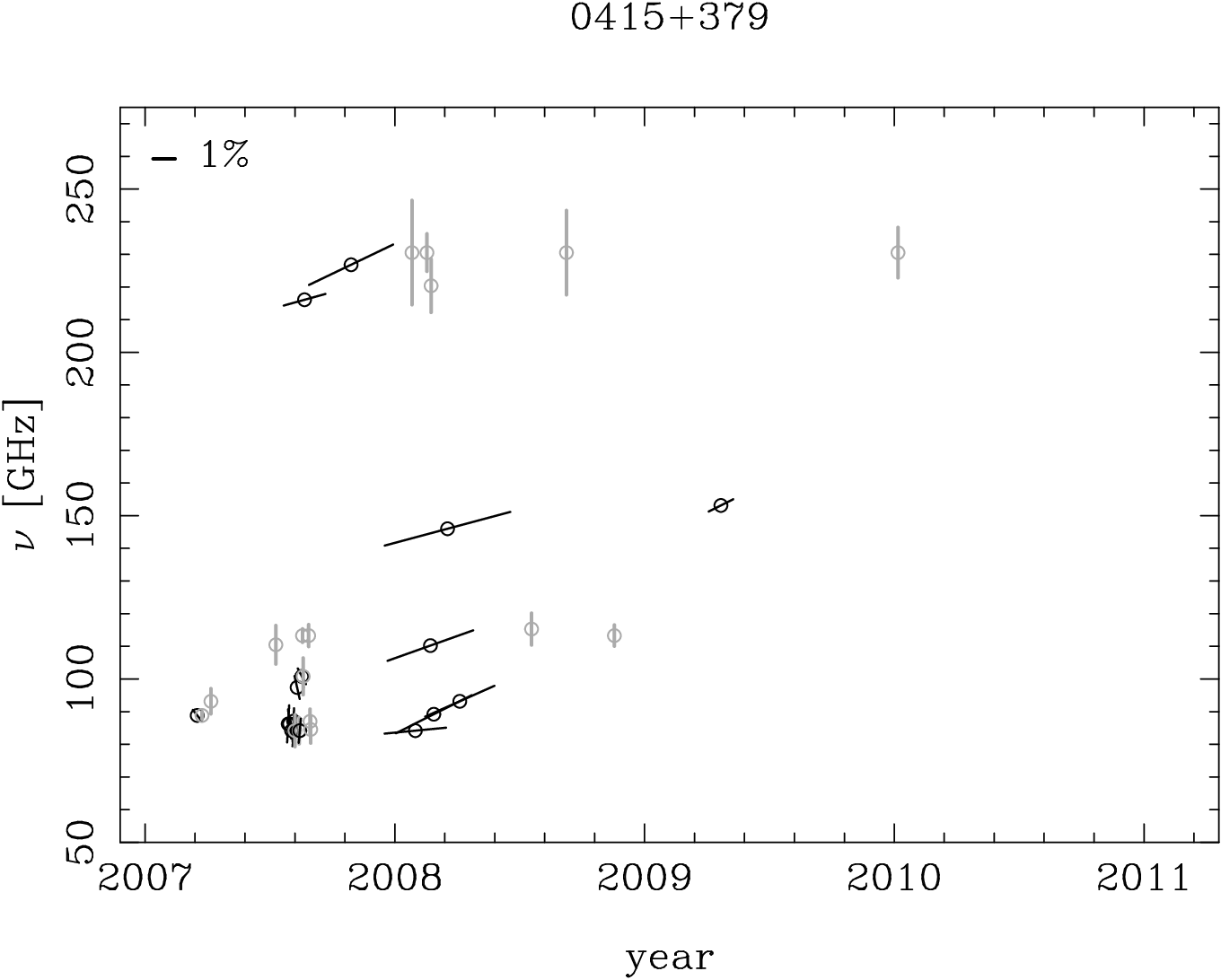}
\hspace{3mm}
\includegraphics[height=8.9cm,angle=-90]{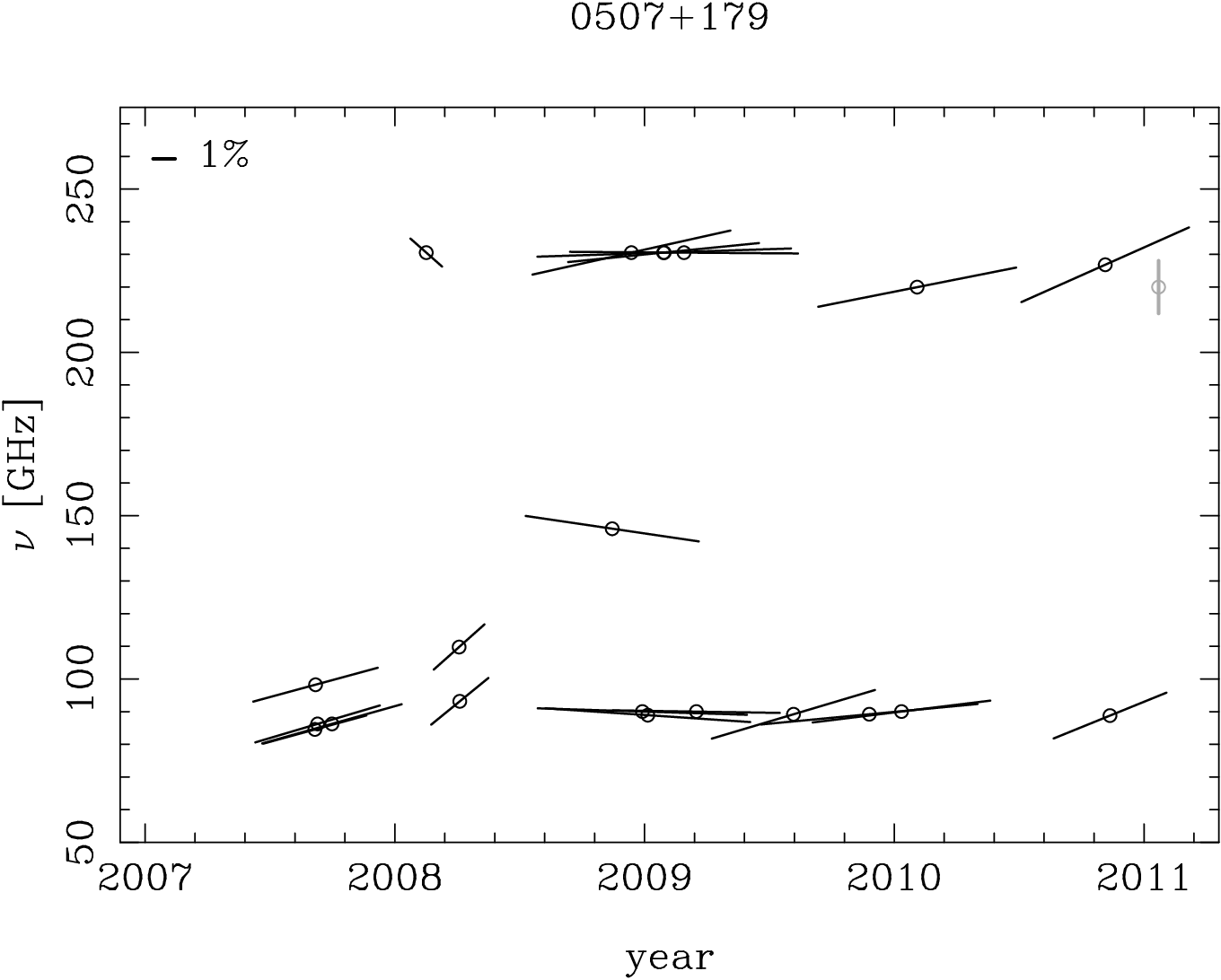} \\
\vspace{5mm}
\includegraphics[height=8.9cm,angle=-90]{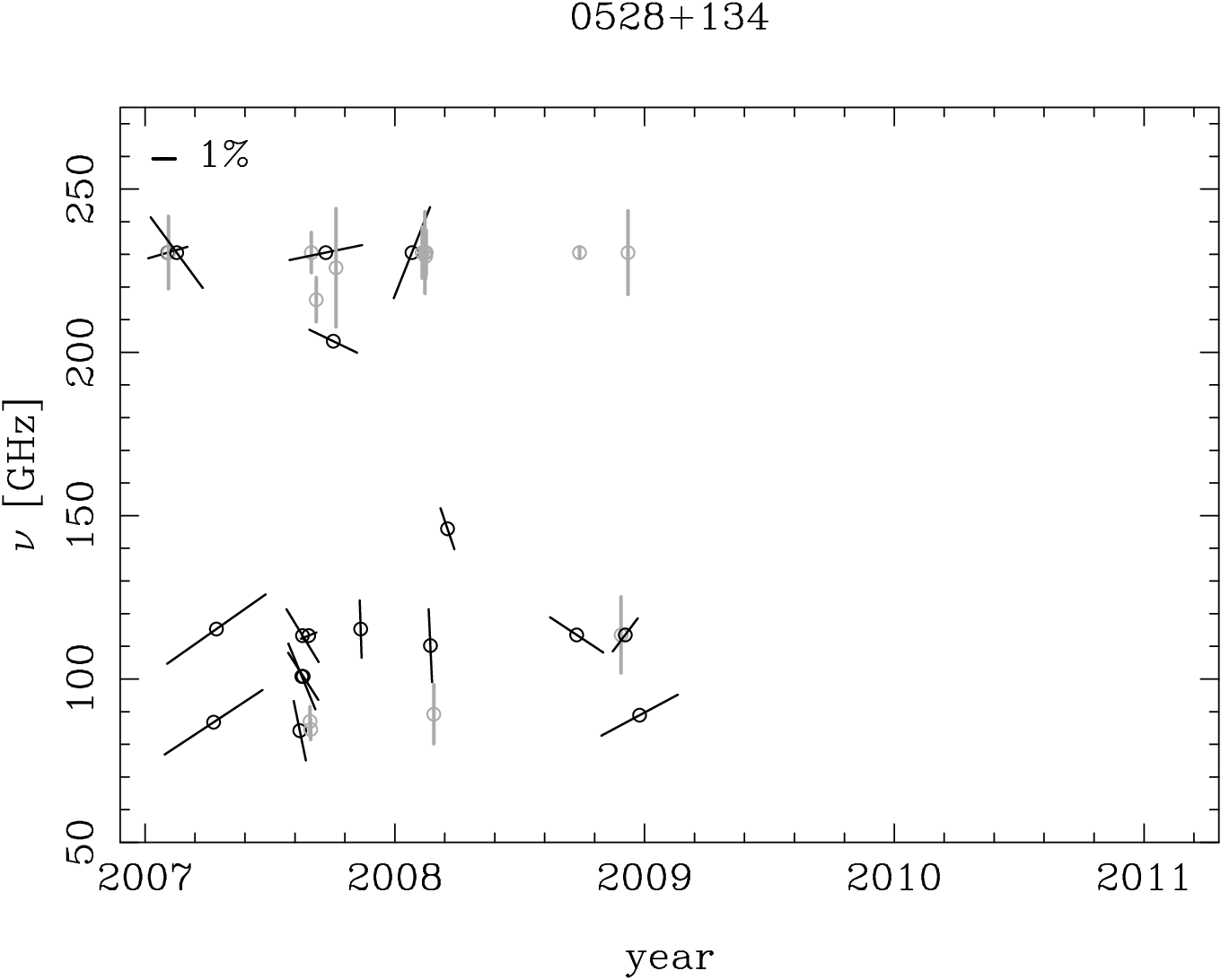}
\hspace{3mm}
\includegraphics[height=8.9cm,angle=-90]{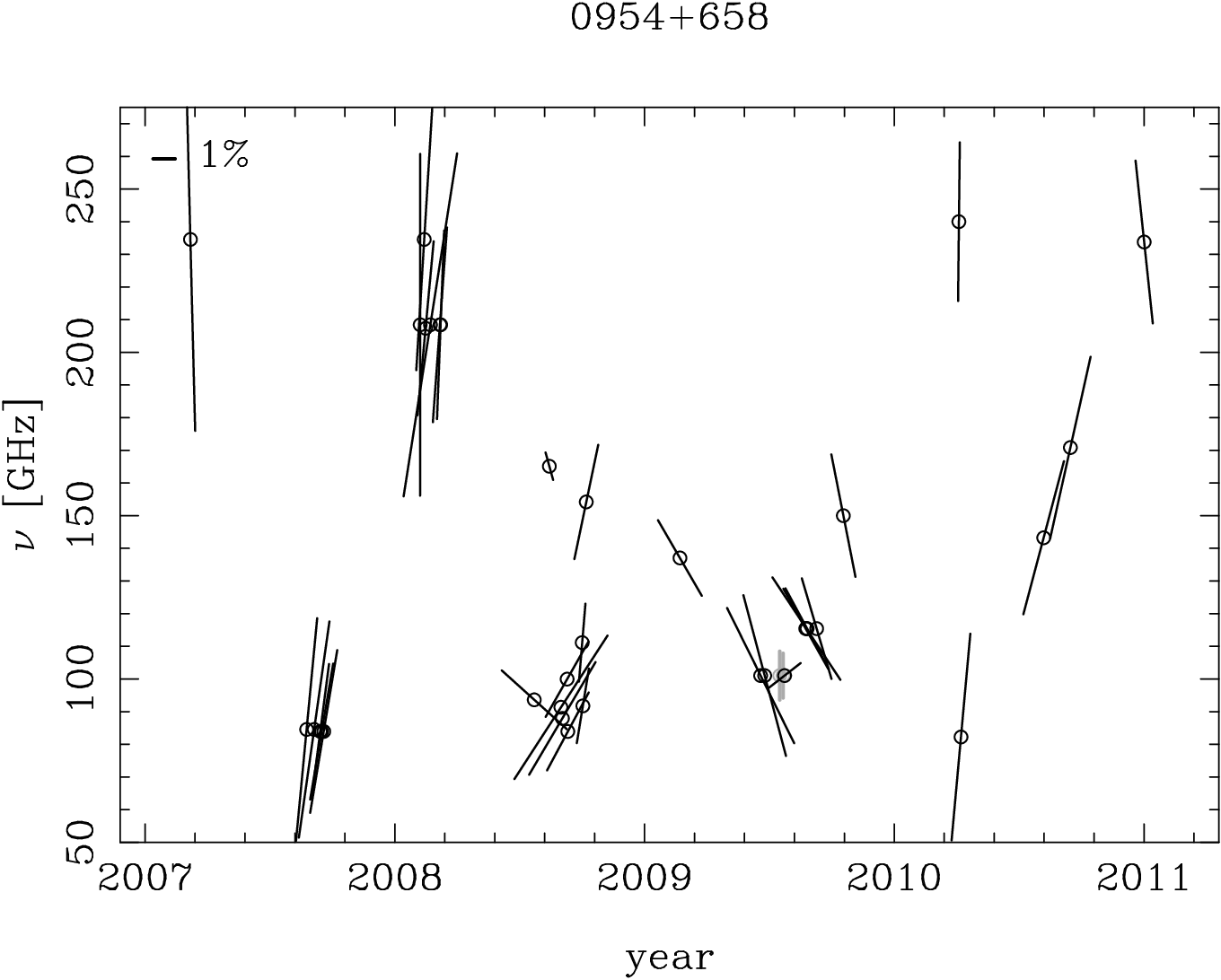} \\
\vspace{5mm}
\includegraphics[height=8.9cm,angle=-90]{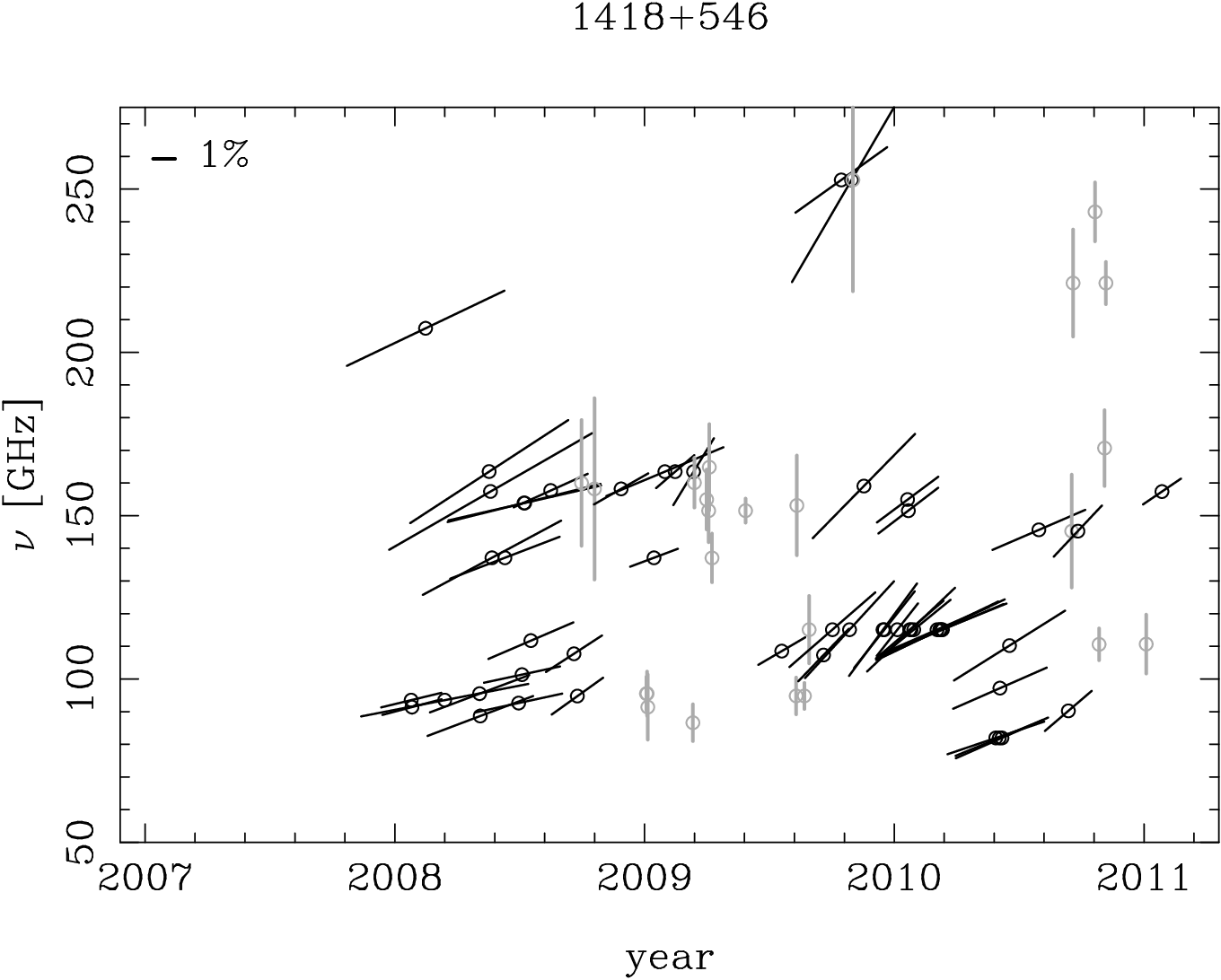}
\hspace{3mm}
\includegraphics[height=8.9cm,angle=-90]{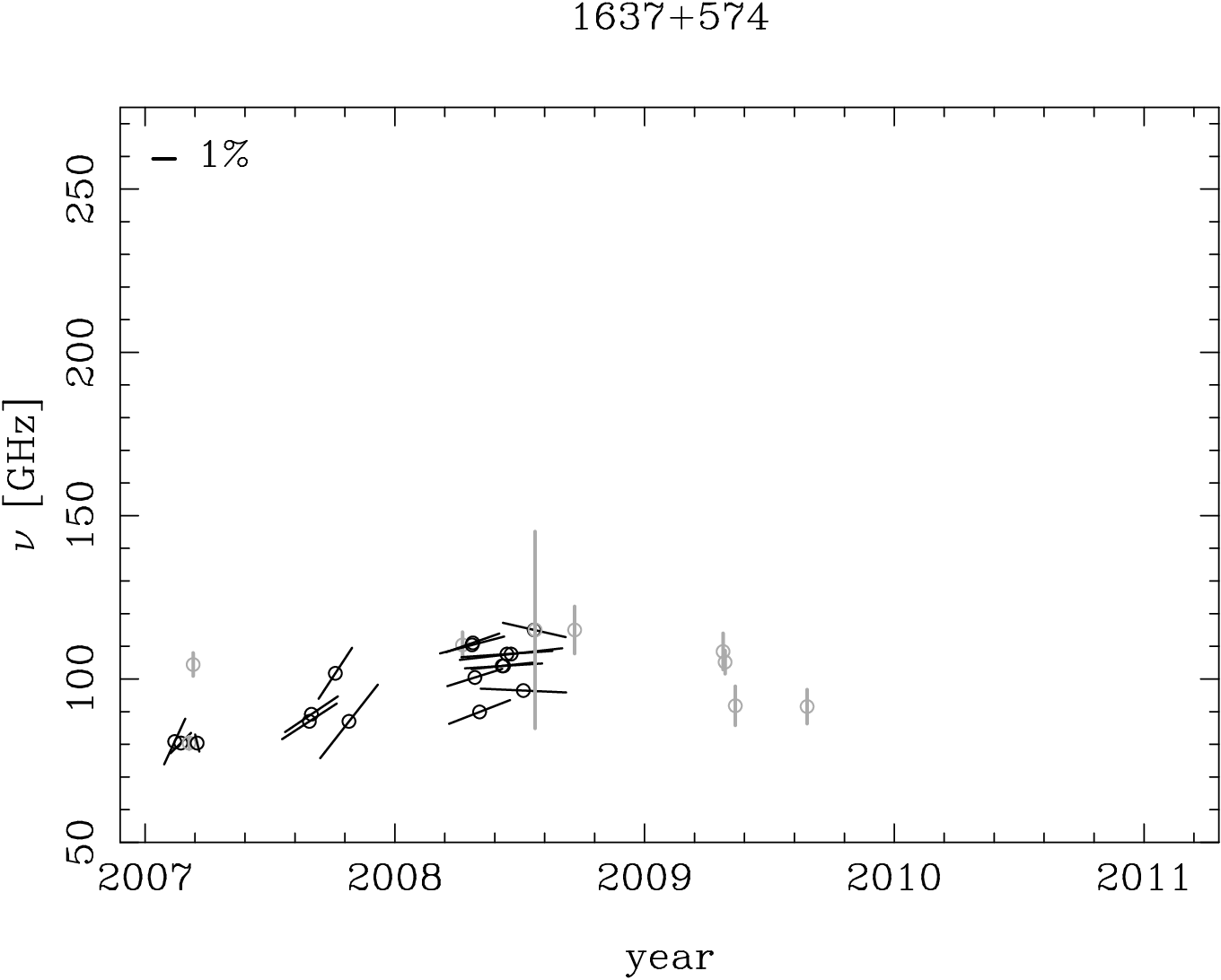}
\caption{Linear polarization state as function of time $t$ and observation frequency $\nu$. We give the result of each polarization measurement via a black bar in the $t-\nu$ plane. The length of each bar corresponds to the degree of linear polarization according to the scales in the upper left corners of the corresponding panels (in units of \%). The orientation of the bar indicates the polarization angle; angles are counted counter-clockwise from the 12-o'clock position. Points in the $t-\nu$ plane where a measurement was carried out but no polarization was detected are indicated by gray bars. The length of each gray hand corresponds to a $3\sigma$ upper limit on the polarization level. As upper limits do not have meaningful polarization angles, all gray bars are fixed at the 12-o'clock position.}
\label{fig_polplane}
\end{figure*}

Our analysis focuses on the properties of AGN for which polarization has been observed multiple ($\geq$15) times. We take into account that the polarization is, a priori, a function of two independent parameters: time and frequency. Accordingly, we may present the polarization of our six target AGN in two-dimensional diagrams with the two dimensions being observing time and observing frequency, respectively. The result is shown in Fig~\ref{fig_polplane}. The diagrams provide degrees of polarization and polarization angles for each measurement that detected significant polarization and upper limits on polarization levels otherwise.

\begin{figure*}
\centering
\includegraphics[height=8.9cm,angle=-90]{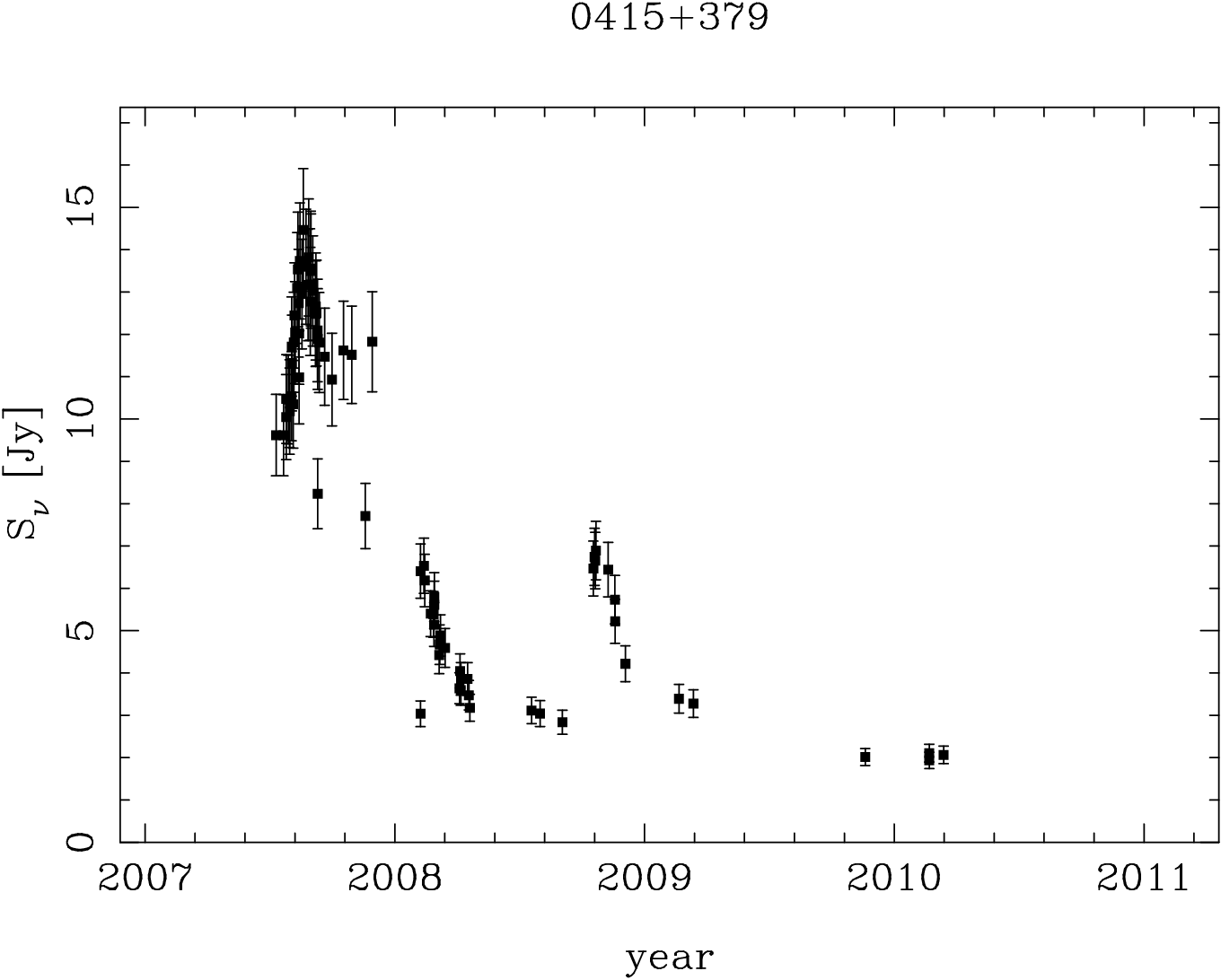}
\hspace{3mm}
\includegraphics[height=8.9cm,angle=-90]{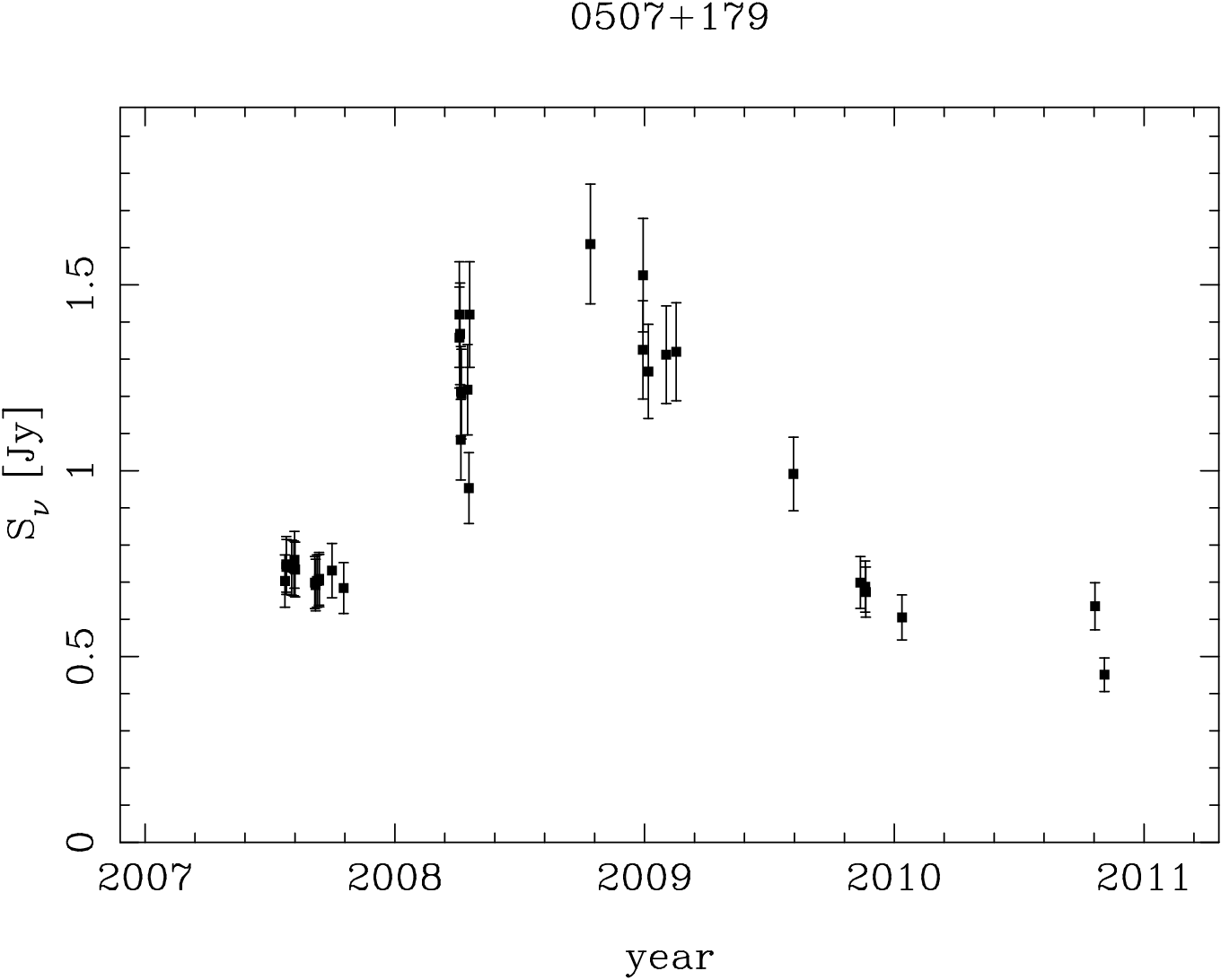} \\
\vspace{5mm}
\includegraphics[height=8.9cm,angle=-90]{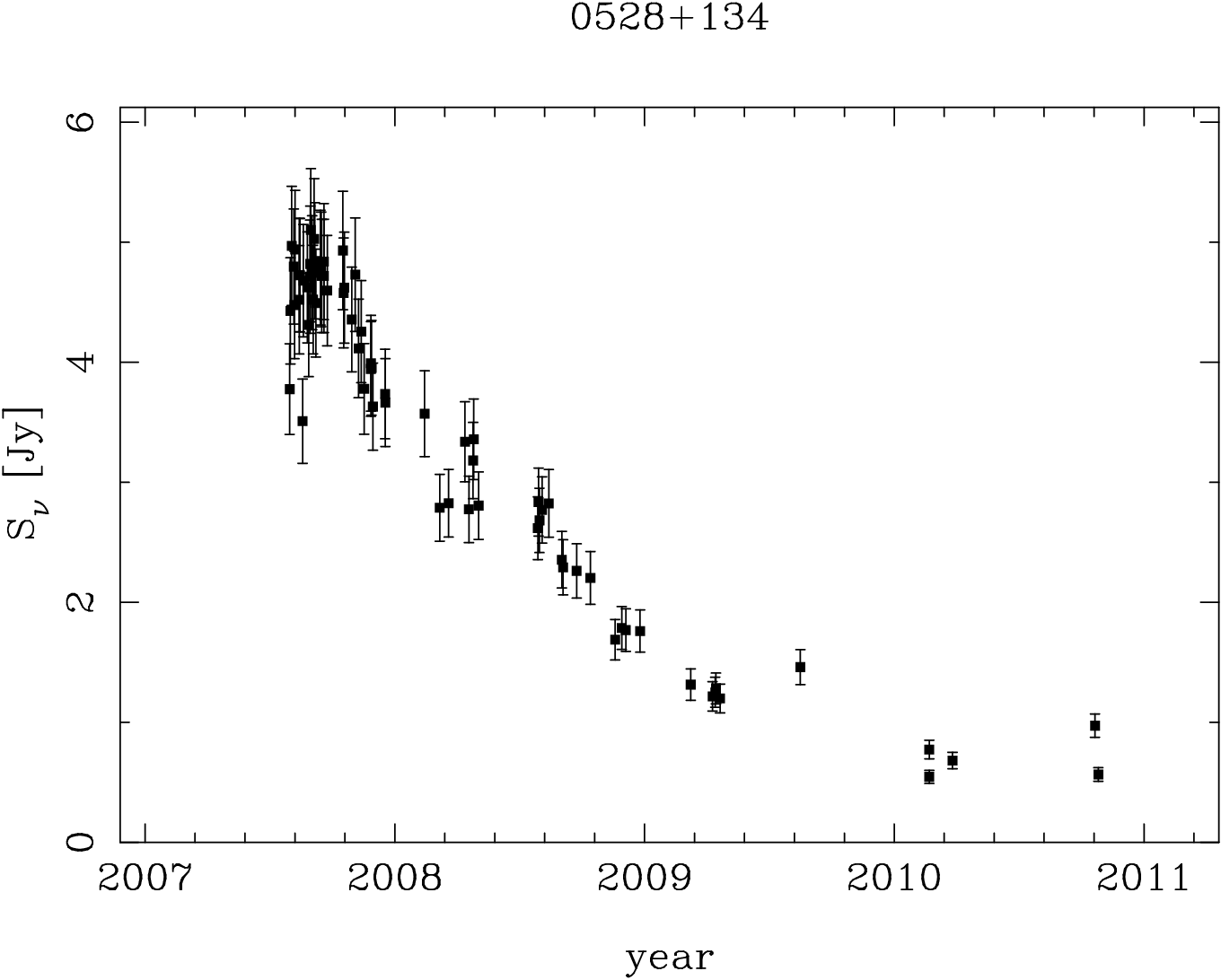}
\hspace{3mm}
\includegraphics[height=8.9cm,angle=-90]{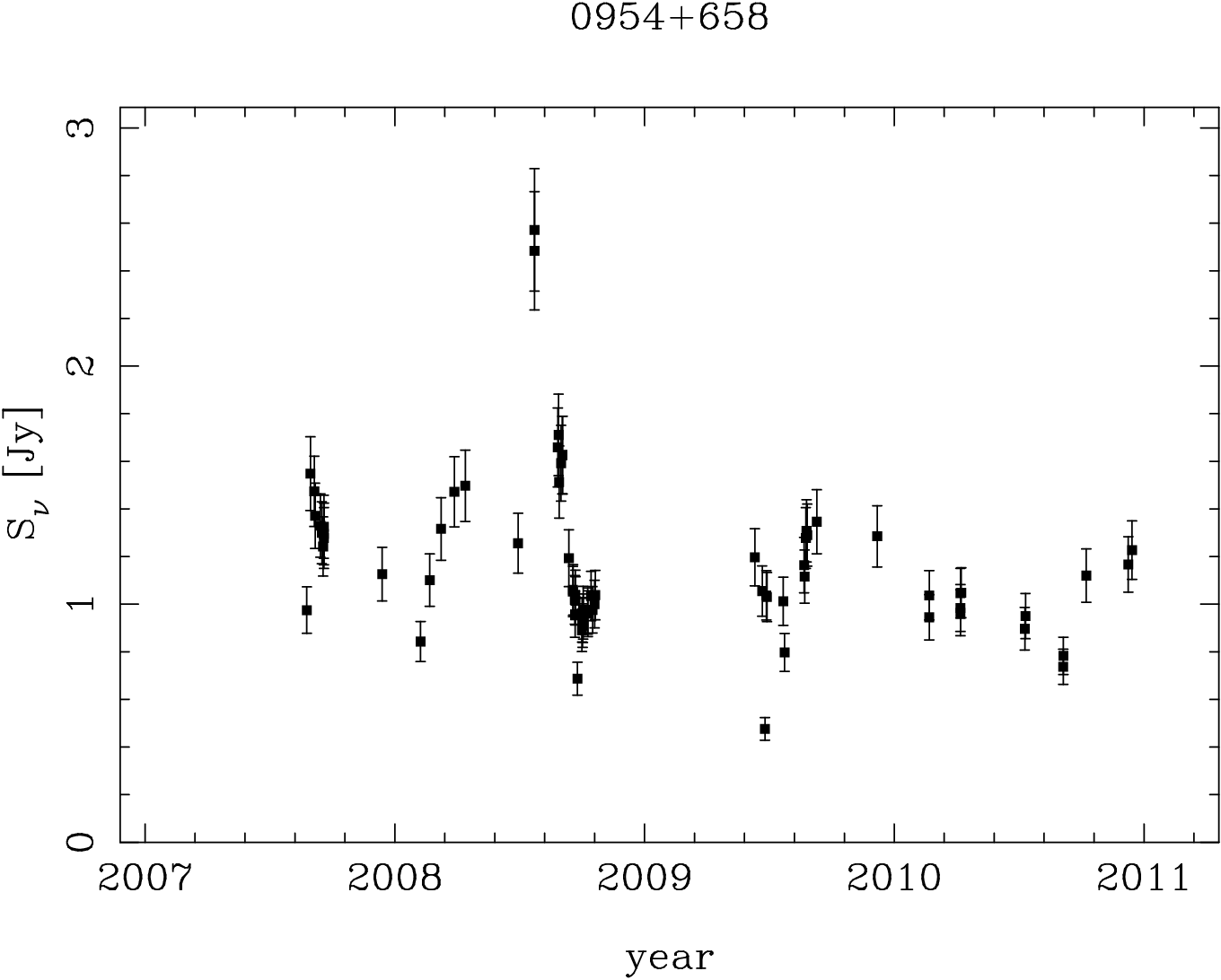} \\
\vspace{5mm}
\includegraphics[height=8.9cm,angle=-90]{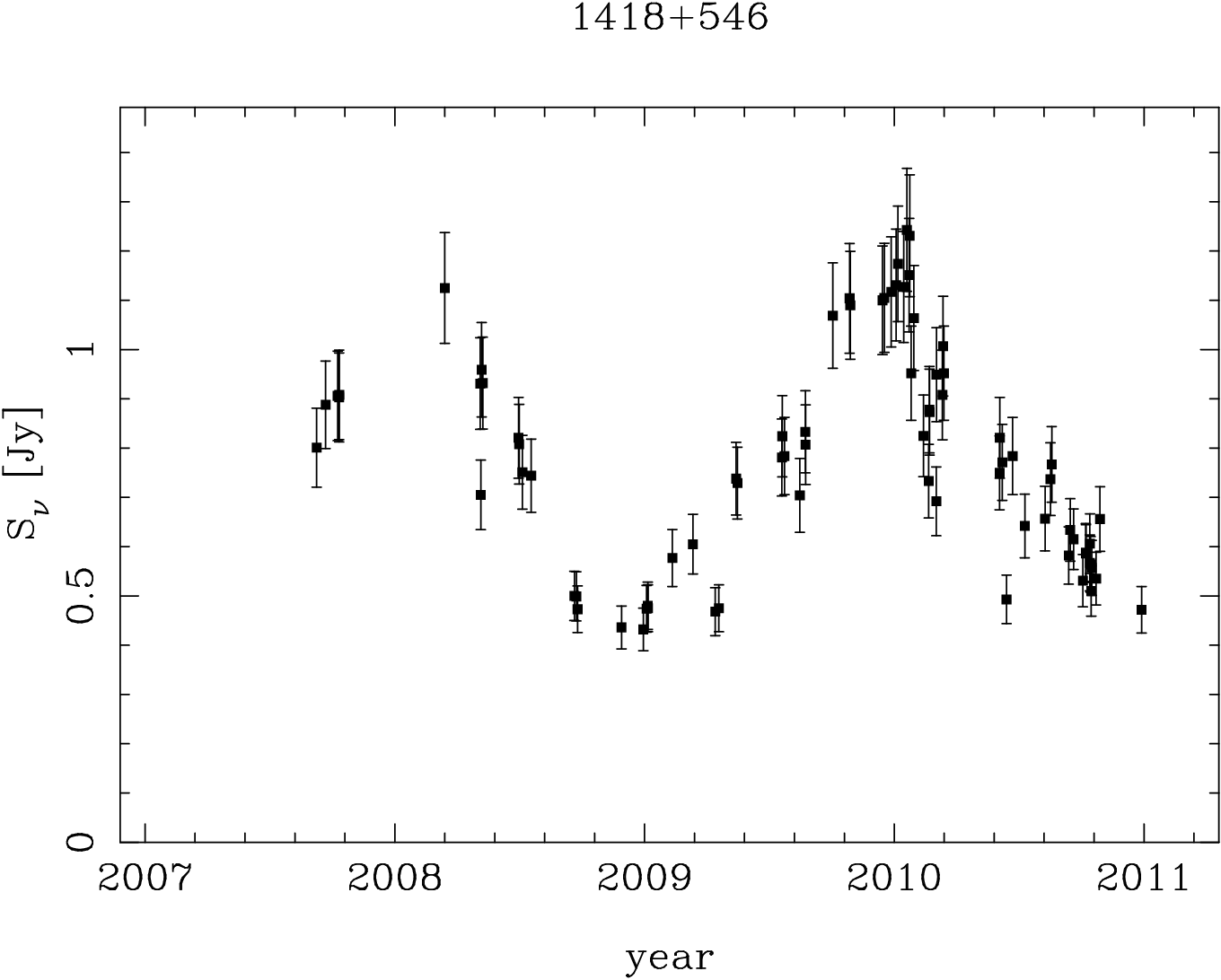}
\hspace{3mm}
\includegraphics[height=8.9cm,angle=-90]{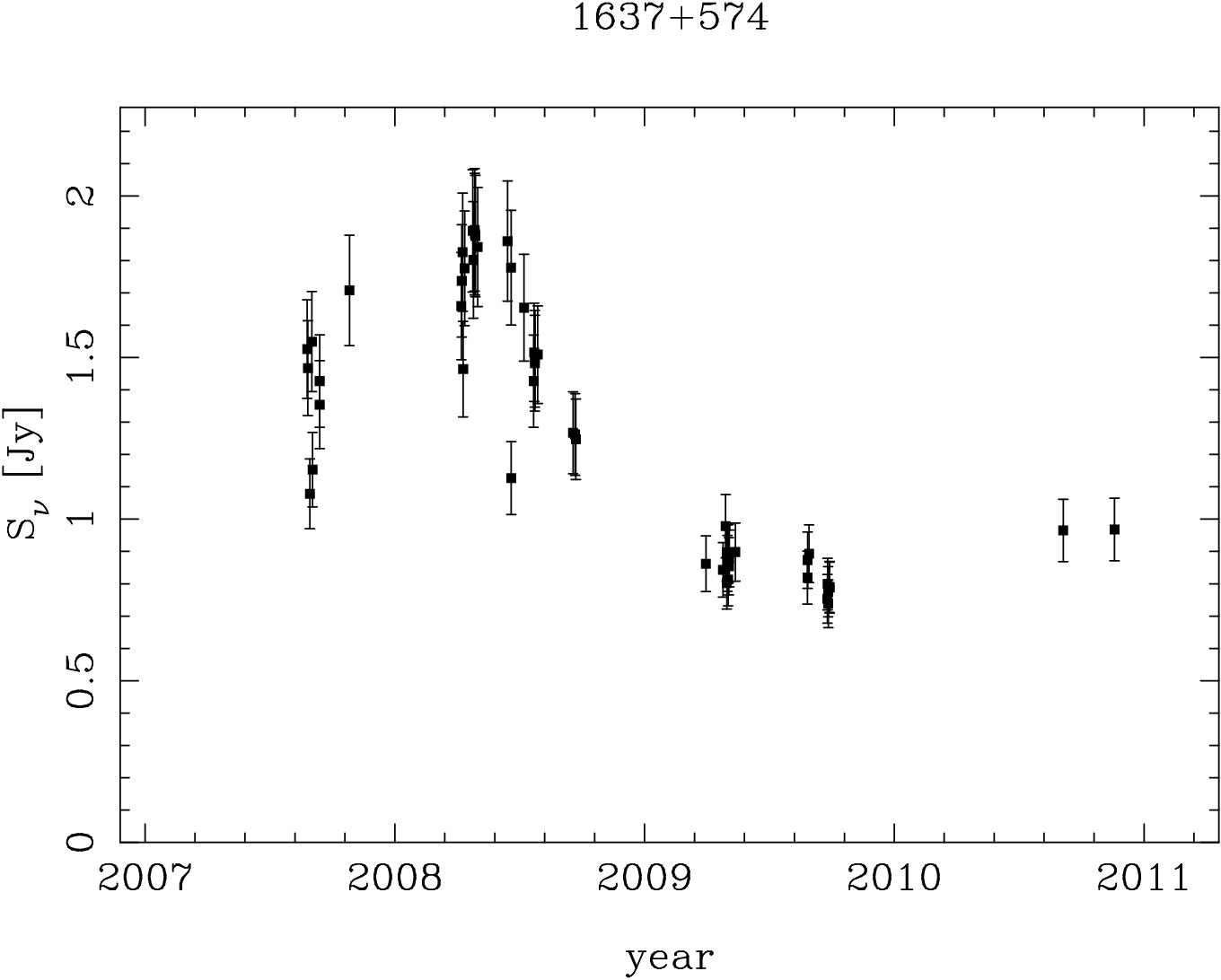}
\caption{3-mm flux densities of our target AGN as function of time. Please note the different axis scales. The errors given are dominated by a $\approx$10\% systematic uncertainty of the flux scale calibration; therefore error bars tend to be smaller for lower fluxes.}
\label{fig_lightcurve}
\end{figure*}

In Fig.~\ref{fig_lightcurve}, we present the 3-mm lightcurves of our target sources. To prevent systematic errors caused spectral slope effects, we limit our variability analysis to flux measurements obtained in the 3-mm band.

To assess variabilities quantitatively, we define the redshift-corrected \emph{fluctuation rate} of a time-dependent parameter $x$ like

\begin{equation}
\xi_x = \frac{1}{N}(1+z)\sum_{i=1}^{N-1}\left|\frac{x_{i+1}-x_i}{t_{i+1}-t_i}\right|  ,
\label{eq_xi}
\end{equation}

\noindent
where $z$ is the redshift, $x_i$ is the $i$th ($i=1,2,3,...$) value of the dataset, $t_i$ the time when $x_i$ was obtained, and $N$ the total number of measurements; $|...|$ denotes the absolute value of the enclosed function. By construction, $\xi_x$ is a measure of the rapidity of variations of the parameter $x$, taking explicitly into account the time axis of the dataset. Thus our parametrization provides additional information compared to commonly used, simpler variability estimators such as the $\sigma_V$ parameter proposed by Hook et al. (\cite{hook1994}). We note that our $\xi$ is similar to a parameter introduced by Villforth et al. (\cite{villforth2009}), which they named ``violence''. 

In Table~\ref{tab_stat}, we present fluctuation rates for the degree of linear polarization $m_L$, $\xi_m$, as well as for the 3-mm flux densities $S_{\nu}$, $\xi_S$. In addition, we show normalized fluctuation rates $\xi_m/\langle m_L\rangle$ and $\xi_S/\langle S_{\nu}\rangle$, with $\langle ...\rangle$ denoting the time average of the corresponding parameter. Finally, we also present the values of the fluctuation rate ratio

\begin{equation}
\rho = \frac{\xi_m/\langle m_L\rangle}{\xi_S/\langle S_{\nu}\rangle}  .
\label{eq_rho}
\end{equation}

\noindent
For the two fluctuation rates, $\xi_m$ and $\xi_S$, we calculated statistical errors numerically. Because all data points are affected by measurement errors, the calculated fluctuation rates deviate from zero even for signals with zero ``true'' variability. Our error analysis is based on a Monte Carlo scheme: we repeated each calculation 1000 times with different realizations of a given data set. The measurement error is defined as the central 68\% (1$\sigma$) interval of the resulting distribution of $\xi_x$ values. A realization of a data set is drawn by adding random, Gaussian errors -- which are in accord with the individual statistical uncertainties -- to all elements of a data set. For the $\xi_m$, errors are small compared to the parameter values and are quoted explicitly in Table~\ref{tab_stat}. For the $\xi_S$, uncertainty ranges are asymmetric and reach the magnitudes of the parameter values; we therefore do not quote errors individually but restrict ourselves to the statement that the $\xi_S$ values are accurate to factors $\approx$2.

\subsection{Correlations between flux and polarization}

The interplay between source polarizations and fluxes promises new insights into the emission mechanisms. We therefore study the relation between fluxes $S_{\nu}$ on the one hand and polarization levels $m_L$, polarized fluxes $P$, and polarization angles $\chi$ on the other hand. Our observational procedures measure $m_L$ according to Eq.~\ref{eq_polformula} but not $P$. Thus we calculate linearly polarized fluxes like $P = m_L S_{\nu}$ by combining those -- initially separate -- polarization and flux measurements obtained simultaneously at the same frequency. Because flux densities are a function of observation frequency $\nu$ like $S_{\nu}\propto\nu^{-\alpha}$, we apply spectral rescaling:

\begin{equation}
S^*_{\nu} = S_{\nu}\left(\frac{\nu}{\rm 90~GHz}\right)^{\alpha}  .
\label{eq_rescale}
\end{equation}

\noindent
This procedure means that we scale all observed flux densities to an observing frequency $\nu=90$~GHz. We used the spectral index information from Paper~I (see also Table~\ref{tab_journal}). Accordingly, we computed rescaled polarized fluxes $P^* = m_L S^*_{\nu}$. For the special case of 0415+379, we included only 3-mm flux densities into our calculations. For this AGN our flux data are affected by a combination of rapid flux variability and irregular sampling that introduce additional systematic errors when combining several wavelength bands; see also Trippe et al. (\cite{trippe2011}). For the remaining five sources we included all data available regardless of frequency. We note that rescaling is based on spectral slopes derived from an ensemble of flux measurements. Accordingly, rescaled parameters can only be used when analyzing the same ensemble of flux data collectively, though not for the analysis of individual or differences between individual measurements as in Sect.~3.1. Our subsequent correlation analysis is based on the rescaled parameters $S^*_{\nu}$ and $P^*$.

In general, we quantify the correlation between two parameters $x$, $y$ by means of the Pearson correlation coefficient

\begin{equation}
r(x,y) = \frac{\sum_i (x_i-\langle x\rangle)(y_i-\langle y\rangle)}{\sqrt{\sum_i (x_i-\langle x\rangle)^2\sum_i (y_i-\langle y\rangle)^2}}
\label{eq_corr}
\end{equation}

\begin{figure*}
\centering
\includegraphics[height=8.9cm,angle=-90]{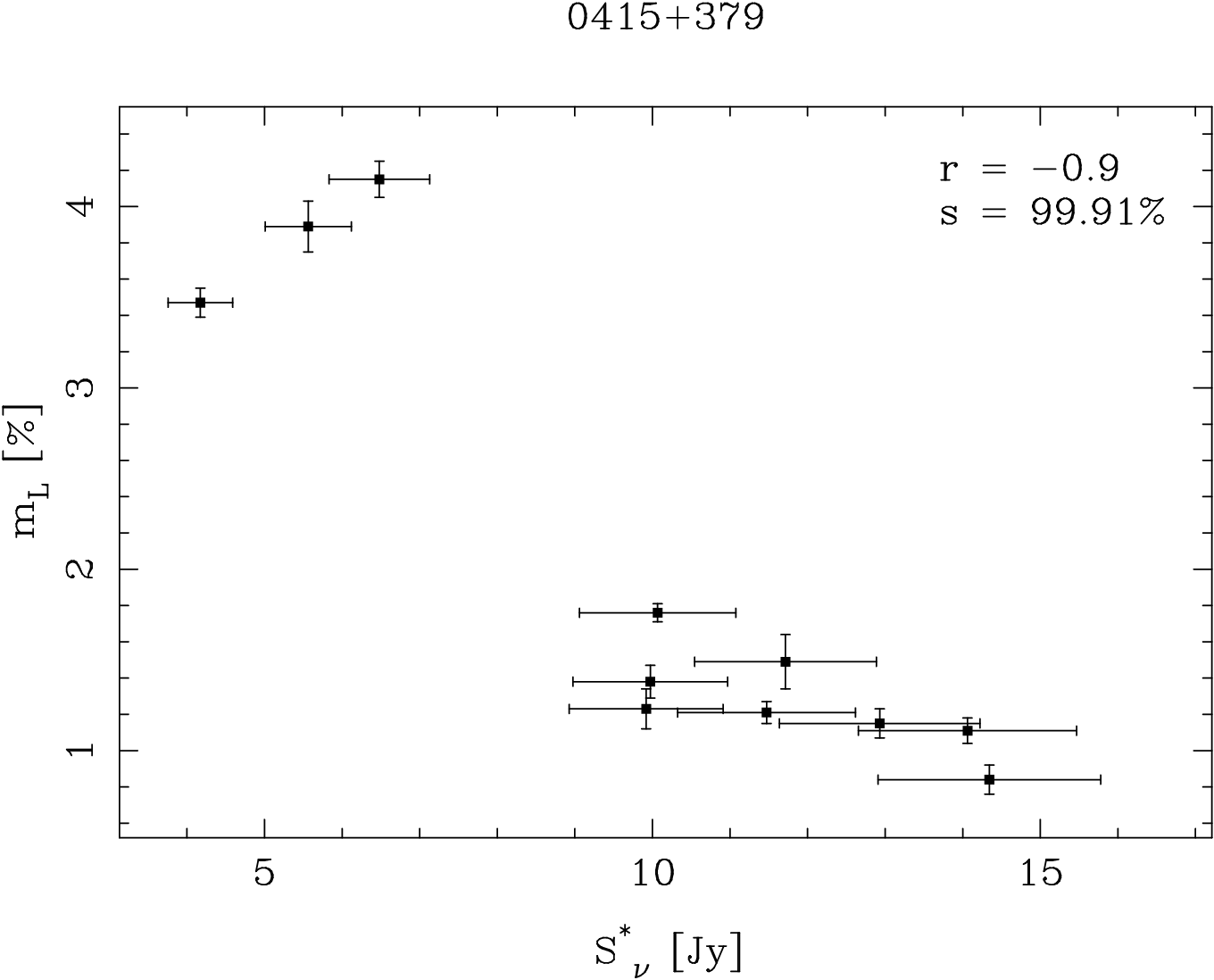}
\hspace{3mm}
\includegraphics[height=8.9cm,angle=-90]{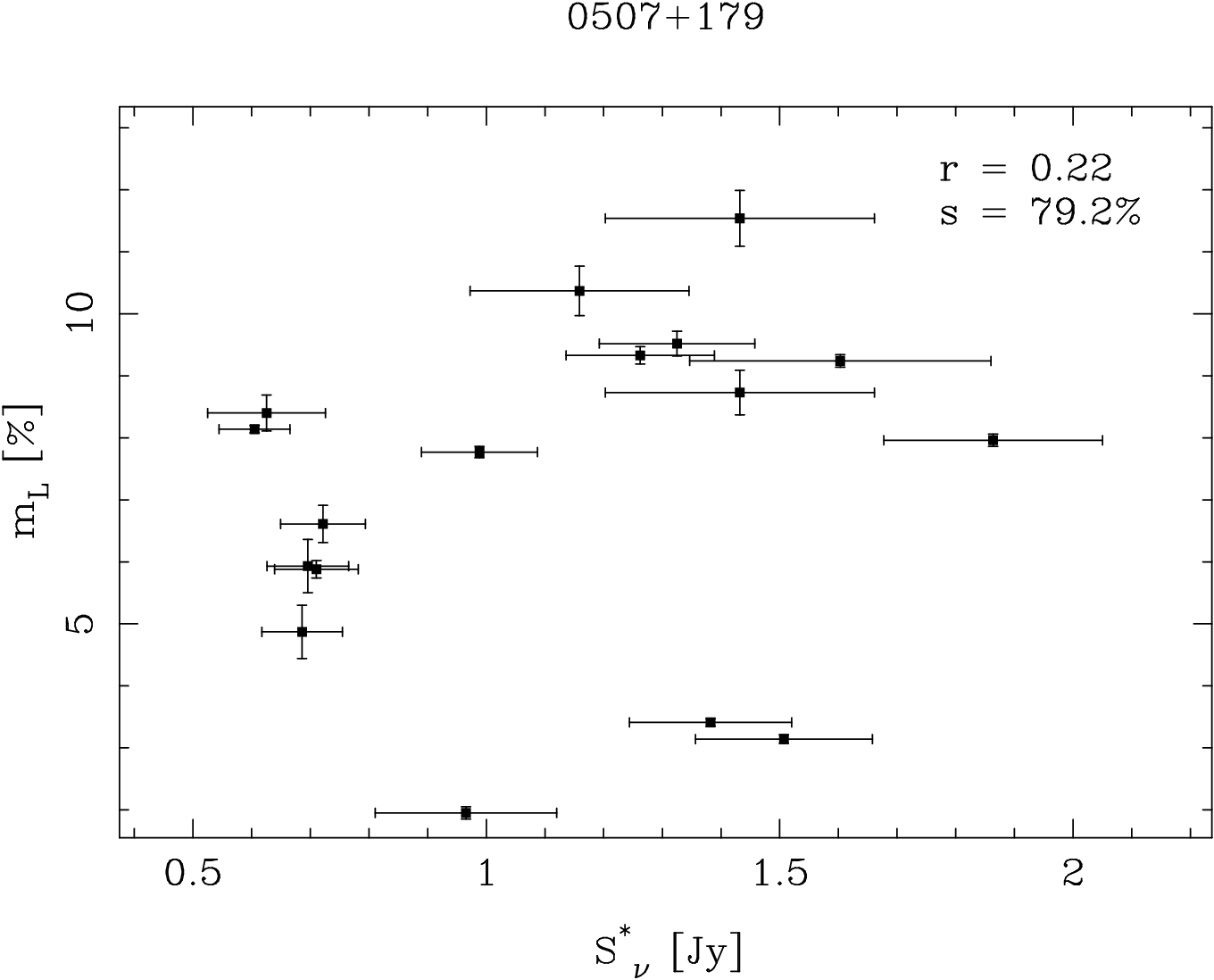} \\
\vspace{5mm}
\includegraphics[height=8.9cm,angle=-90]{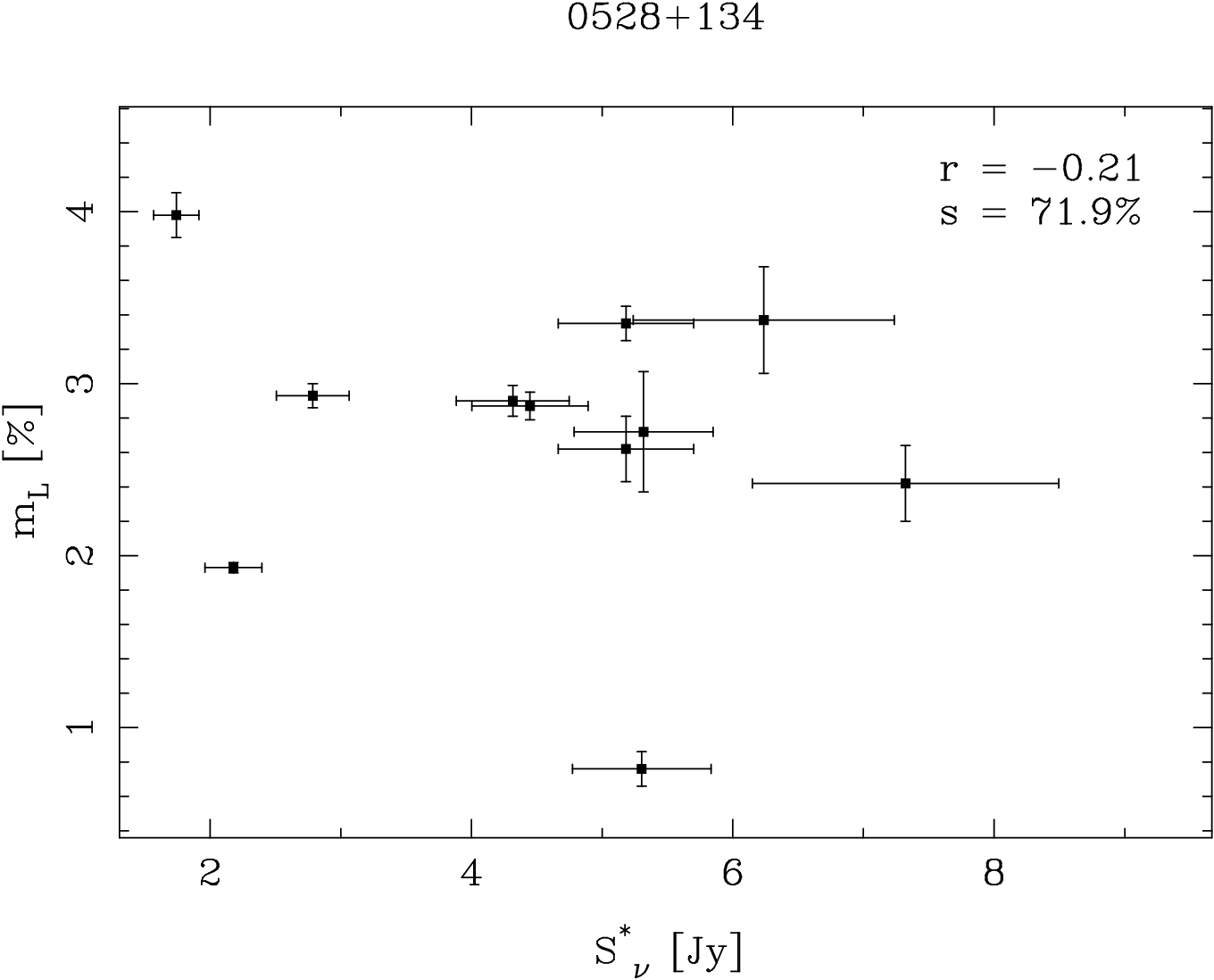}
\hspace{3mm}
\includegraphics[height=8.9cm,angle=-90]{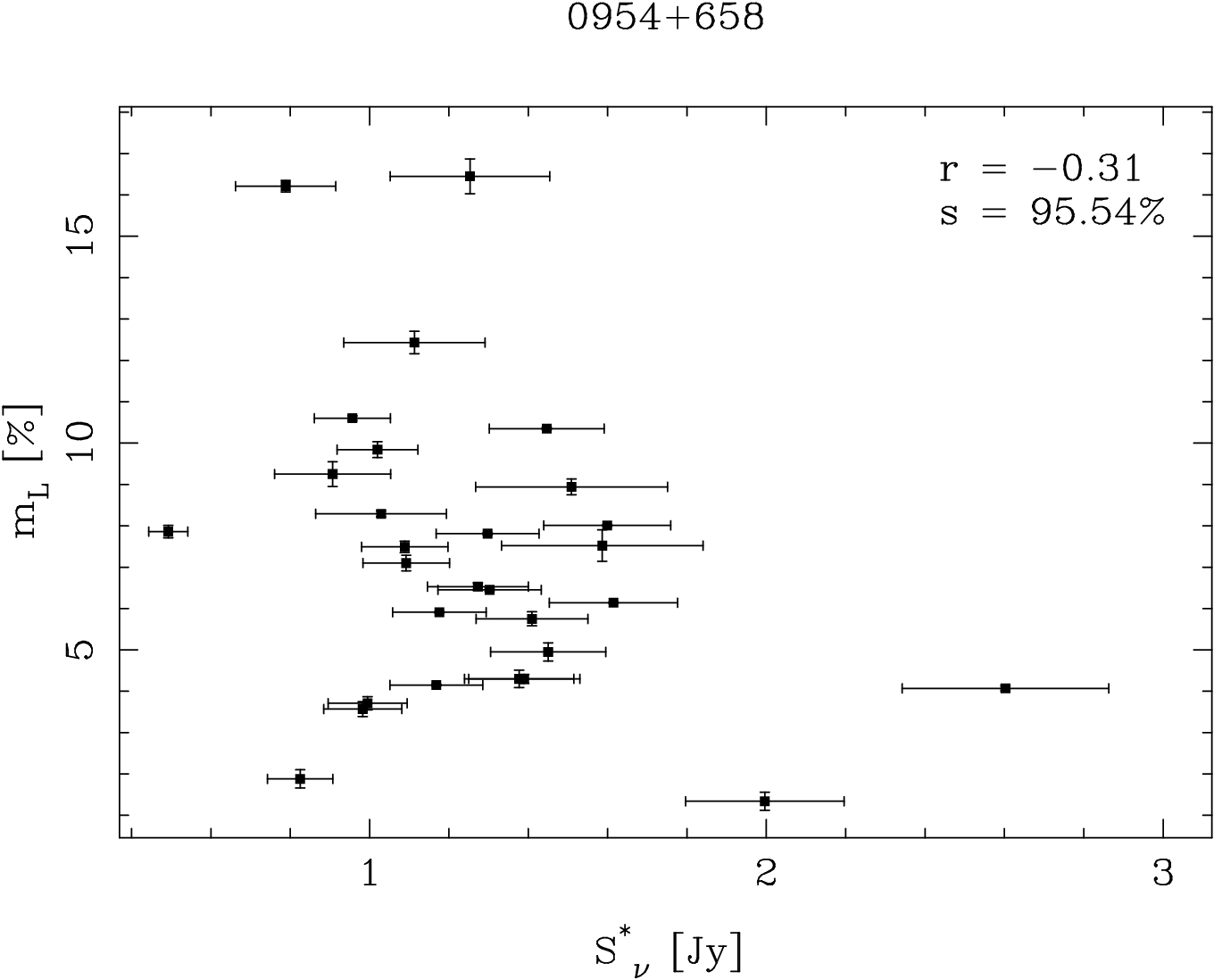} \\
\vspace{5mm}
\includegraphics[height=8.9cm,angle=-90]{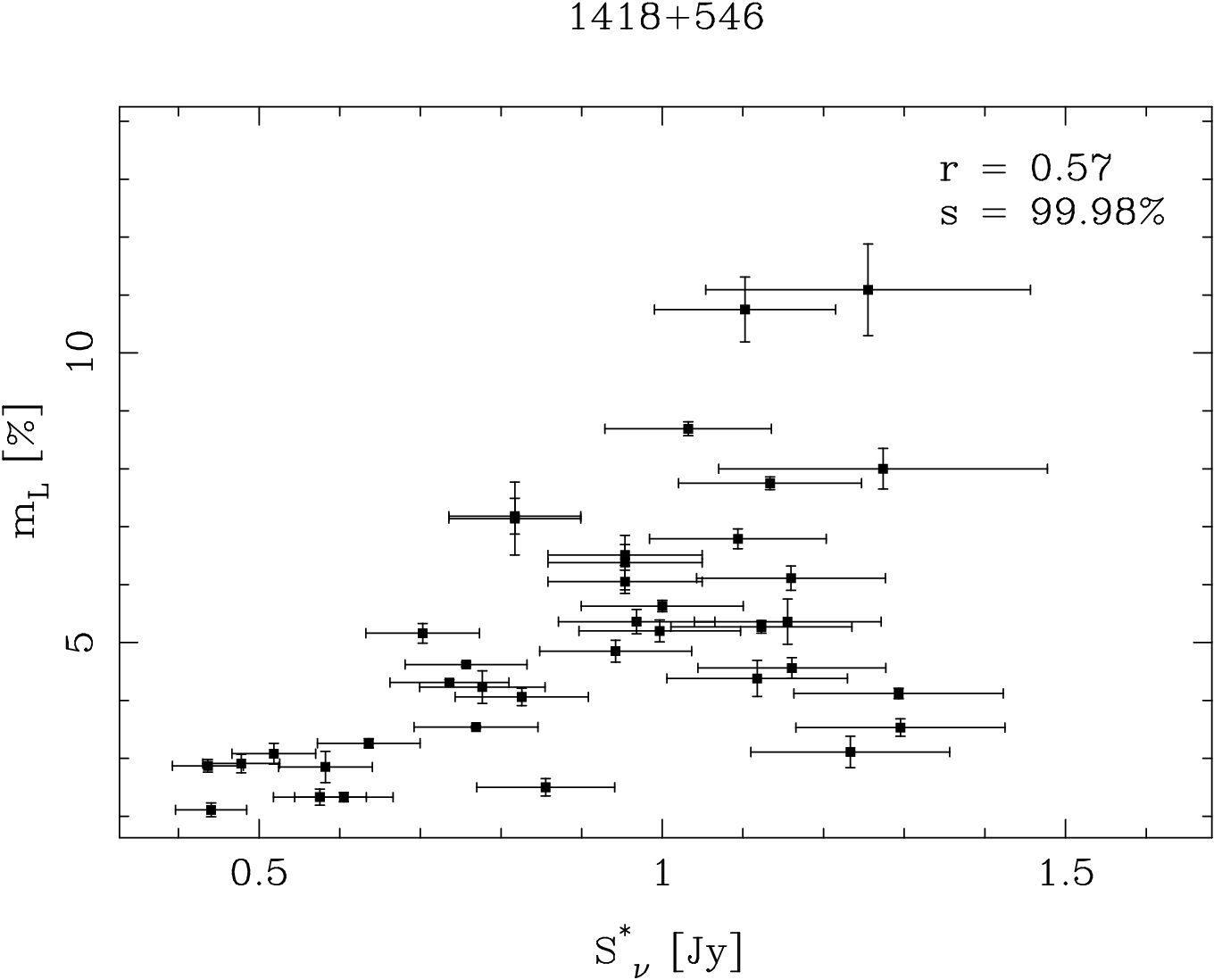}
\hspace{3mm}
\includegraphics[height=8.9cm,angle=-90]{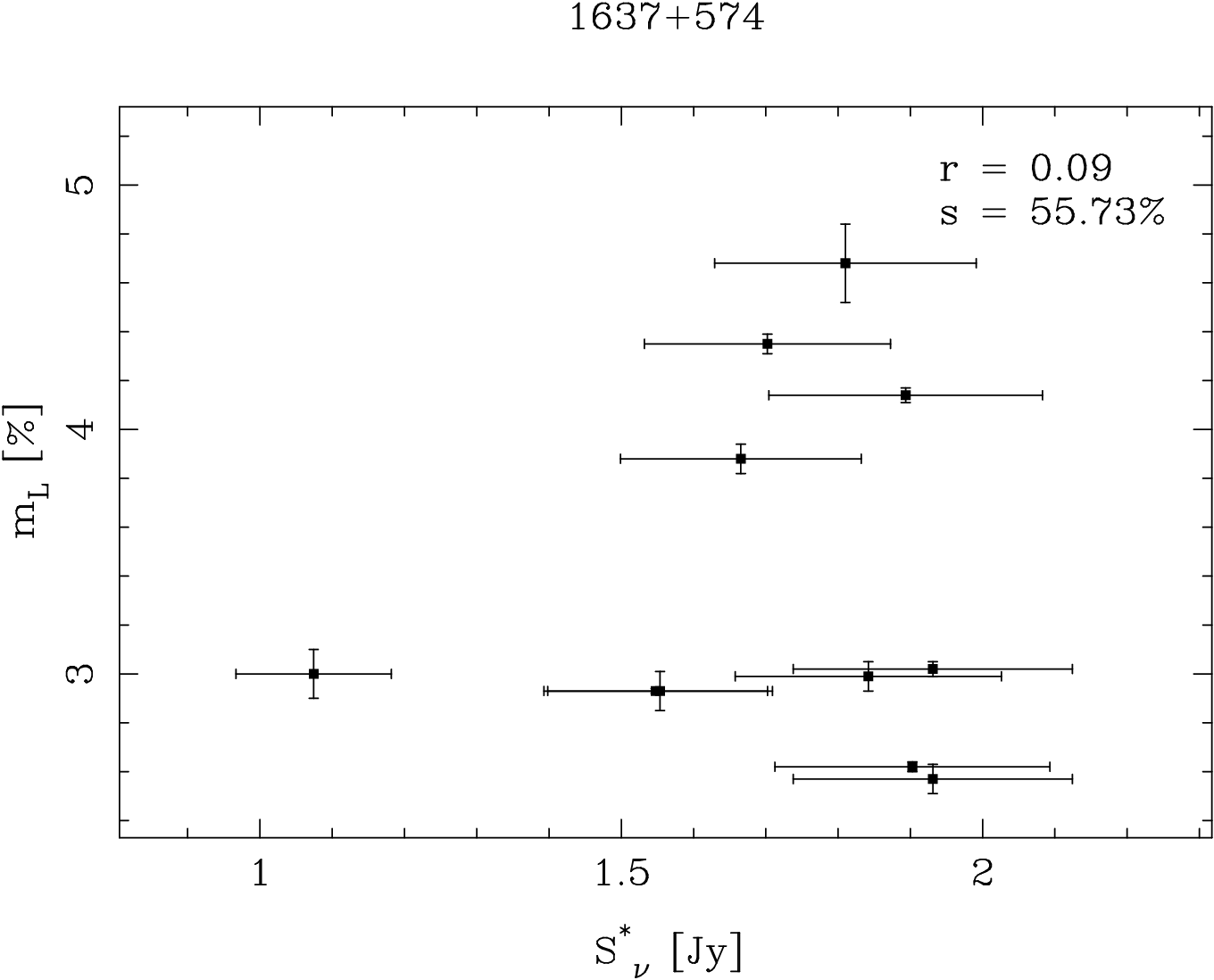}
\caption{Degree of linear polarization $m_L$ as function of rescaled flux density $S^{*}_{\nu}$. Please note the different axis scales. Errorbars denote $1\sigma$ errors. In each $m_L-S^*_{\nu}$ diagram we give the Pearson correlation coefficient $r$. We also quote (in units of \%) for each value of $r$ the significance $s=1-p$ with $p$ being the false-alarm probability for the null hypothesis ``the true value of $r$ is zero''.}
\label{fig_pol-vs-flux}
\end{figure*}

\begin{figure*}
\centering
\includegraphics[height=8.9cm,angle=-90]{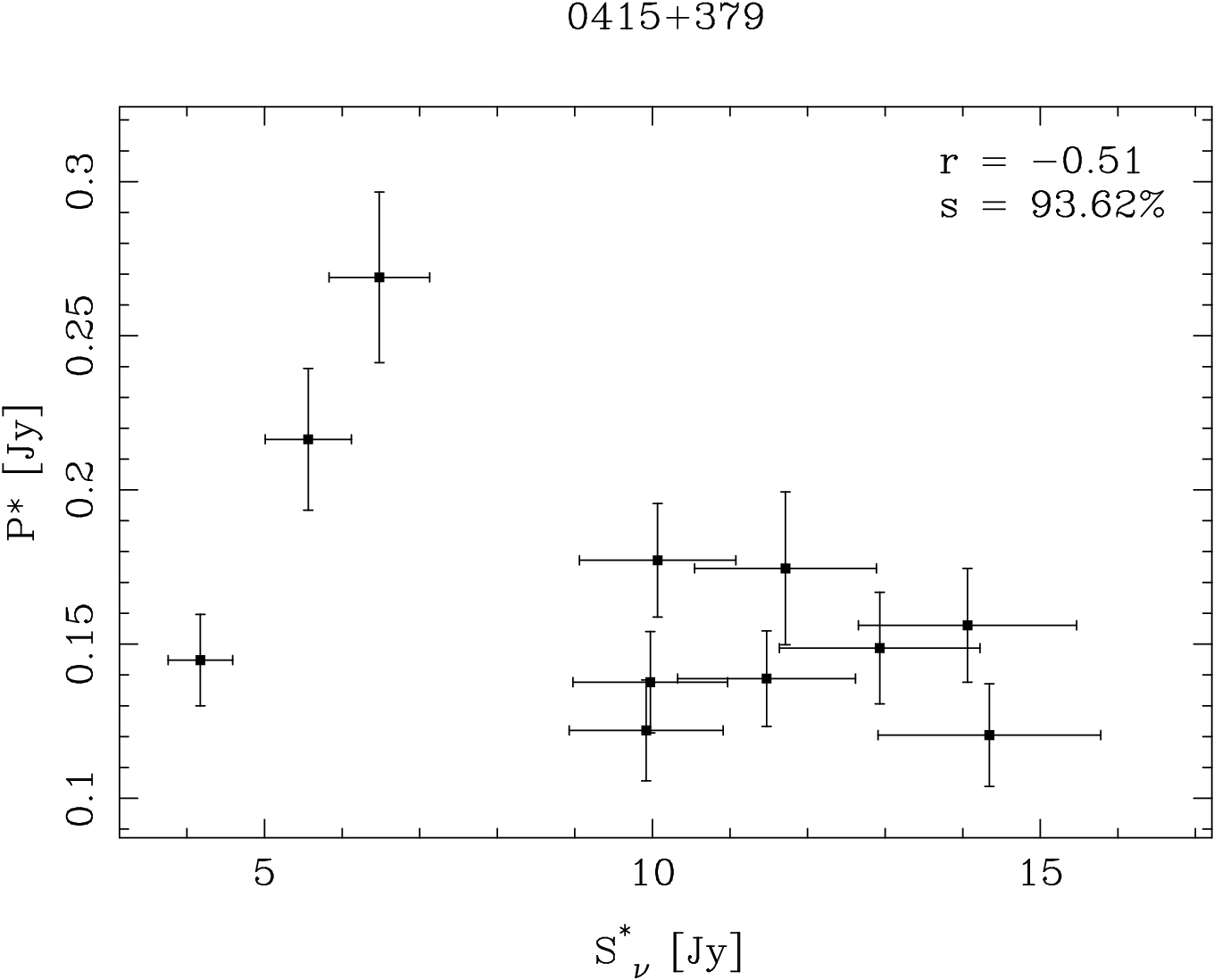}
\hspace{3mm}
\includegraphics[height=8.9cm,angle=-90]{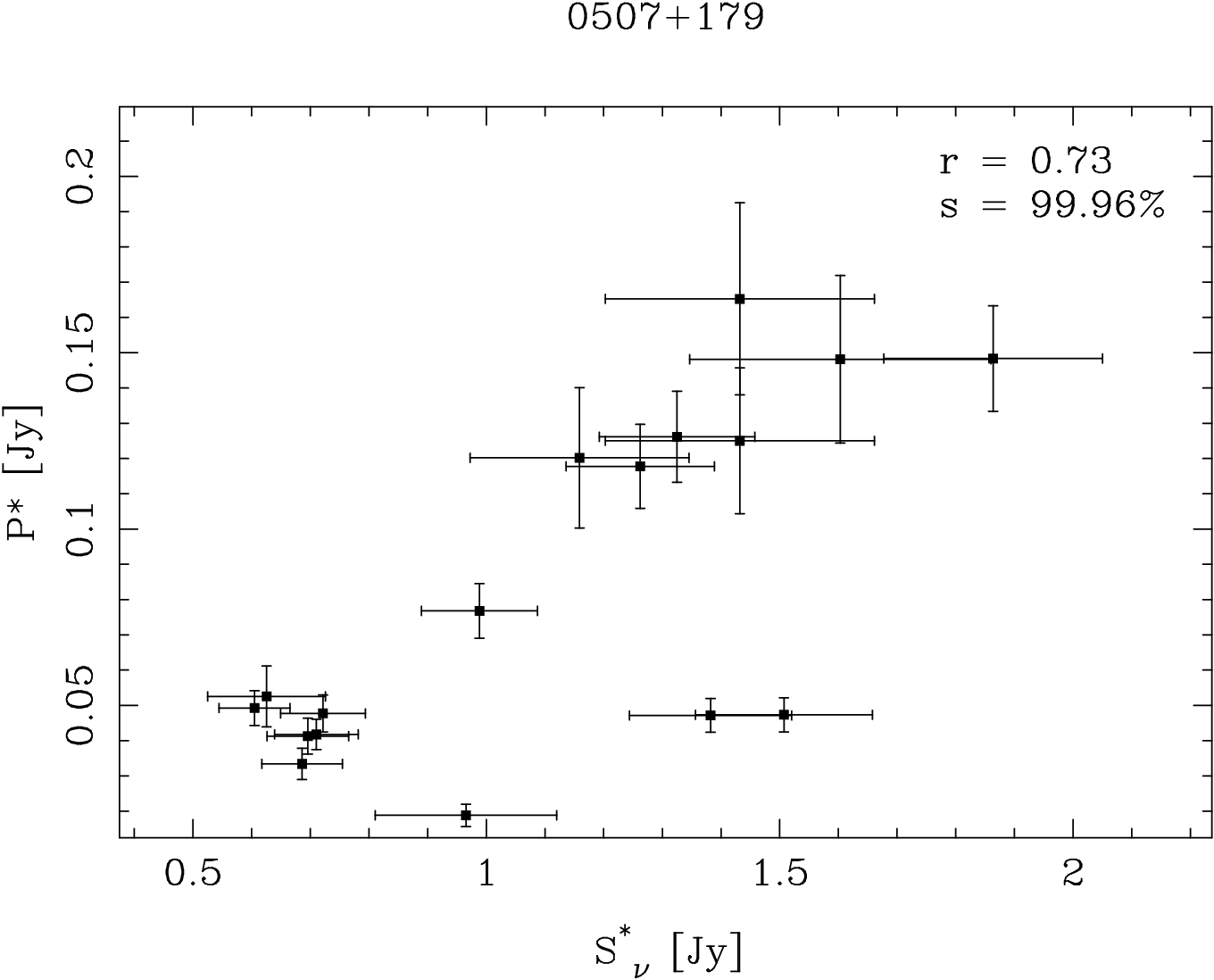} \\
\vspace{5mm}
\includegraphics[height=8.9cm,angle=-90]{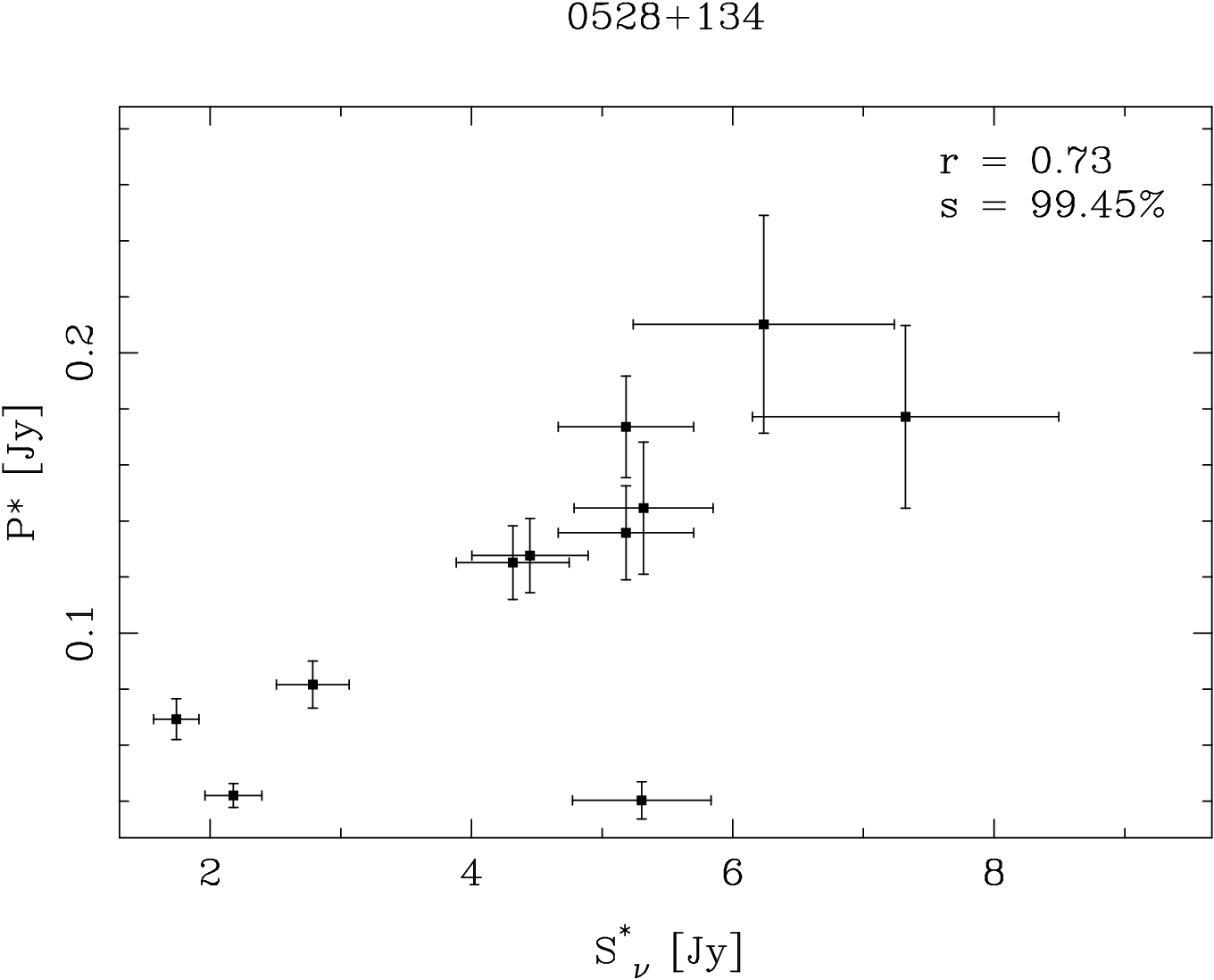}
\hspace{3mm}
\includegraphics[height=8.9cm,angle=-90]{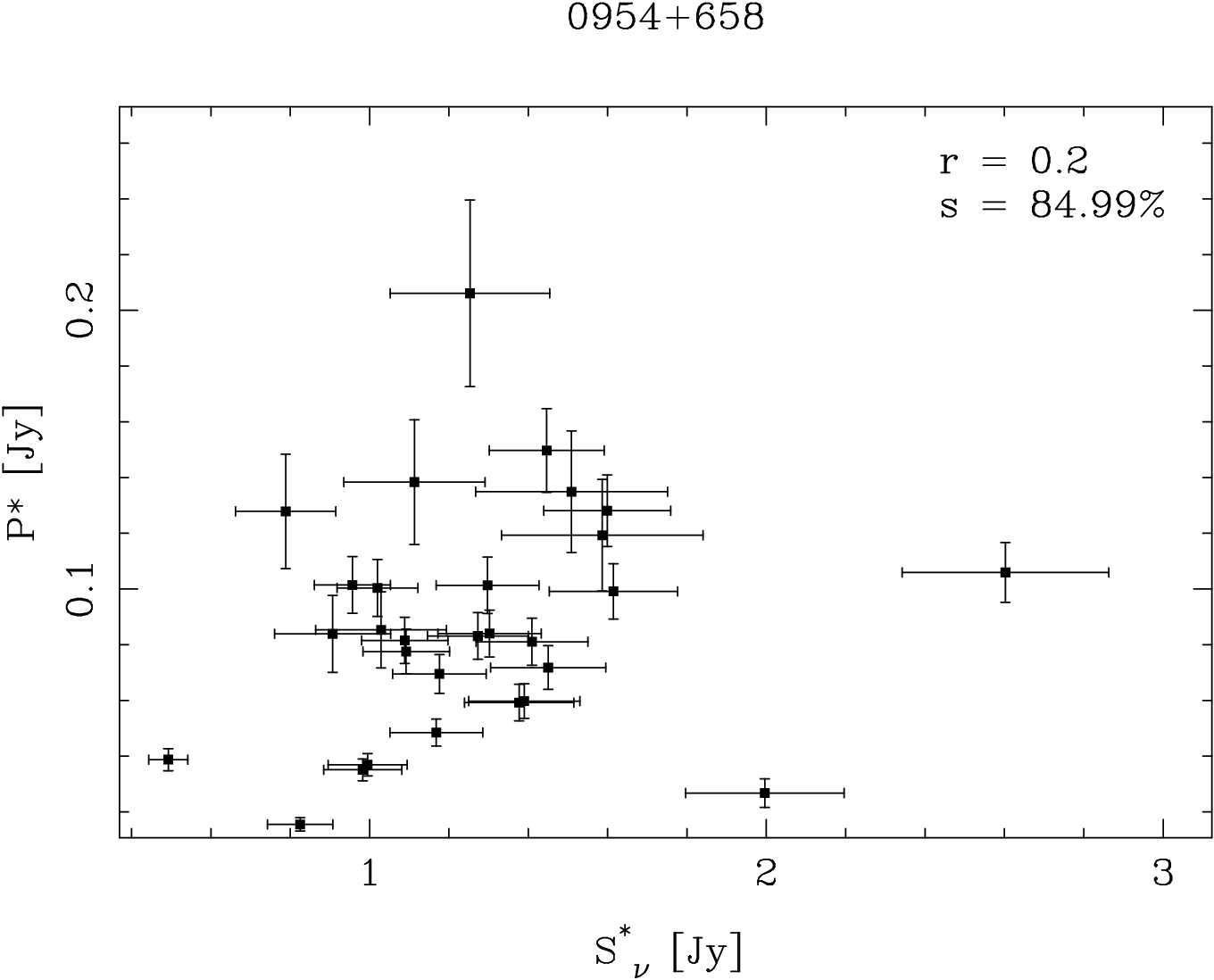} \\
\vspace{5mm}
\includegraphics[height=8.9cm,angle=-90]{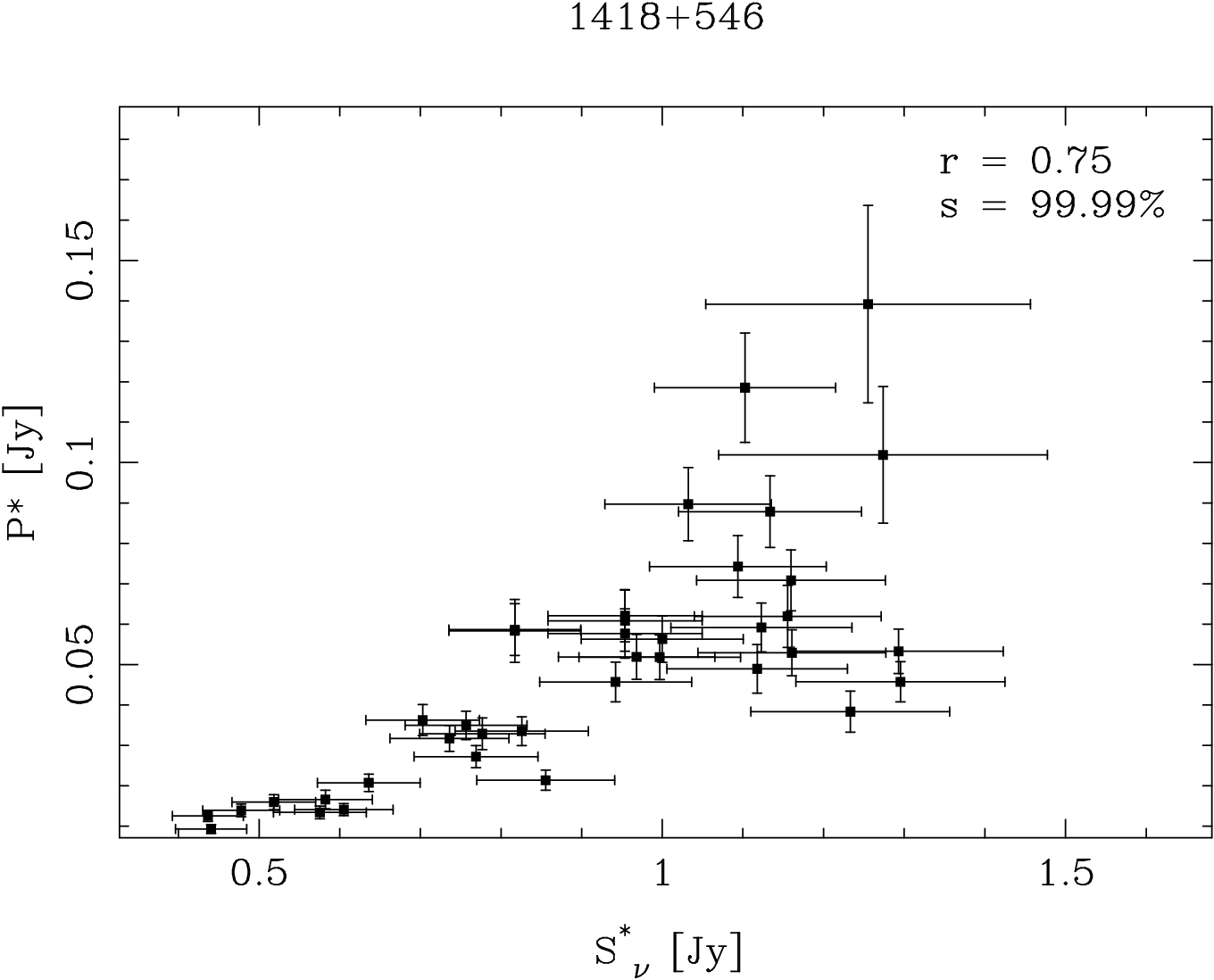}
\hspace{3mm}
\includegraphics[height=8.9cm,angle=-90]{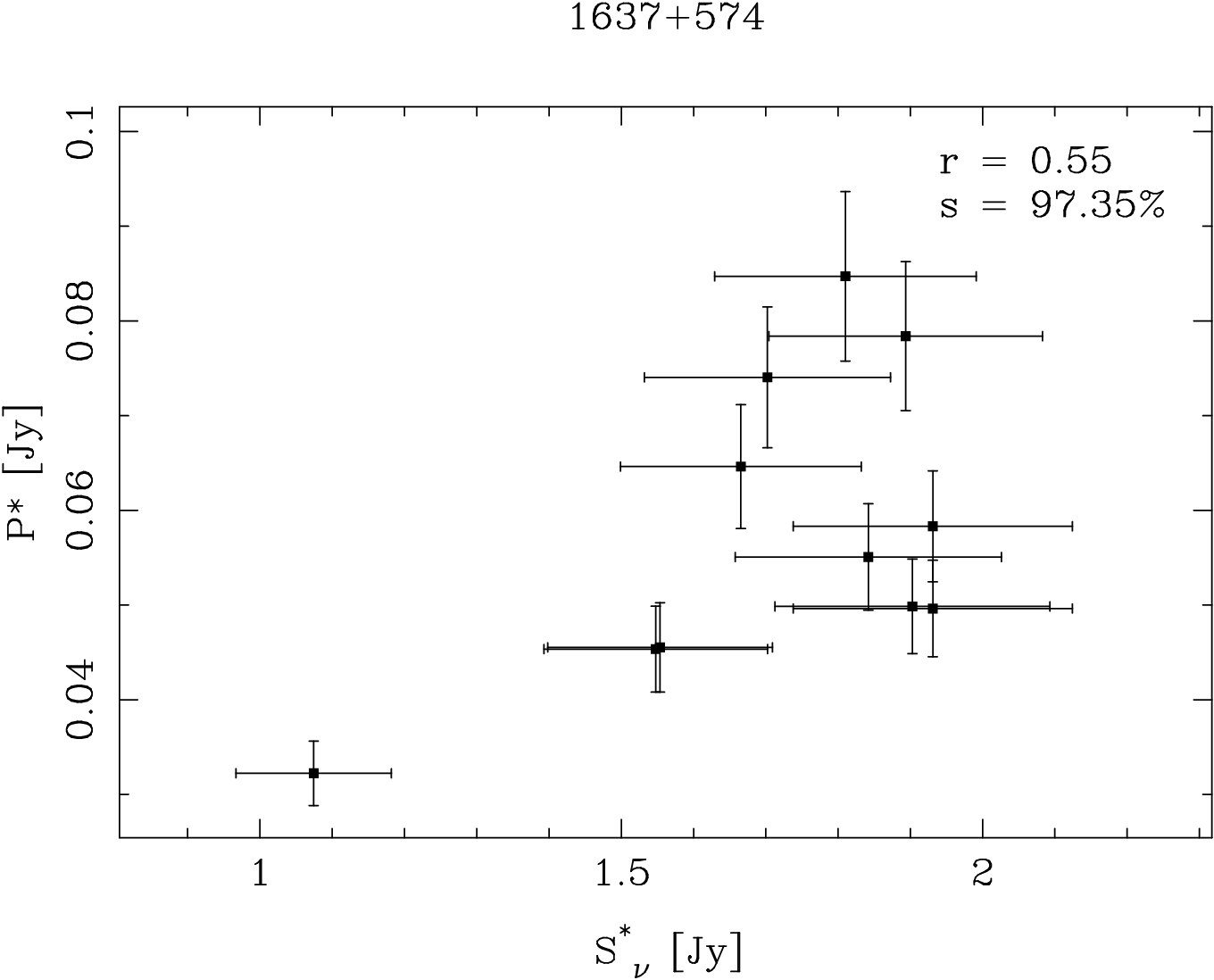}
\caption{Rescaled linearly polarized flux $P^*$ vs. rescaled flux density $S^*_{\nu}$. Please note the different axis scales. Errorbars denote $1\sigma$ errors. We quote correlation coefficients $r$ and significance levels $s$ like in Fig.~\ref{fig_pol-vs-flux}. }
\label{fig_polflux-vs-flux}
\end{figure*}

\begin{figure*}
\centering
\includegraphics[height=8.9cm,angle=-90]{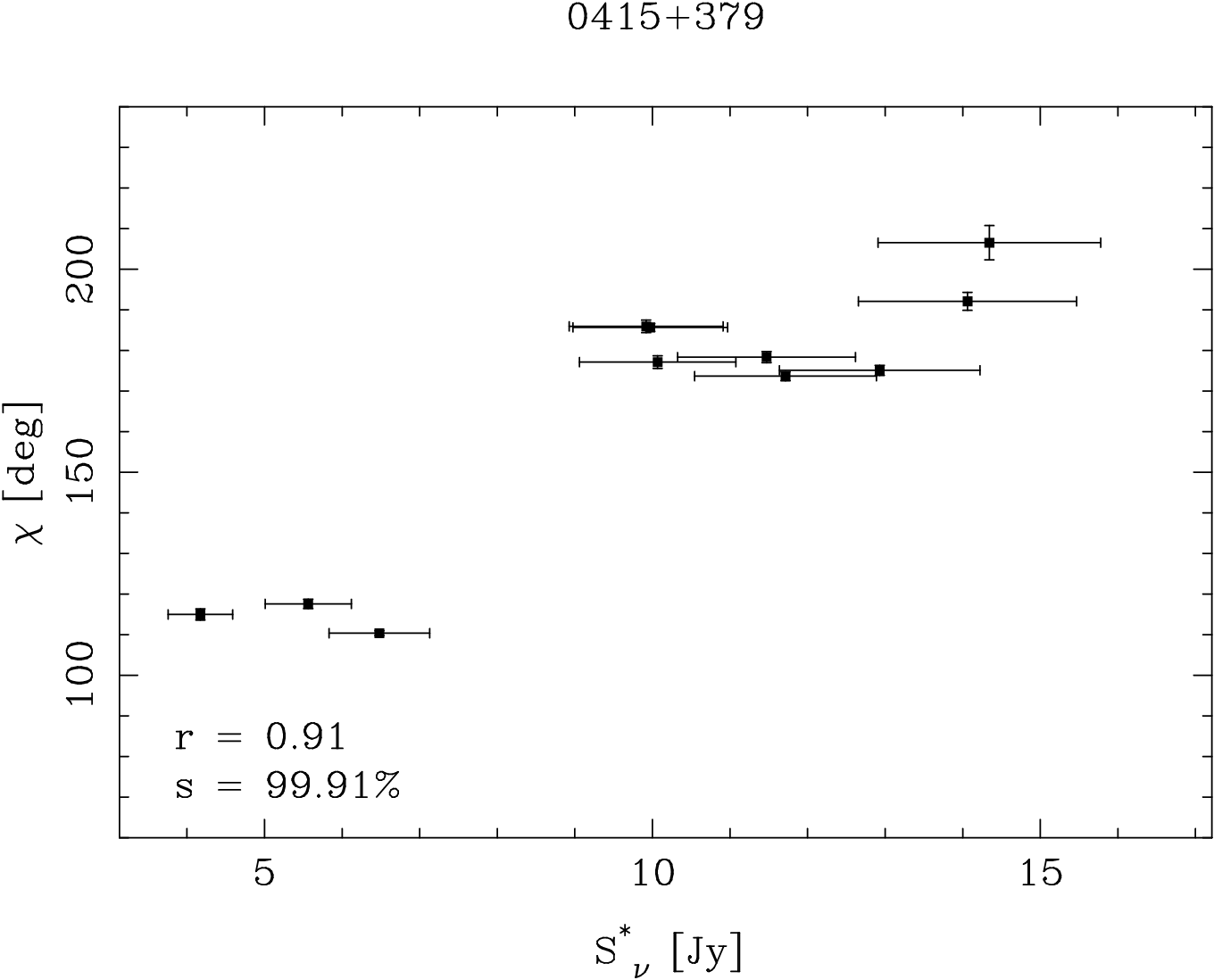}
\hspace{3mm}
\includegraphics[height=8.9cm,angle=-90]{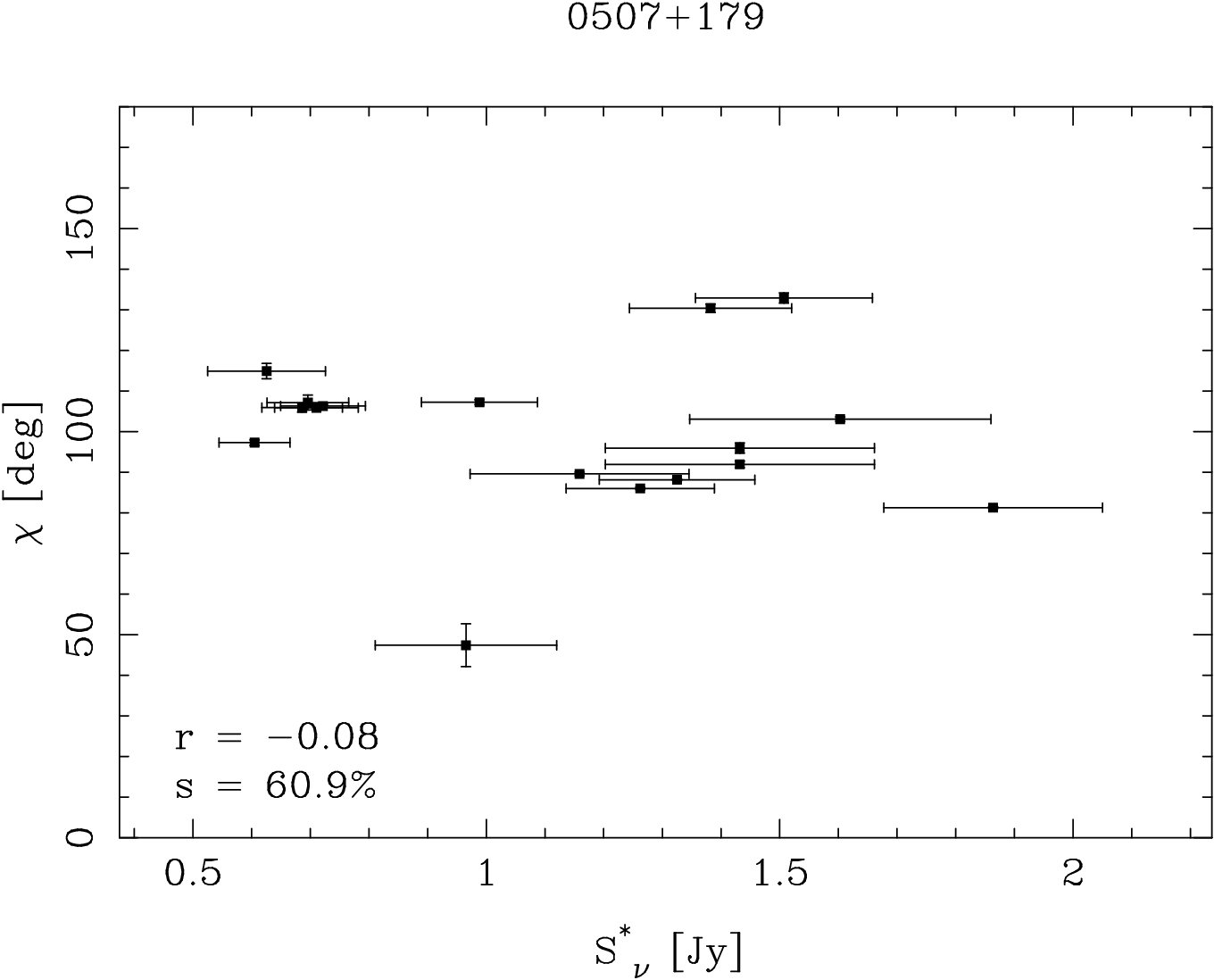} \\
\vspace{5mm}
\includegraphics[height=8.9cm,angle=-90]{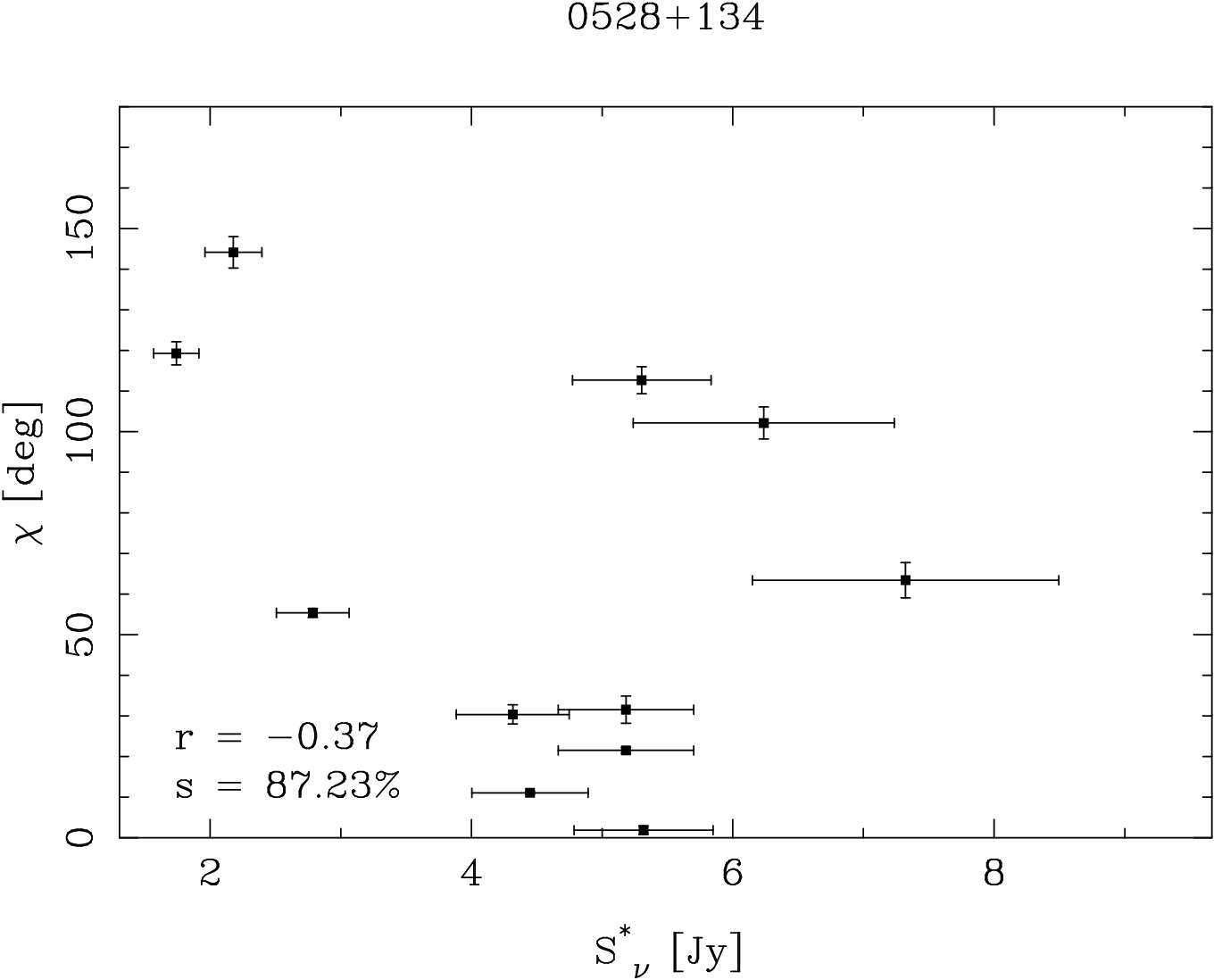}
\hspace{3mm}
\includegraphics[height=8.9cm,angle=-90]{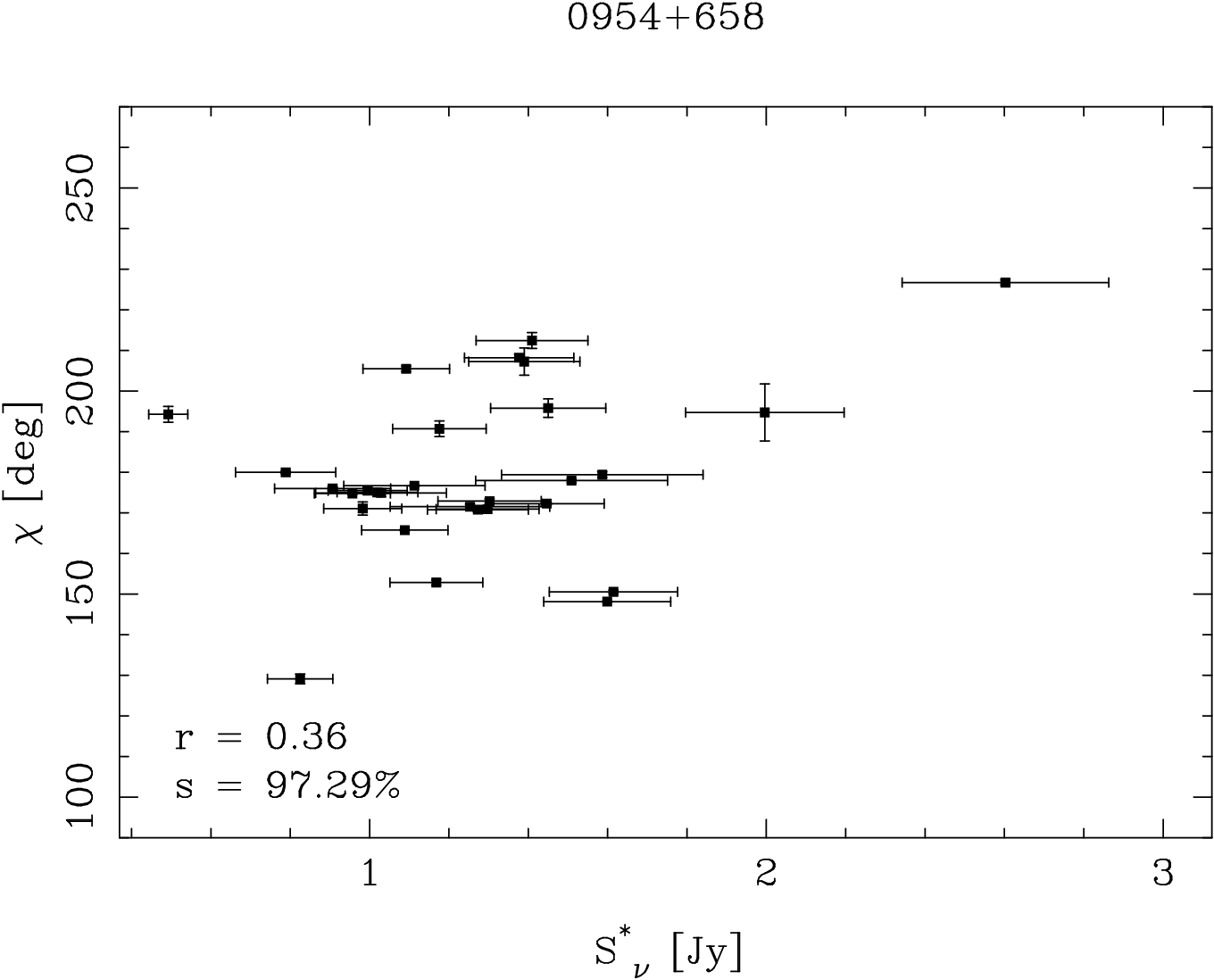} \\
\vspace{5mm}
\includegraphics[height=8.9cm,angle=-90]{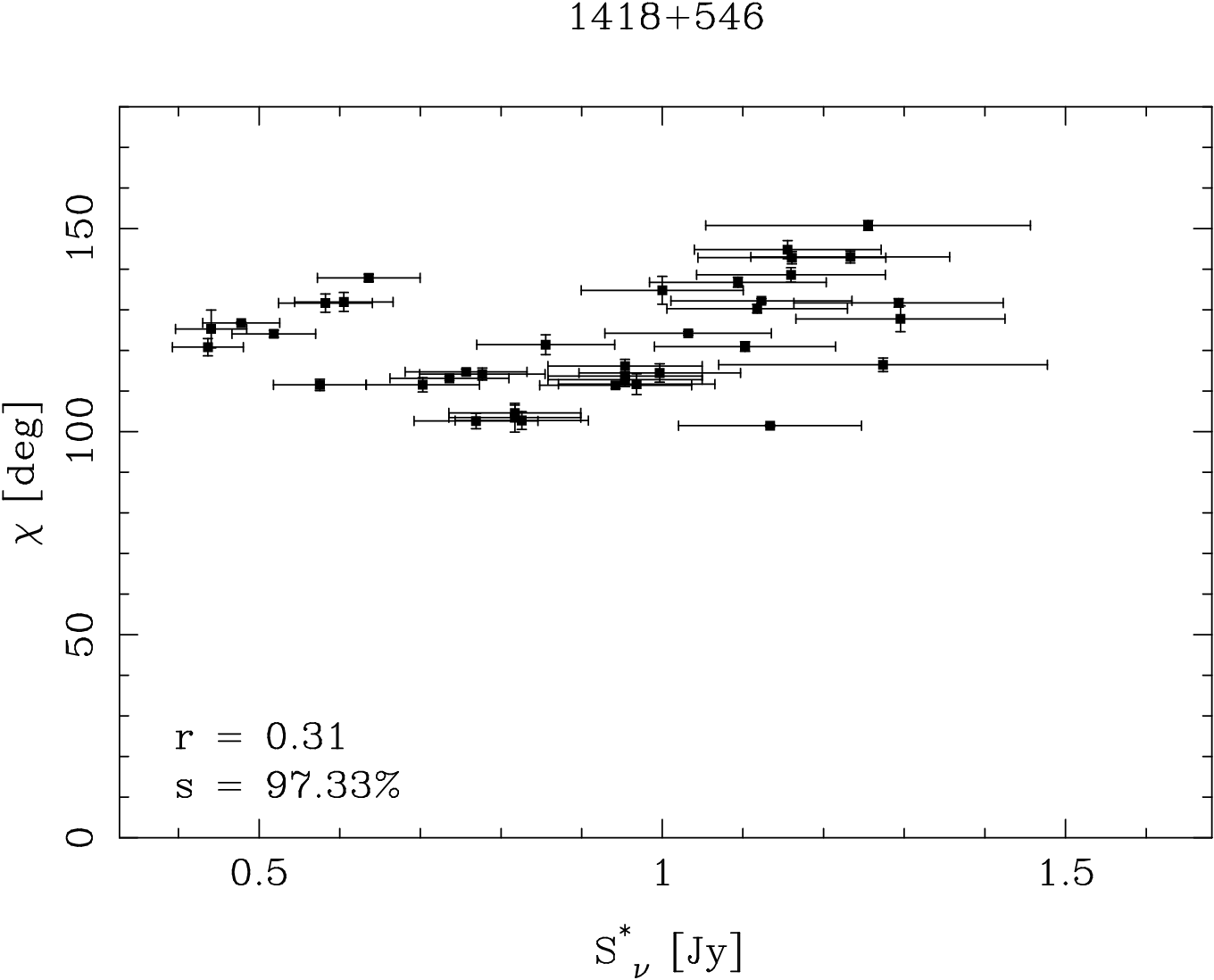}
\hspace{3mm}
\includegraphics[height=8.9cm,angle=-90]{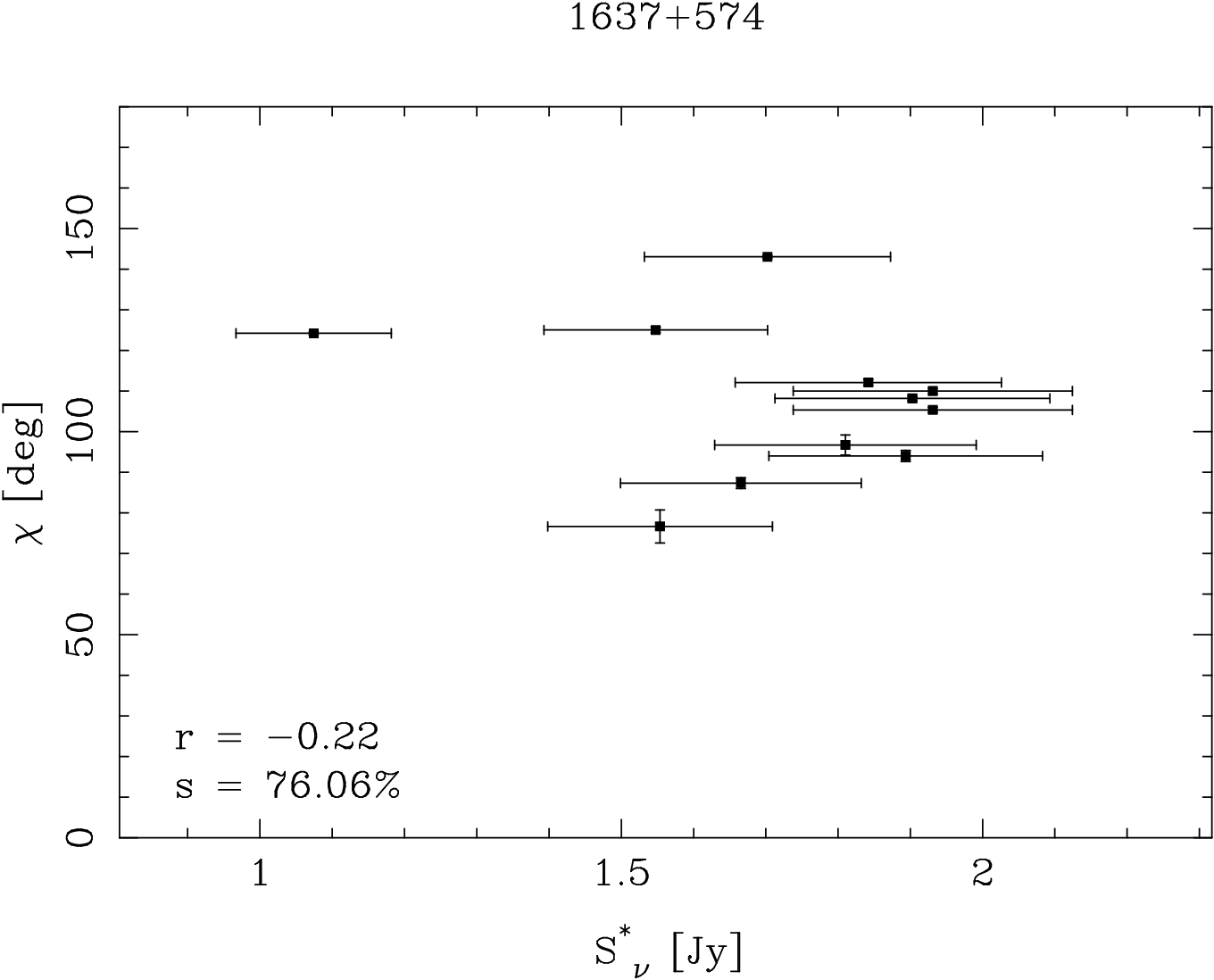}
\caption{Polarization angle $\chi$ vs. rescaled flux density $S^*_{\nu}$. Please note the different axis scales. Polarization angles are restricted to the interval $[0^{\circ},180^{\circ}]$; for 0415+379 and 0954+658 these intervals have been shifted to resolve $0^{\circ}/180^{\circ}$ ambiguities. \rm Errorbars denote $1\sigma$ errors. We quote correlation coefficients $r$ and significance levels $s$ like in Fig.~\ref{fig_pol-vs-flux}. }
\label{fig_chi-vs-flux}
\end{figure*}

\noindent
(e.g., Snedecor \& Cochran \cite{snedecor1980}). By construction, $r$ values of $(-1)$ +1 correspond to perfect (anti)correlation, a value of 0 means absence of any correlation. For each AGN, we computed correlation coefficients with $x$ being $S^*_{\nu}$ and $y$ being either $m_L$ or $P^*$ or $\chi$, thus providing\footnote{We note that, mathematically, $r(S^*_{\nu},m_L)$ and $r(S^*_{\nu},P^*)$ are degenerate because $P^* = m_L S^*_{\nu}$. However, the scatter in our data makes it necessary to analyze both parameters because all values of $r$ come with substantial uncertainties.} $r(S^*_{\nu},m_L)$, $r(S^*_{\nu},P^*)$, and $r(S^*_{\nu},\chi)$. We present the distributions and correlations of our parameters in Figs. \ref{fig_pol-vs-flux}, \ref{fig_polflux-vs-flux}, and \ref{fig_chi-vs-flux}.

Inspecting the distributions in Figs. \ref{fig_pol-vs-flux}, \ref{fig_polflux-vs-flux}, and \ref{fig_chi-vs-flux} shows that the presence or absence of correlations is far from obvious due to the substantial scatter within our data. To permit quantitative conclusions, we calculated significances by means of a permutation test. For each correlation coefficient $r(x,y)$ we computed a sample of modified coefficients $r(x,y')$ with $y'$ being a copy of the initial parameter vector $y$ with its elements rearranged in random order. Because the pairs $(x,y')$ are drawn randomly, the expected value for $r(x,y')$ is zero. From a total of 10\,000 random realizations we determined the fraction $p$ for which
\\

$r(x,y')\geq r(x,y)$  ~~~ if $r(x,y)>0$ ,

$r(x,y')\leq r(x,y)$  ~~~ if $r(x,y)\leq0$ .
\\
\\
The fraction $p$ is thus the false alarm probability, i.e. the probability that a value $r(x,y)\neq0$ is produced  by a statistical fluctuation. Accordingly, our randomization test probes the null hypothesis ``the parameters $x$ and $y$ are intrinsically uncorrelated''. In the following discussion as well as in Figs. \ref{fig_pol-vs-flux}, \ref{fig_polflux-vs-flux}, and \ref{fig_chi-vs-flux}, we quote significance levels $s=1-p$ in units of \%.

\subsection{Relations between polarization and frequency}

\begin{figure*}
\centering
\includegraphics[height=8.9cm,angle=-90]{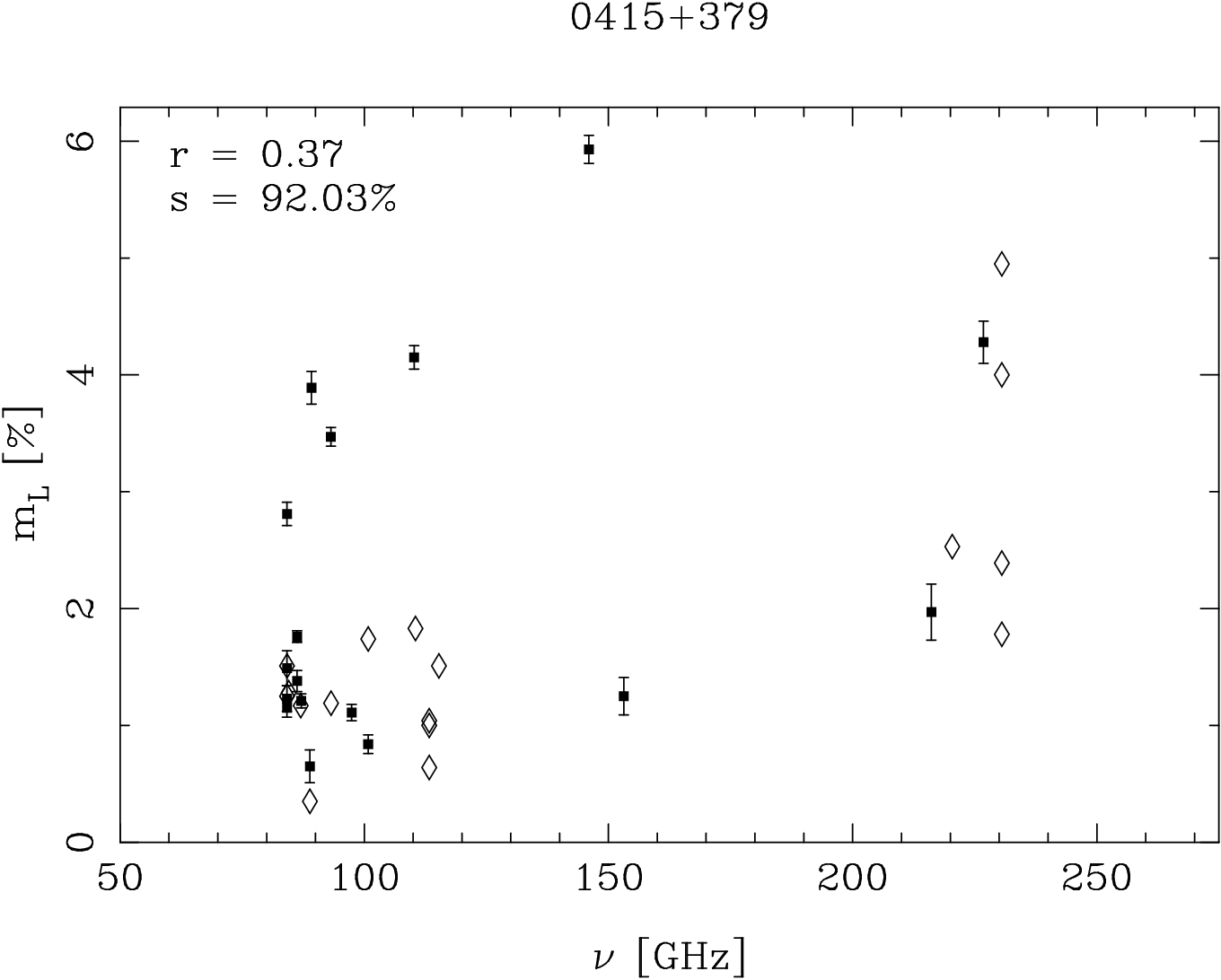}
\hspace{3mm}
\includegraphics[height=8.9cm,angle=-90]{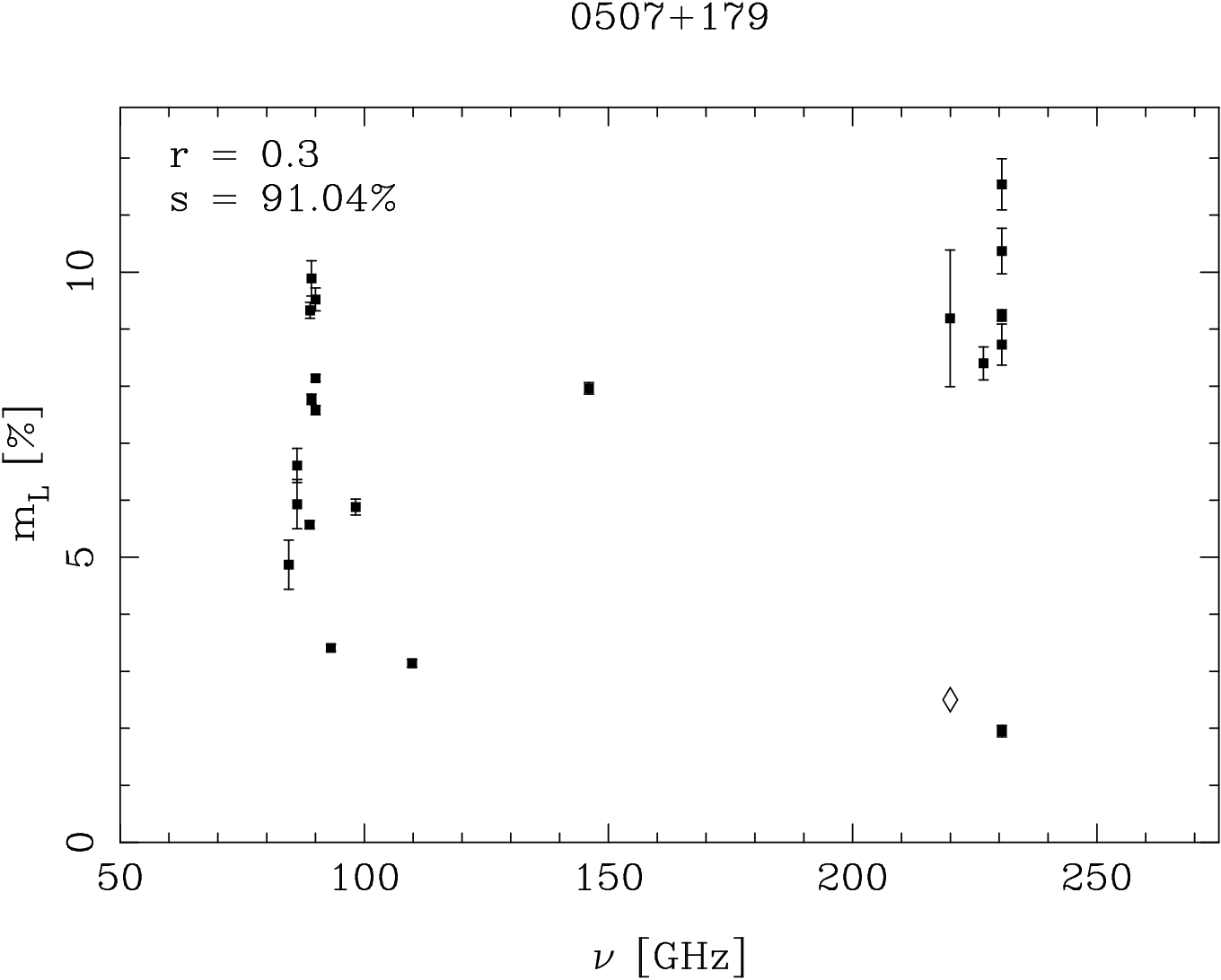} \\
\vspace{5mm}
\includegraphics[height=8.9cm,angle=-90]{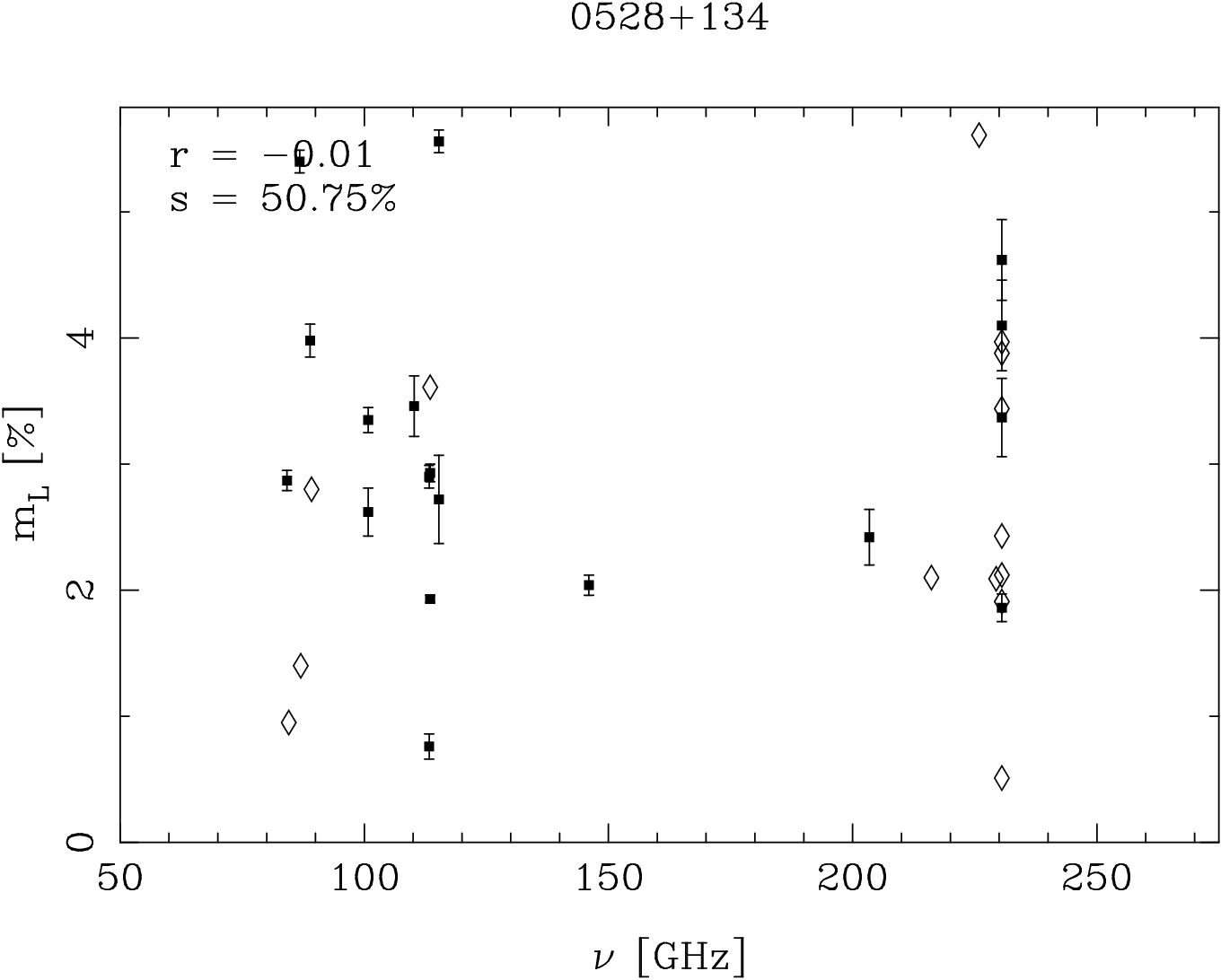}
\hspace{3mm}
\includegraphics[height=8.9cm,angle=-90]{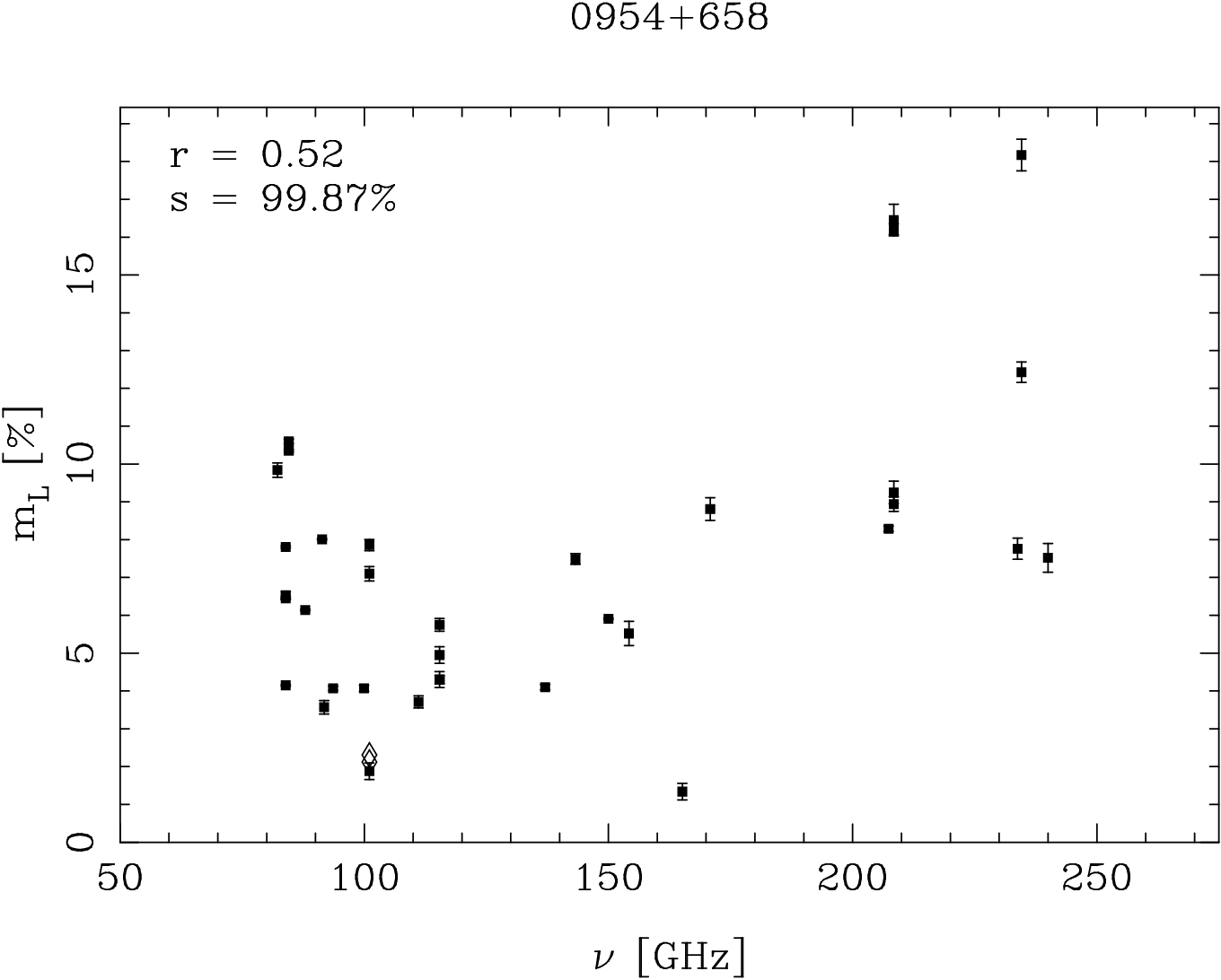} \\
\vspace{5mm}
\includegraphics[height=8.9cm,angle=-90]{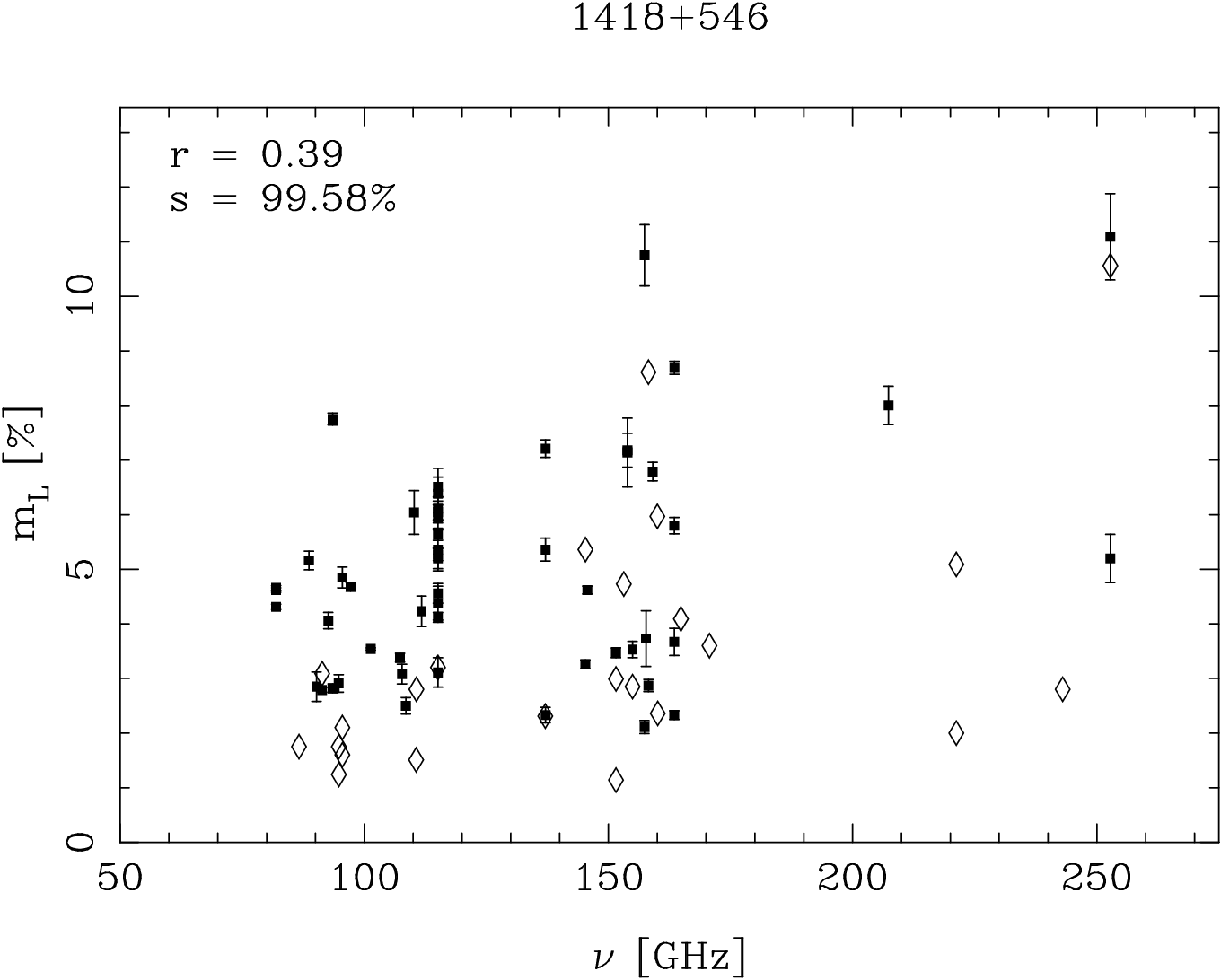}
\hspace{3mm}
\includegraphics[height=8.9cm,angle=-90]{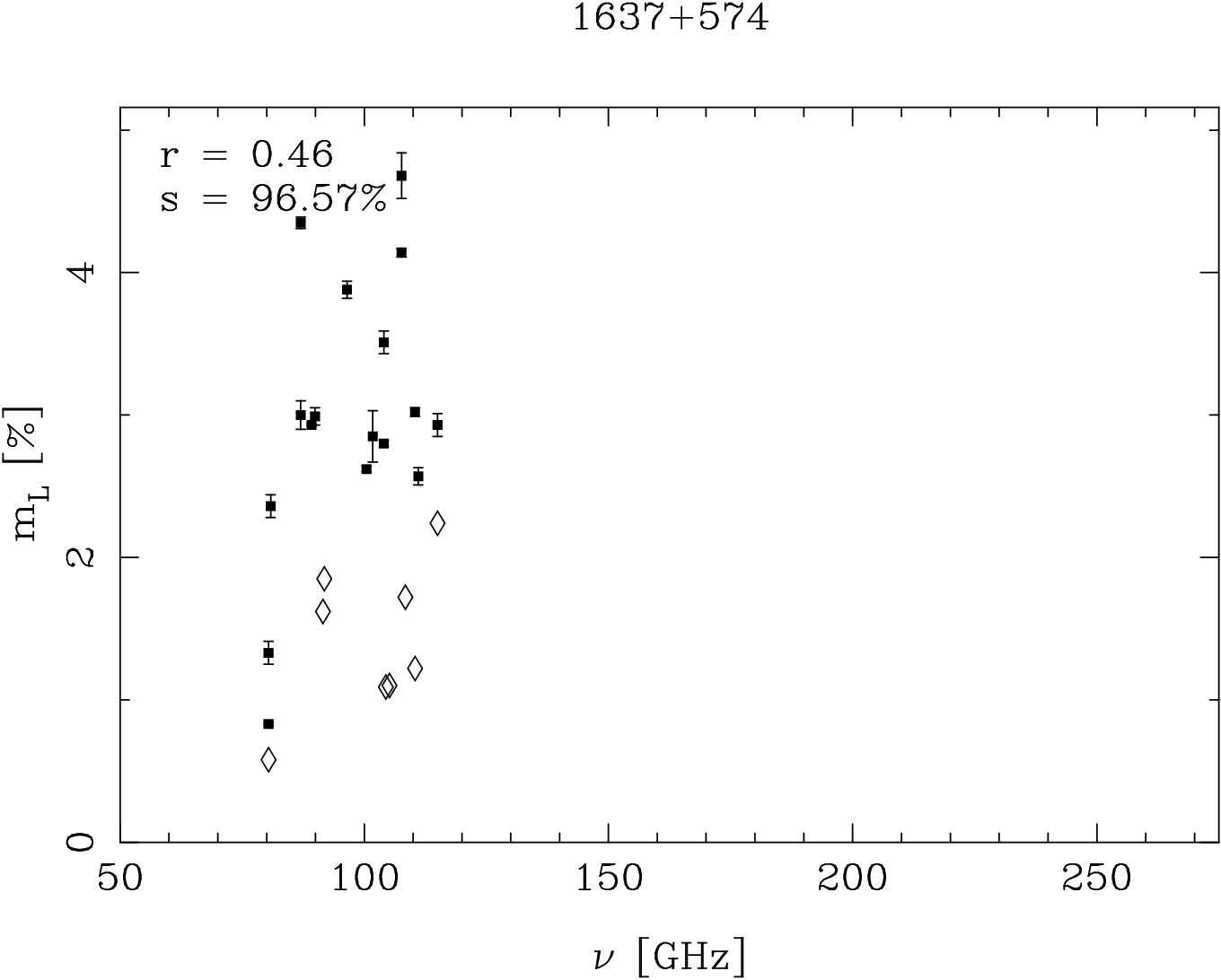}
\caption{Degree of linear polarization $m_L$ as function of observing frequency $\nu$. Squares with errorbars denote polarization measurements; diamonds indicate $3\sigma$ upper limits. Please note the different $m_L$ axis scales. Errorbars denote $1\sigma$ errors. We quote correlation coefficients $r$ and significance levels $s$ like in Fig.~\ref{fig_pol-vs-flux}.}
\label{fig_pol-vs-nu}
\end{figure*}

Because our observations are resolved in frequency, our data permit probing correlations between polarization properties and frequency. In Fig.~\ref{fig_pol-vs-nu}, we present the degree of linear polarization $m_L$ as function of observing frequency $\nu$. We quote Pearson correlation coefficients and their significances computed analogously to the procedure outlined in Sect.~3.2. As already noted in Paper~I (see especially the top panel of Fig.~6 therein), a positive correlation -- if any -- may be expected already because of observational bias: measurements obtained at higher frequencies tend to suffer from reduced signal-to-noise (S/N) ratios. This effect reduces the probability of detecting low values of $m_L$ at high frequencies, resulting in an apparent increase of the average $m_L$ with frequency.

A more promising approach is the analysis of the polarization angle $\chi$. Other than for $m_L$, we are not aware of any mechanism that could introduce a bias affecting the observed values of $\chi$. This is a convenient situation because it permits an analysis of Faraday rotation. For linearly polarized light propagating through a plasma permeated by a magnetic field, the polarization angle is related to the rest-frame wavelength\footnote{At this point, we imply that any Faraday rotation potentially observed is predominantly caused by material located within or nearby the emitter. A justification is provided in Sect.~5.4.} $\lambda_0$ like

\begin{equation}
\chi = {\rm RM} \times \lambda_0^2  ,
\label{eq_chi}
\end{equation}

\noindent
where RM is the rotation measure

\begin{equation}
{\rm \frac{RM}{rad\,m^{-2}}} = 8.1\times10^5 \int_{\rm l.o.s.} \left(\frac{B_{||}}{\rm G}\right) \left(\frac{n_e}{\rm cm^{-3}}\right) {\rm d}\left(\frac{l}{\rm pc}\right)  ,
\label{eq_rm}
\end{equation}

\noindent
with $B_{||}$ being the strength of the magnetic field parallel to the line of sight (l.o.s.), $n_e$ being the electron number density, and $l$ being the coordinate directed along the l.o.s. (e.g., Wilson, Rohlfs \& H\"uttemeister \cite{wilson2010}). Thus, deriving RM permits constraining the l.o.s.-integrated magnetic field strength and particle density of an active galaxy.

\begin{figure*}
\centering
\includegraphics[height=8.9cm,angle=-90]{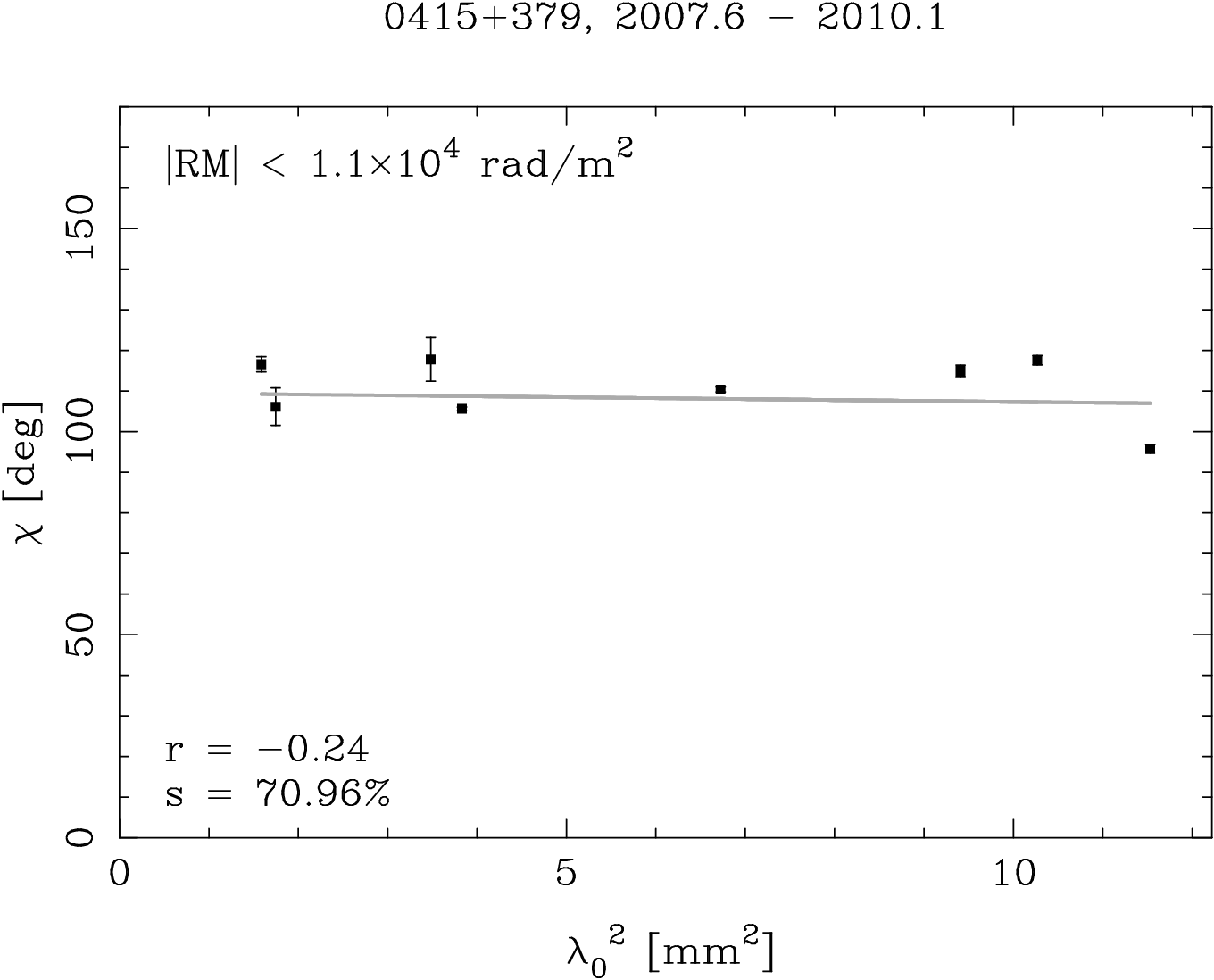}
\hspace{3mm}
\includegraphics[height=8.9cm,angle=-90]{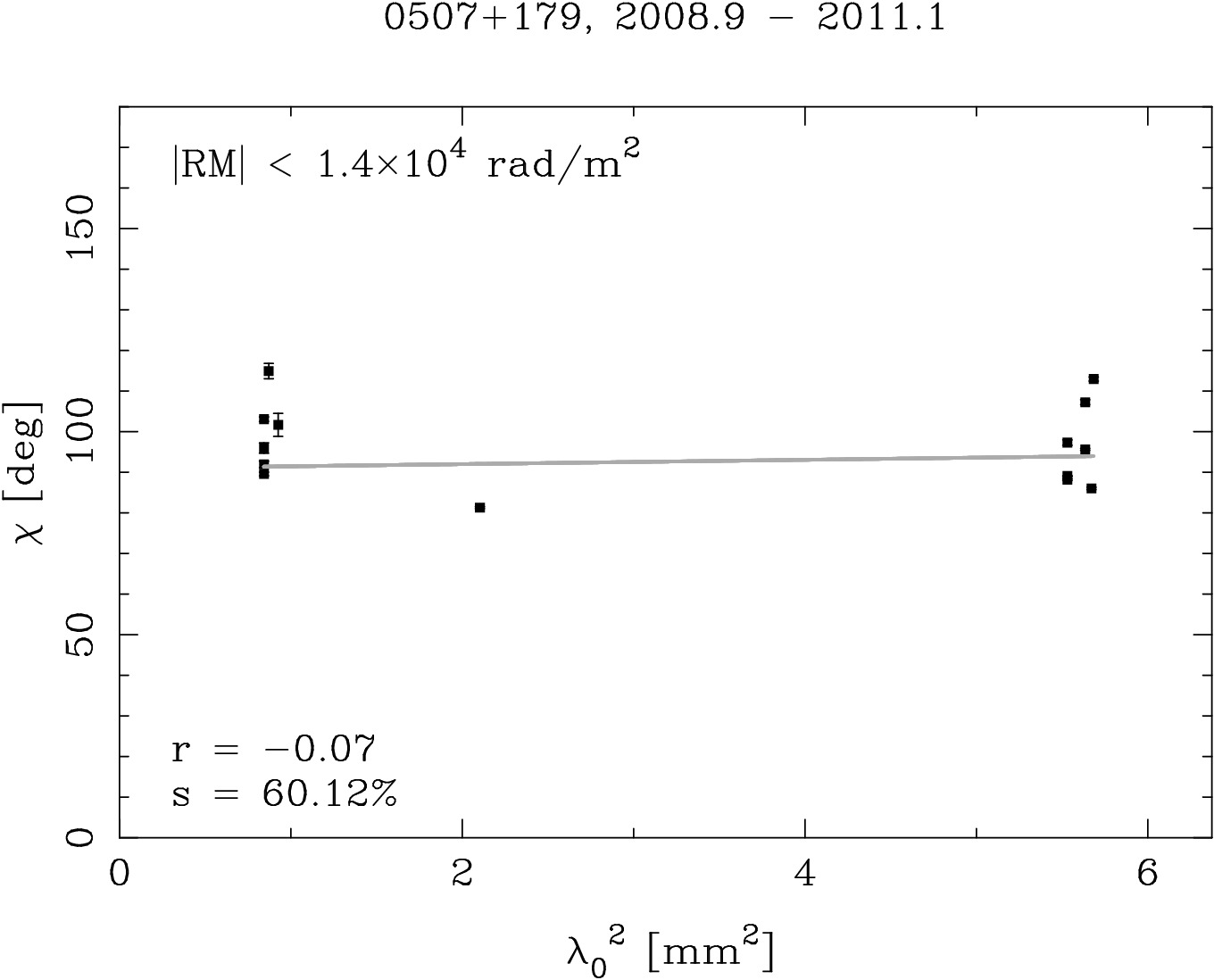} \\
\vspace{5mm}
\includegraphics[height=8.9cm,angle=-90]{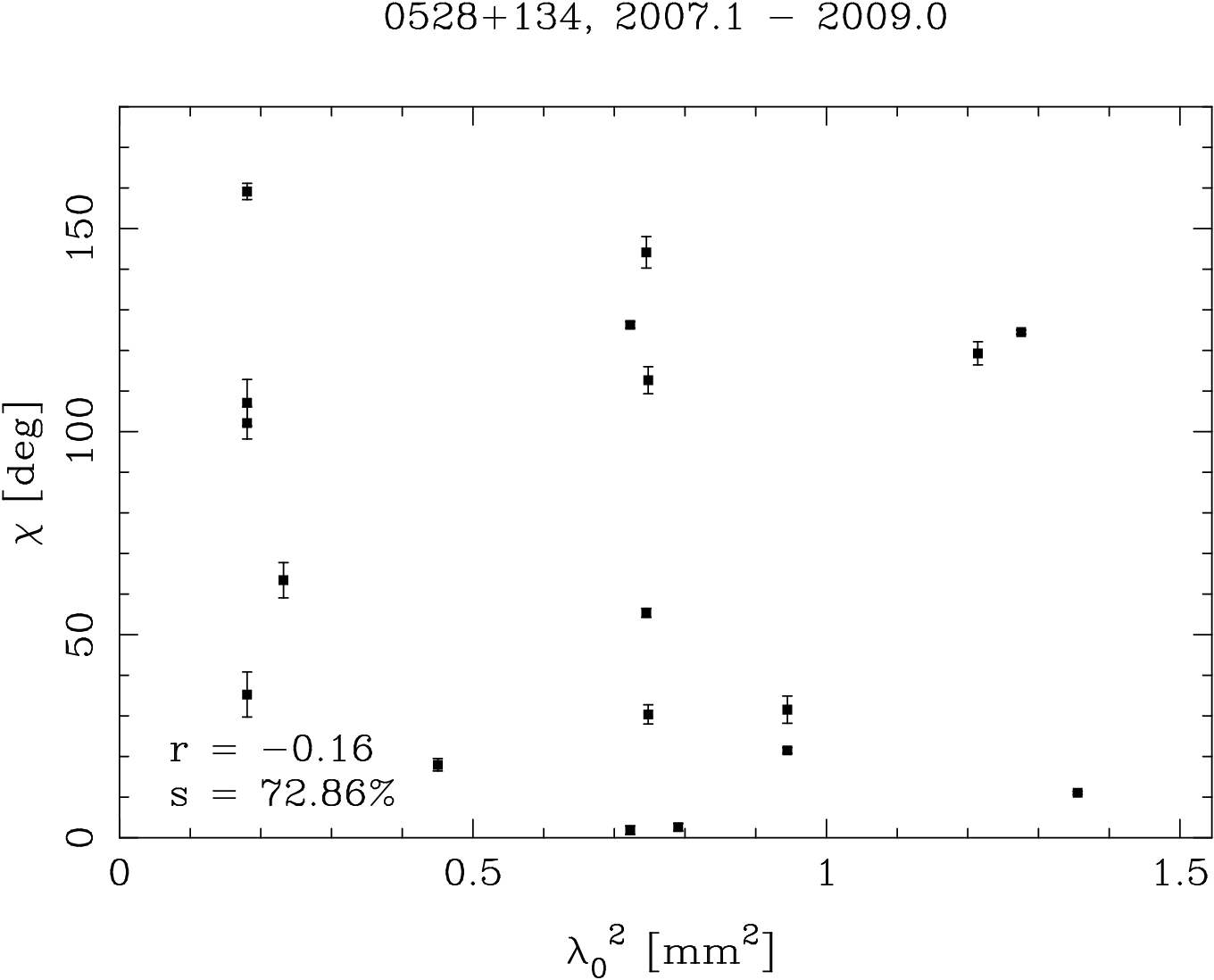}
\hspace{3mm}
\includegraphics[height=8.9cm,angle=-90]{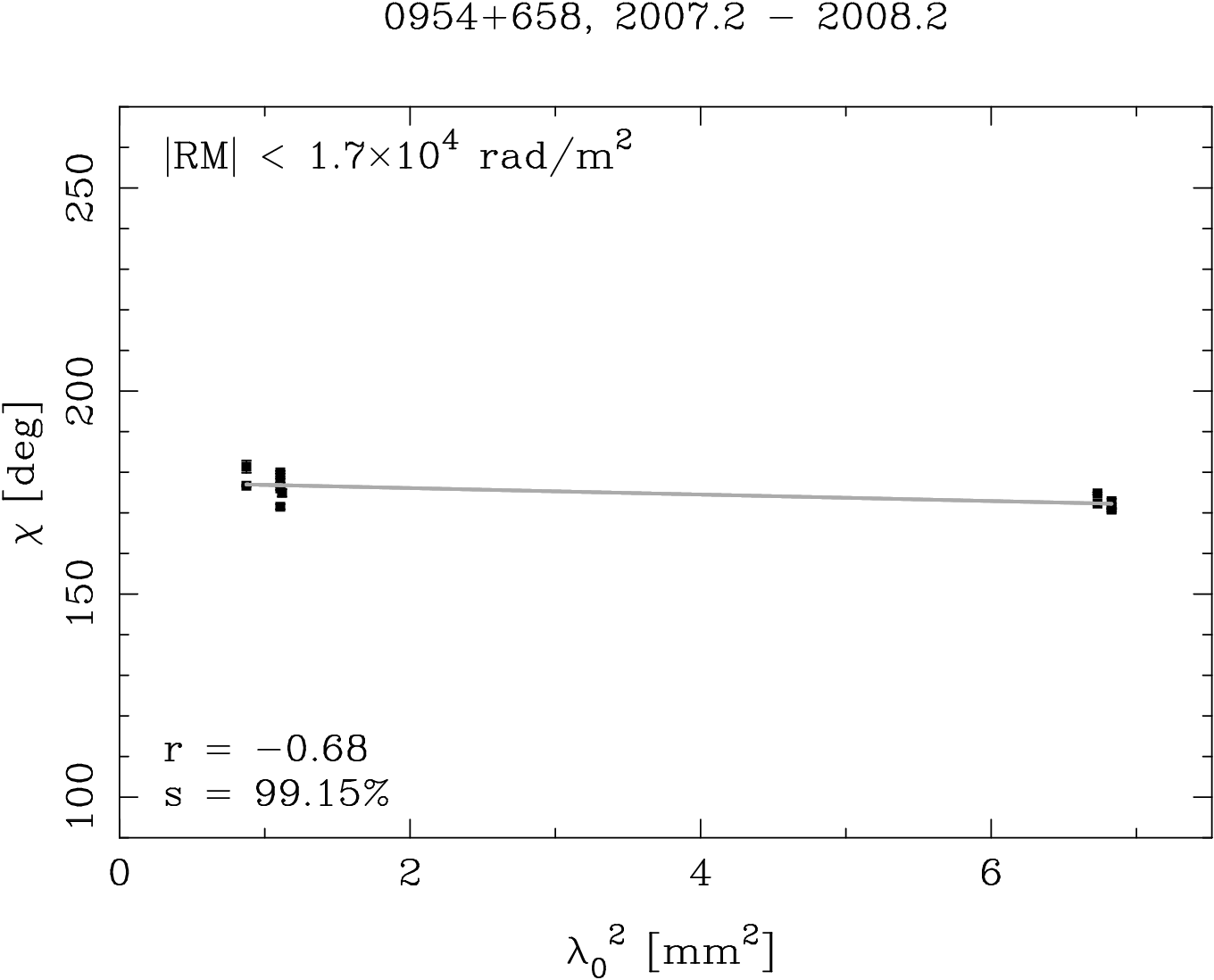} \\
\vspace{5mm}
\includegraphics[height=8.9cm,angle=-90]{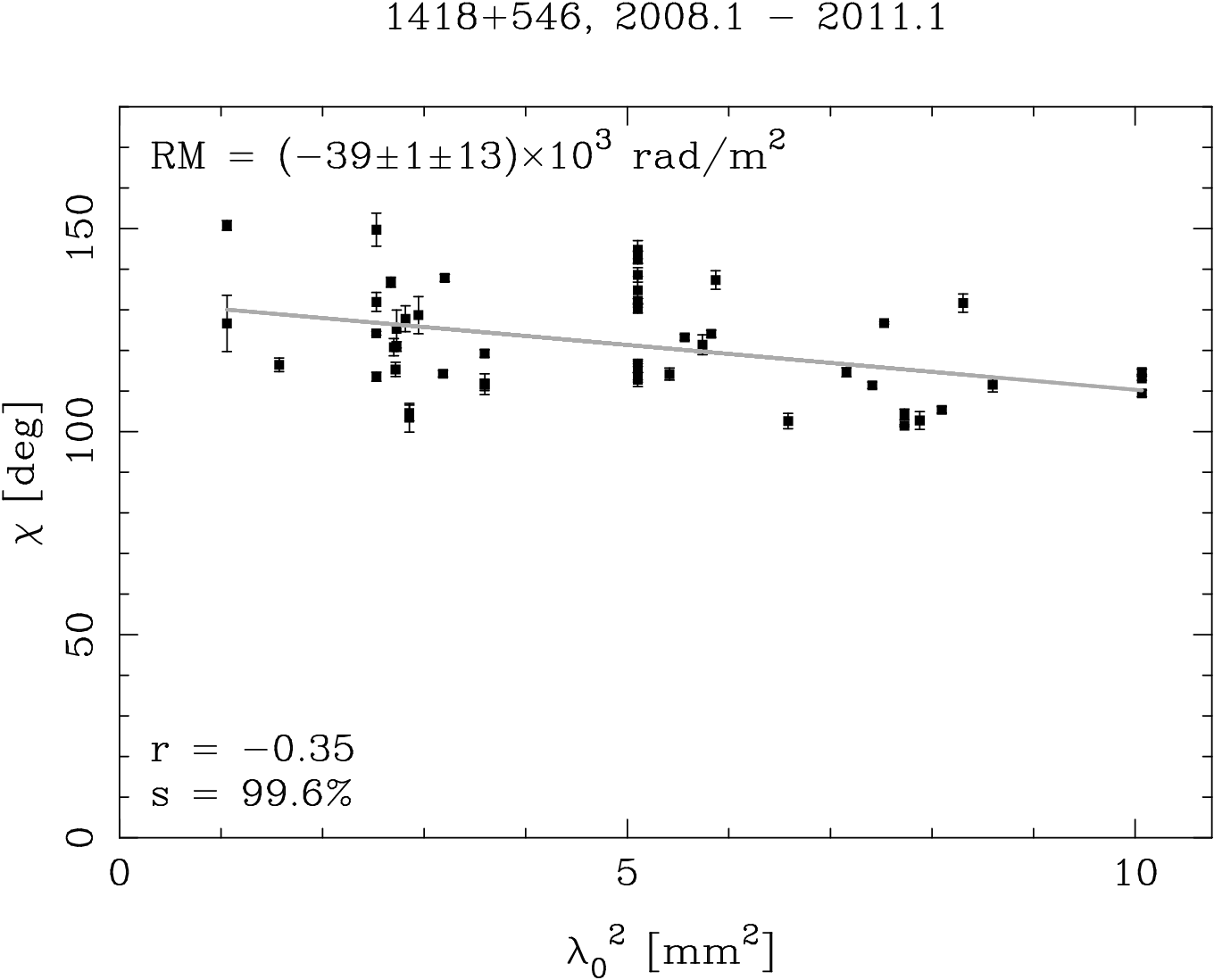}
\hspace{3mm}
\includegraphics[height=8.9cm,angle=-90]{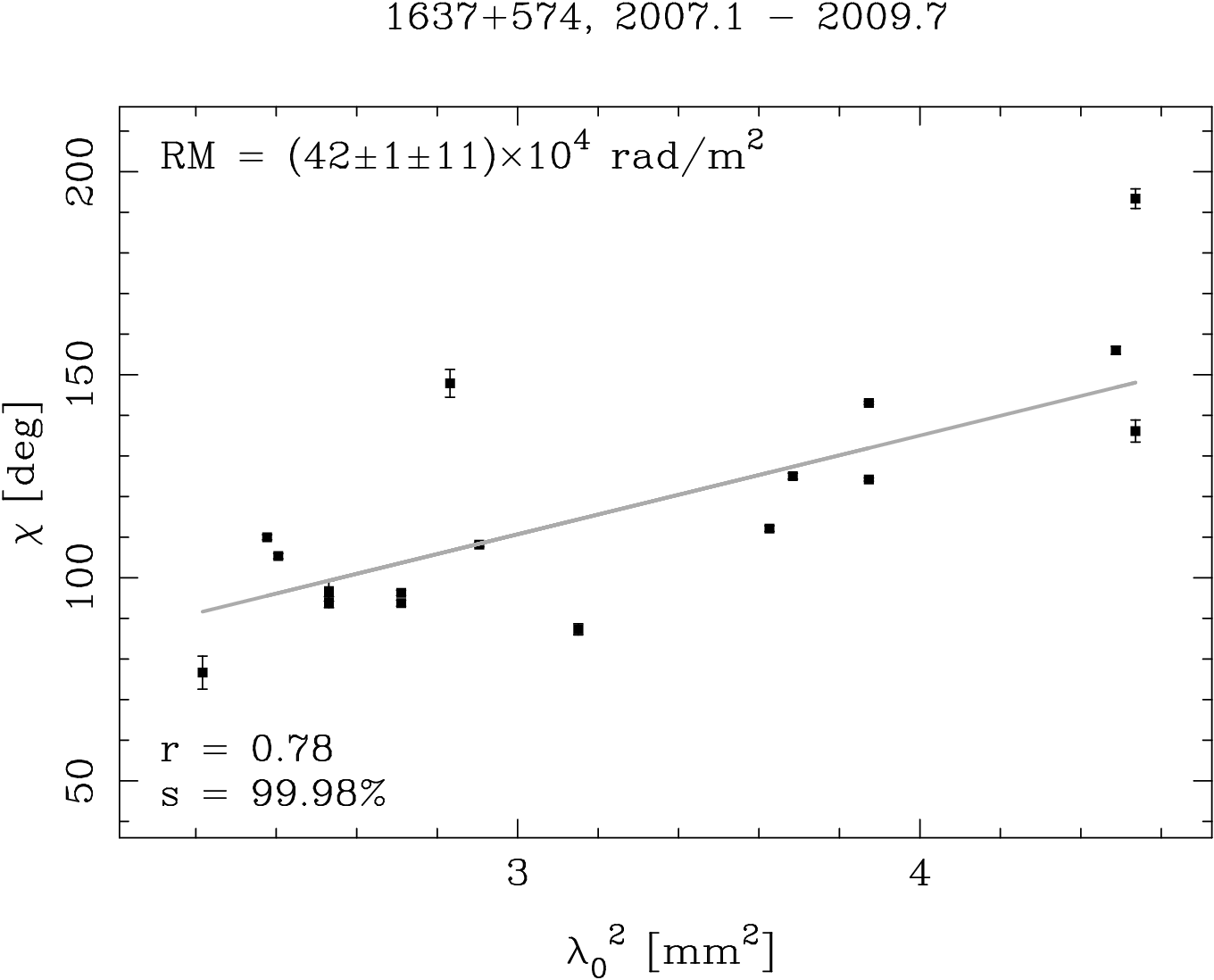}
\caption{Polarization angle $\chi$ as function of the square of the rest-frame wavelength $\lambda_0$. Polarization angles are restricted to the interval $[0^{\circ},180^{\circ}]$; for 0954+658 and 1637+574 these intervals have been shifted to resolve $0^{\circ}/180^{\circ}$ ambiguities. The selection of data is limited to the epoch ranges quoted on top of each plot to maximize the significance of a rotation measure signal. We quote correlation coefficients $r$ and significance levels $s$ like in Fig.~\ref{fig_pol-vs-flux}. For 0415+379, 0507+179, and 0954+658, we derive $3\sigma$ upper limits on $|{\rm RM}|$. For 1418+546 and 1637+574, we quote the rotation measures together with their statistical and systematic errors. Rotation measures are derived via linear fits (gray lines) to the data.}
\label{fig_chi-vs-lambda}
\end{figure*}

In Fig.~\ref{fig_chi-vs-lambda} we present for each source observed polarization angles $\chi$ as function of squared rest-frame wavelength $\lambda_0$. Our data are affected by strong temporal variability. Where possible, we therefore limit our analysis to time windows for which the polarization appears coherent (compare Fig.~\ref{fig_polplane}, Table~\ref{tab_obsjournal}). We selected those time windows such that a minimum number of data points is excluded and the significance of a potential rotation measure signal is enhanced. We note that this ``windowing'' does not bias our results because it applies a selection in time but not in polarization angle or frequency. We quote Pearson correlation coefficients and their significance levels computed analogously to the procedure outlined in Sect.~3.2. For all sources except 0528+134 the observed polarization angles (after windowing) are distributed over ranges notably smaller than the permitted $[0,\pi]$ intervals. For these AGN we attempted to derive rotation measures via linear fits to the data in accord with Eq.~\ref{eq_chi}. Subsequently, we assumed that the observed distributions of $\chi$ vs. $\lambda_0^2$ are superpositions of two principal effects: (1) A linear trend due to Faraday rotation, and (2) scatter due to temporal variability of the AGN polarization that is independent of wavelength. In this case the RM is given by the slope of the straight line fitting the data best. We claim a successful rotation measure measurement if, and only if, the data show a linear correlation $r$ between $\chi$ vs. $\lambda_0^2$ that is significantly different from zero.

We note that the statistical errors of the RM appear to be very small when regarding the scatter of the data. This is a consequence of our ``two-effect assumption'' outlined above. Our RM measurements used a $\chi^2$ minimization algorithm\footnote{In this paragraph, $\chi$ always denotes the polarization angle, whereas $\chi^2$ always denotes the weighted sum of the squares of the differences between data and model. This is an unfortunate collision of common nomenclature standards.} that fits linear models to the data. For models with few parameters, the $1\sigma$ confidence intervals of the parameters are given by variations of $\chi^2$ around its minimum of order unity. In our case however, the scatter of the data is much larger than the statistical error of any individual data point; therefore even the best-fitting solutions have $\chi^2$ in the order of several thousands. From this it follows that the statistical errors we derive are meaningful if, and only if, our initial ``two-effect assumption'' is applicable. Obviously, the accuracy of our results is limited by systematic errors due to intrinsic temporal variability; statistical errors due to random measurement errors are hardly relevant.
We therefore eventually quote for each RM value two errors: the statistical fit error and a systematic error. The systematic error essentially quantifies the uncertainty due to the temporal variability of $\chi$. We derived systematic errors from the confidence levels $s$ of the Pearson correlation coefficients, expressed in units of Gaussian $\sigma$. For a significance level $s$ that corresponds to $n\times\sigma$ in Gaussian terms (with $n$ being a positive real number), we quote ${\rm RM}/n$ as an estimate of the systematic error. Our results are summarized in Table~\ref{tab_rm}.

\begin{table*}
\caption{Estimates on the variability of fluxes and polarization levels. For each source we give the number of polarimetric observations $N_{\rm obs}$, the number of measurements that actually detected significant polarization $N_{\rm det}$, the time averaged degree of linear polarization $\langle m_L\rangle$, the standard deviation of the degree of linear polarization $\sigma_m$, the polarization fluctuation rate $\xi_m$, the relative polarization fluctuation rate $\xi_m/\langle m_L\rangle$, the time averaged 3-mm flux density $\langle S_{\nu}\rangle$, the flux fluctuation rate $\xi_S$, the relative flux fluctuation $\xi_S/\langle S_{\nu}\rangle$, and the fluctuation rate ratio $\rho = (\xi_m/\langle m_L\rangle)/(\xi_S/\langle S_{\nu}\rangle)$. For the polarization fluctuation rates we give the statistical $1\sigma$ errors. The flux fluctuation rates, and thus also $\rho$, are accurate to factors $\approx$2.}
\label{tab_stat}
\centering
\begin{tabular}{l c c c c c c c c c c}
\hline\hline
Object & $N_{\rm obs}$ & $N_{\rm det}$ & $\langle m_L\rangle$ [\%] & $\sigma_m$ [\%] & $\xi_m$ [\%\,d$^{-1}$] & $\xi_m/\langle m_L\rangle$ [d$^{-1}$] & $\langle S_{\nu}\rangle$ [Jy] & $\xi_S$ [Jy\,d$^{-1}$] & $\xi_S/\langle S_{\nu}\rangle$ [d$^{-1}$] & $\rho$ \\
\hline
0415+379 & 34 & 17 & 2.3 & 1.5 & 0.09$\pm$0.01 & 0.041$\pm$0.005 & 7.5 & 0.7  & 0.1  & 0.4 \\
0507+179 & 22 & 21 & 7.4 & 2.5 & 0.36$\pm$0.05 & 0.048$\pm$0.007 & 1.0 & 0.07 & 0.07 & 0.7 \\
0528+134 & 32 & 18 & 3.2 & 1.2 & 0.33$\pm$0.04 & 0.104$\pm$0.014 & 3.5 & 1.1  & 0.3  & 0.3 \\
0954+658 & 37 & 35 & 7.4 & 3.9 & 0.50$\pm$0.02 & 0.067$\pm$0.003 & 1.2 & 0.1  & 0.08 & 0.8 \\
1418+546 & 76 & 52 & 5.0 & 2.0 & 0.28$\pm$0.02 & 0.056$\pm$0.004 & 0.8 & 0.04 & 0.05 & 1.1 \\
1637+574 & 26 & 17 & 3.0 & 1.0 & 0.16$\pm$0.01 & 0.053$\pm$0.003 & 1.2 & 0.2  & 0.1  & 0.4 \\
\hline
\end{tabular}
\end{table*}

\section{Results}

\subsection{0415+379 (3C~111)}

For this Seyfert~1 galaxy located at $z\approx0.05$, 17 out of 34 polarimetric observations, obtained between March 2007 and January 2010, detected linearly polarized emission. The average degree of polarization is $\langle m_L \rangle\approx2.3$\%. Initially, low polarization levels below 2\% and polarization angles around $0^{\circ}$ (modulo $0^{\circ}/180^{\circ}$ ambiguities) are predominant. In August 2007, the polarization changed to $m_L\approx4-6$\% and $\chi\approx100^{\circ}$. After April 2008, polarization could only be detected at one occasion (April 2009, $m_L\approx1\%, \chi\approx120^{\circ}$). The degree of polarization fluctuates at (normalized) rates of $\xi_m\approx0.09$~\%\,d$^{-1}$ and $\xi_m/\langle m_L\rangle\approx0.04$~d$^{-1}$, respectively.

The 3-mm flux density was highly variable, with the average value being $\langle S_{\nu}\rangle\approx7.5$~Jy. In mid-2007, flux measurements commenced at values $S_{\nu}\approx10$~Jy, rising to $S_{\nu}\approx15$~Jy shortly thereafter. Since then, the flux declined to $S_{\nu}\approx2$~Jy, with a short-lived intermediate high of $S_{\nu}\approx7$~Jy observed at the end of 2008. The flux density fluctuates with $\xi_S\approx0.7$~Jy\,d$^{-1}$, the normalized fluctuation rate being $\xi_S/\langle S_{\nu}\rangle\approx0.1$~d$^{-1}$. From a comparison of the fluctuation rates in flux and polarization, one obtains a fluctuation rate ratio of $\rho\approx0.4$.

Degree of polarization and rescaled flux show a strong ($r=-0.9$) anticorrelation. With $s=99.91$\% (corresponding to $\approx$3.3$\sigma$ in terms of a Gaussian distribution), this anticorrelation is significant. An anticorrelation ($r=-0.51$) is apparent also when comparing rescaled polarized flux $P^*$ and $S^*_{\nu}$; however, as $s\approx94$\% (corresponding to $\approx$1.9$\sigma$), this signal is statistically insignificant. Additionally, polarization angle $\chi$ and $S^*_{\nu}$ show a strong ($r=0.91$) correlation, which is significant at a level of $s=99.91$\% ($\approx$3.3$\sigma$).

A search for a correlation between $m_L$ and observing frequency $\nu$ provides a null result ($r=0.37$, $s=92$\%). When comparing polarization angle $\chi$ versus squared rest-frame wavelength $\lambda_0^2$, we formally derive a rotation measure ${\rm RM}=(-4.0\pm2.3)\times10^3$~rad\,m$^{-2}$ (statistical error only). As $s=71$\%, this signal is insignificant. Accordingly, we quote a $3\sigma$ upper limit on $|{\rm RM}|$ given by the absolute value of the RM formally derived plus three times its statistical error. Eventually, we find $|{\rm RM}|<1.1\times10^4$~rad\,m$^{-2}$.

\subsection{0507+179}

This quasar, located at $z\approx0.42$, shows significant polarization in 21 out of 22 measurements obtained between September 2007 and January 2011. The average polarization level is $\langle m_L \rangle\approx7.4$\%. In 2007, the degree of polarization is $m_L\approx6$\%, declines to $m_L\approx3$\% in the first half of 2008, rises again to levels up to $m_L\approx11$\%, and decreases again to $m_L\approx8$\%. In 2007 and 2008, the polarization angle varies in the range $\chi\approx50...110^{\circ}$, remaining around $\chi\approx100^{\circ}$ thereafter. The degree of polarization fluctuates at rates of $\xi_m\approx0.4$~\%\,d$^{-1}$ and $\xi_m/\langle m_L\rangle\approx0.05$~d$^{-1}$, respectively.

The 3-mm flux evolved smoothly, with values $S_{\nu}\approx0.7$~Jy in 2007, peaking at $S_{\nu}\approx1.5$~Jy at the beginning of 2009, and decreasing again to $S_{\nu}\approx0.5$~Jy. The average flux density is found to be $\langle S_{\nu}\rangle\approx1$~Jy. It fluctuates with $\xi_S\approx0.1$~Jy\,d$^{-1}$, corresponding to a normalized fluctuation rate of $\xi_S/\langle S_{\nu}\rangle\approx0.1$~d$^{-1}$. The polarization -- flux fluctuation rate ratio is $\rho\approx0.7$.

Polarization level and rescaled flux density appear to be uncorrelated ($r=0.22$, $s\approx79$\%). When comparing rescaled polarized flux and rescaled flux density, we find a positive correlation ($r=0.73$), which is statistically significant at a level of $s=99.96$\%, corresponding to $\approx$3.5$\sigma$. Polarization angle and rescaled flux are uncorrelated ($r=-0.08$, $s\approx61$\%).

From a comparison of $m_L$ and observing frequency $\nu$ we do not find any significant correlation ($r=0.3$, $s\approx91$\%). We find a slightly positive correlation ($r=0.3$) between polarization angle $\chi$ and the square of the rest-frame wavelength $\lambda_0^2$. We can formally derive a rotation measure of ${\rm RM}=(9.3\pm1.6)\times10^3$~rad\,m$^{-2}$ (statistical error only). However, as the significance of the correlation is only $s=60$\% we have to reject this result as insignificant. Accordingly, we quote a $3\sigma$ upper limit of $|{\rm RM}|<1.4\times10^4$~rad\,m$^{-2}$.

\subsection{0528+134 (OG+134)}

For this high-polarization quasar located at $z\approx2.1$, 18 out of 32 polarimetric observations, obtained between February 2007 and December 2008, detected linearly polarized emission. The average degree of polarization is $\langle m_L \rangle\approx3.2$\%. During the observation timeline, polarization levels varied in the range $m_L\approx2...6$\%, whereas the polarization angles varied in the range $\chi\approx0...140^{\circ}$. The degree of polarization fluctuates at rates of $\xi_m\approx0.3$~\%\,d$^{-1}$ and $\xi_m/\langle m_L\rangle\approx0.1$~d$^{-1}$, respectively.

The 3-mm flux density was at its peak of $S_{\nu}\approx5$~Jy when observations commenced in 2007 and decreased steadily to levels of $S_{\nu}\approx1$~Jy in 2010. The average of the flux density is $\langle S_{\nu}\rangle\approx3.5$~Jy. The flux density fluctuates with $\xi_S\approx1.1$~Jy\,d$^{-1}$, the normalized fluctuation rate being $\xi_S/\langle S_{\nu}\rangle\approx0.3$~d$^{-1}$. By comparing the fluctuation rates in flux and polarization, we obtain a fluctuation rate ratio of $\rho\approx0.3$.

The degree of polarization and the rescaled flux $S^*_{\nu}$ do not show a correlation ($r=-0.21$, $s\approx72$\%). When analyzing rescaled polarized flux $P^*$ versus $S^*_{\nu}$, we see a positive correlation ($r=0.73$). However, as $s=99.45$\%, being equivalent to $\approx$2.8$\sigma$, this signal is only marginally significant. For the correlation between polarization angle and rescaled flux we find $r=-0.37$ and $s\approx87$\%.

Polarization level and observing frequency $\nu$ are uncorrelated ($r=-0.01$). A comparison of polarization angle $\chi$ and squared rest-frame wavelength $\lambda_0^2$ finds that the values observed for $\chi$ occupy the entire range $[0,\pi]$; evidently, strong temporal variability renders our rotation measure analysis inconclusive even when using ``windowing''.

\subsection{0954+658}

This BL Lacertae object, located at $z\approx0.37$, shows significant polarization in 35 out of 37 measurements obtained between March 2007 and January 2011. The average polarization level is $\langle m_L \rangle\approx7.4$\%. 0954+658 shows by far the highest degrees of polarization among our sources, ranging up to $m_L\approx18$\% in 2007. In 2009, the polarization drops to $m_L\approx5$\% and increases again up to $m_L\approx10$\% in 2010/2011. The polarization angles remain around $\chi\approx0^{\circ}$ for most of the observing timeline; the only notable exception occurs in 2009 when values of $\chi\approx10...30^{\circ}$ occur. The degree of polarization fluctuates at a rate of $\xi_m\approx0.5$~\%\,d$^{-1}$ and a normalized rate $\xi_m/\langle m_L\rangle\approx0.07$~d$^{-1}$, respectively.

The average of the 3-mm flux density is $\langle S_{\nu}\rangle\approx1.2$~Jy. The flux varies remains within the range $S_{\nu}\approx0.5...1.5$~Jy throughout the observing timeline, the only exception being a short peak with $S_{\nu}\approx2.5$~Jy occurring mid-2008. The flux fluctuation rate is $\xi_S\approx0.1$~Jy\,d$^{-1}$, corresponding to a normalized fluctuation rate of $\xi_S/\langle S_{\nu}\rangle\approx0.1$~d$^{-1}$. The fluctuation rate ratio of polarization and flux is $\rho\approx0.8$.

Polarization level $m_L$ and rescaled flux $S^*_{\nu}$ are uncorrelated ($r=-0.31$, $s\approx96$\%). Likewise, rescaled polarized flux $P^*$ and $S^*_{\nu}$ are uncorrelated ($r=0.2$, $s\approx85$\%). Polarization angle $\chi$ and rescaled flux $S^*_{\nu}$ are uncorrelated ($r=0.36$, $s\approx97\%$).

A correlation analysis of $m_L$ versus observing frequency $\nu$ unveils a positive correlation with $r=0.52$. As $s\approx99.9$\% ($\approx$3.2$\sigma$), the correlation is statistically significant. We find a slightly negative correlation ($r=-0.21$) between polarization angle $\chi$ and the square of the rest-frame wavelength $\lambda_0^2$. Regarding the polarization angle, we can formally derive a rotation measure of ${\rm RM}=(-14\pm1)\times10^3$~rad\,m$^{-2}$ (statistical error only). However, as the significance of the correlation is $s\approx99.2$\% ($\approx$2.6$\sigma$), the correlation falls short of being statistically significant. Accordingly, we quote a $3\sigma$ upper limit of $|{\rm RM}|<1.7\times10^4$~rad\,m$^{-2}$. \rm

\subsection{1418+546 (OQ+530)}

For this BL Lacertae object located at $z\approx0.15$, 52 out of 76 polarimetric observations, obtained between January 2008 and January 2011, detected linearly polarized emission. The average degree of polarization is $\langle m_L \rangle\approx5.0$\%. Polarization levels vary in the range $m_L\approx3...10$\% with a tendency for the highest values to occur at the beginning and the lowest values to be found at the end of the observed period. The degree of polarization fluctuates at rates of $\xi_m\approx0.3$~\%\,d$^{-1}$ and $\xi_m/\langle m_L\rangle\approx0.06$~d$^{-1}$, respectively. The angles of polarization are located in the range $\chi\approx110...150^{\circ}$. From inspecting the corresponding time--frequency diagram (Fig.~\ref{fig_polplane}) we note a ``swing pattern'' for the polarization angles, with the highest values of $\chi$ to be found around the beginning of 2010.

The 3-mm lightcurve shows a double-peak morphology, with values ranging from $S_{\nu}\approx0.5$~Jy at the beginnings of 2009 and 2011 to $S_{\nu}\approx1.3$~Jy at the beginnings of 2008 and 2010. The average value of the flux is $\langle S_{\nu}\rangle\approx0.8$~Jy. The flux density fluctuates with $\xi_S\approx0.04$~Jy\,d$^{-1}$, the normalized fluctuation rate being $\xi_S/\langle S_{\nu}\rangle\approx0.05$~d$^{-1}$. From comparison of the fluctuation rates in flux and polarization, we obtain a fluctuation rate ratio of $\rho\approx1.1$.

When comparing $m_L$ and $S^*_{\nu}$, we find a positive correlation ($r=0.57$), which is significant as $s\approx99.98\%$ (i.e. $\approx$3.7$\sigma$). An analysis of rescaled polarized flux $P^*$ versus $S^*_{\nu}$ indicates a strong ($r=0.75$) and significant ($s\approx99.99$\%, corresponding to $\approx$3.9$\sigma$) positive correlation. Polarization angle and flux are uncorrelated ($r=0.31$, $s\approx97$\%).

Polarization level and observing frequency $\nu$ show a positive correlation ($r=0.48$), which appears to be significant ($s\approx99.6$\%, i.e. $\approx$2.9$\sigma$). We find a significant -- with $s\approx99.6$\%, meaning $\approx$2.9$\sigma$ -- negative correlation ($r=-0.35$) between polarization angle $\chi$ and the square of the rest-frame wavelength $\lambda_0^2$. We are able to derive a rotation measure of ${\rm RM}=(-39\pm1_{\rm stat}\pm13_{\rm sys})\times10^3$~rad\,m$^{-2}$.

\subsection{1637+574 (OS+562)}

This Seyfert~1 galaxy, located at $z\approx0.75$, shows significant polarization in 17 out of 26 measurements obtained between February 2007 and August 2009. The average polarization level is $\langle m_L \rangle\approx3.0$\%. Degrees of polarization are found in the range $m_L\approx1...5$\%, with the lowest values occurring at the beginning of the observed period. The angles of polarization swing from $\chi\approx150^{\circ}$ to $\chi\approx90^{\circ}$. The degree of polarization fluctuates at a rate of $\xi_m\approx0.2$~\%\,d$^{-1}$ and a normalized rate $\xi_m/\langle m_L\rangle\approx0.05$~d$^{-1}$, respectively.

The 3-mm flux density is initially observed at values $S_{\nu}\approx1.2$~Jy, increases up to  $S_{\nu}\approx2$~Jy in 2008 and decreases again to $S_{\nu}\approx1$~Jy in 2009. The average flux is $\langle S_{\nu}\rangle\approx1.2$~Jy. The flux fluctuation rate is $\xi_S\approx0.17$~Jy\,d$^{-1}$, corresponding to a normalized fluctuation rate of $\xi_S/\langle S_{\nu}\rangle\approx0.12$~d$^{-1}$. The fluctuation rate ratio of polarization and flux is $\rho\approx0.4$.

Polarization level $m_L$ and rescaled flux $S^*_{\nu}$ are uncorrelated ($r=0.1$). When comparing rescaled polarized flux $P^*$ and $S^*_{\nu}$, we find a positive correlation with $r=0.55$; however, as $s\approx97$\%, the correlation is insignificant. Likewise, polarization angle and flux are uncorrelated ($r=-0.22$, $s\approx76$\%).

An analysis of $m_L$ versus observing frequency finds a positive correlation ($r=0.46$), which is insignificant as $s\approx97$\%. We find a strong, positive correlation ($r=0.78$) between polarization angle $\chi$ and the square of the rest-frame wavelength $\lambda_0^2$. As $s\approx99.98$\%, corresponding to $\approx$3.7$\sigma$, the correlation is significant. The rotation measure we derive is ${\rm RM}=(42\pm1_{\rm stat}\pm11_{\rm sys})\times10^4$~rad\,m$^{-2}$.

\section{Discussion}

\subsection{Polarization levels}

We present the polarization properties of six mm/radio-luminous -- with $S_{\nu}\gtrsim0.8$~Jy in the 3-mm band -- active galactic nuclei. Our sample is a mix of Seyfert galaxies, BL~Lacertae objects, and quasars. An inspection of their mm/radio spectral indices (as listed in Table~\ref{tab_journal}) indicates that the radiation of two sources -- 0415+379 and 0528+134 -- is optically thin and outflow-dominated, whereas the remaining four AGN show optically thick, core-dominated emission when applying the conventional division between the two cases placed at $\alpha=0.5$ (e.g., Krolik~\cite{krolik1999}).

The average degrees of polarization $m_L$ we observe are located between $\approx$2\% and $\approx$7\%. This is substantially lower than the values one can derive from the theory of synchrotron emission from homogeneous and isotropic ensembles of relativistic electrons moving in uniform magnetic fields (e.g., Ginzburg \& Syrovatskii \cite{ginzburg1965}; Pacholczyk~\cite{pacholczyk1970}; see also the corresponding discussion in Paper~I). Given the range of $\alpha$ observed, we may expect theoretical polarization levels from $m_L\approx13$\% (for $\alpha=0.3$, optically thick emission) to $m_L\approx74$\% (for $\alpha=0.9$, optically thin emission). The most probable explanation is provided by geometric averaging: our observations do not resolve our targets spatially, whereas extended outflows (jets or halos) have been observed in all of them.
More specifically, Lee et al. (\cite{lee2008}) have been able to map four sources -- 0415+379, 0528+134, 0954+658, and 1637+574 -- using global VLBI observations at 86~GHz. In addition, Sokolovsky et al. (\cite{sokolovsky2011}) observed an optically thick core as well as an optically thin jet in VLBA radio maps obtained in the range 1.4--15~GHz. These maps show jets for all sources observed, indicating that probably all our targets show optically thin outflows if observed with sufficient angular resolution. Accordingly we have to assume that any of our observations takes the average of multiple emission regions with different optical depths and different magnetic field orientations.
Assuming we observe $N$ emission zones with randomly distributed magnetic field orientations in any given source, we may expect source-integrated degrees of polarization $m_L \rightarrow m'_L \approx m_L / \sqrt{N}$. From our observations (see Tables \ref{tab_journal}, \ref{tab_stat}) we can then estimate values between $N\approx3$ (for 0507+179 and 0954+658, assuming optically thick emission) and $N\approx1000$ (for 0415+379, assuming optically thin emission). The variation in $N$ mirrors the variation in relevant emission region sizes: AGN radio cores are placed at spatial scales of parsecs whereas jets occur on scales of many kiloparsecs ($\approx$80~kpc for the radio jet of 0415+379; Liu \& Xie \cite{liu1992}). However, this simple ``static'' picture is complicated by any dynamical evolution of the emission regions; we discuss the potential signatures of shocks in the following subsections.

An independent check of our results is provided by Agudo et al. (\cite{agudo2010}). This study provides single-epoch measurements of linear polarization obtained between 2005 and 2006 at 86~GHz for five of our targets. The authors found degrees of polarization of $<1.5$\% (i.e., an upper limit only), $2.6\pm0.5$\%, $5.2\pm0.5$\%, $2.8\pm0.6$\%, and $<1.5$\% for 0415+379, 0528+134, 0954+658, 1418+546, and 1637+574, respectively. In addition, Agudo et al. (\cite{agudo2010}) derived polarization angles of $91\pm6^{\circ}$,$8\pm3^{\circ}$, and $108\pm6^{\circ}$ for 0528+134, 0954+658, and 1418+546, respectively. These values may be compared with our results given in Tables \ref{tab_stat} and \ref{tab_obsjournal} as well as Fig.~\ref{fig_polplane}; taking into account the temporal variability, both datasets agree well. This indicates that neither Agudo et al. (\cite{agudo2010}) nor our study seem to have observed a source in a temporary atypical or extreme state. This reduces the probability that any of the two studies is affected by temporal selection effects.
\rm

\subsection{Variability}

Variability of AGN polarization is well-known in the cm/radio regime (e.g., Aller et al. \cite{aller1985}) and has been probed in the mm/radio regime for a few selected sources (e.g., Jorstad et al. \cite{jorstad2005}) as well as in the optical (e.g., Villforth et al. \cite{villforth2009}). Our variability analysis (see Eq.~\ref{eq_xi},\ref{eq_rho} and Table~\ref{tab_stat}) treats variability in source flux $S_{\nu}$ and degree of polarization $m_L$ jointly (see also Figs.~\ref{fig_polplane},\ref{fig_lightcurve}). Both parameters fluctuate at relative rates of few per cent per day. For the flux densities, those fluctuation rates have been observed in cm/radio as well as mm/radio bands and have been discussed in the context of potential intra-day variability (e.g., Fuhrmann et al. \cite{fuhrmann2008}). Rapid flux variability is the basis of the famous historical statement that the size of the (relevant parts of the) individual emission regions cannot exceed a few light days (Smith \& Hoffleit \cite{smith1963}). Our analysis makes it possible to extend this statement to the observed polarization: because the fluctuation rates of flux density and degree of polarization are compatible, the spatial scales of the corresponding emission regions should be compatible as well. We may quantify this by means of the fluctuation rate ratios $\rho$ as given by Eq.~\ref{eq_rho}. Our analysis finds values $\rho\approx0.3-1.1$, permitting two principal conclusions:

\begin{enumerate}

\item  Within their uncertainties of factors $\approx$2, the $\rho$ values are on the order of unity for all our six target sources. This suggests that ratios $\rho\approx1$ are indeed characteristic for AGN emission (at least in mm/radio) -- pointing towards a strong \emph{structural} correlation between total and polarized flux.

\item  Even though all ratios $\rho$ are consistent with being on the order of unity, the individual values tend toward numbers somewhat smaller than one.\footnote{We note that the two sources with steep spectral indices $\alpha\approx0.9$ show the lowest ratios $\rho\approx0.3-0.4$. However, given the low-number statistics and the substantial uncertainties, we cannot claim the discovery of a trend.}  This indicates that the size scales relevant for the polarized emission -- like turbulence or eddy scales of the magnetic fields involved -- are \emph{compatible with, but not smaller than,} the relevant size scales for the overall emission.

\end{enumerate}

Rapid -- referring here to timescales shorter than a few months -- variability of polarization is supposedly caused by shocks propagating through magnetized plasma (e.g., Hughes et al. \cite{hughes1985}). In this case, the degree of linear polarization of synchrotron radiation from a partially compressed plasma is reduced by a factor $\mu$ compared to the case without compression (Hughes et al. \cite{hughes1985}; Cawthorne \& Wardle \cite{cawthorne1988}), where

\begin{equation}
\mu = \frac{\delta}{2 - \delta}
\label{eq_mu}
\end{equation}

\noindent
with the shock parameter

\begin{equation}
\delta = (1 - k^2)\cos^2\epsilon  .
\label{eq_delta}
\end{equation}

\noindent
Here $0\leq k\leq 1$ is the compression factor, meaning the factor by which the length of the affected region is reduced by compression. The angle $\epsilon$ is the angle between the line of sight and the plane of compression in the frame of the emitter. From Eq.~\ref{eq_delta} follows $0\leq\delta\leq1$. Additional information from, e.g., time-resolved radio interferometric maps, is not available, making it impossible to disentangle $k$ and $\epsilon$ in $\delta$; accordingly, $\delta$ is our only observable. As discussed in Sect.~5.1, we probably observe polarization from multiple emission regions; in this case, any attempt to constrain the parameters of Eqs.~\ref{eq_mu} and \ref{eq_delta} actually provides source-averaged, ``typical'' values. Assuming we observe different values for $\mu$ at two points of time ``1'' and ``2'', we may write

\begin{equation}
\Delta\mu = \mu_1 - \mu_2 = \frac{\delta_1}{2 - \delta_1} - \frac{\delta_2}{2 - \delta_2} = \frac{2(\delta_1 - \delta_2)}{(2-\delta_1)(2-\delta_2)}  .
\label{eq_delta-mu}
\end{equation}

\noindent
Our observations do not provide the absolute, intrinsic degrees of polarization -- meaning the polarization levels corresponding to the case when shocks are absent -- of individual emission regions.
Instead, we observe source-averaged (except, possibly, 1418+546; see Sect.~5.3) polarization levels that fluctuate with time. From this we attempt to constrain at least average, ``typical'' shock parameters $\delta$. For a given source, we use the time average of the degree of polarization $\langle m_L\rangle$ as an approximate reference. We then use the normalized standard deviation of the degree of polarization $\sigma_m/\langle m_L\rangle$ as an estimate for $\Delta\mu$, meaning 

\begin{equation}
\Delta\mu \approx \frac{\sigma_m}{\langle m_L\rangle}  .
\label{eq_sigma-m-mu}
\end{equation}

\begin{figure}
\centering
\includegraphics[height=8.8cm,angle=-90]{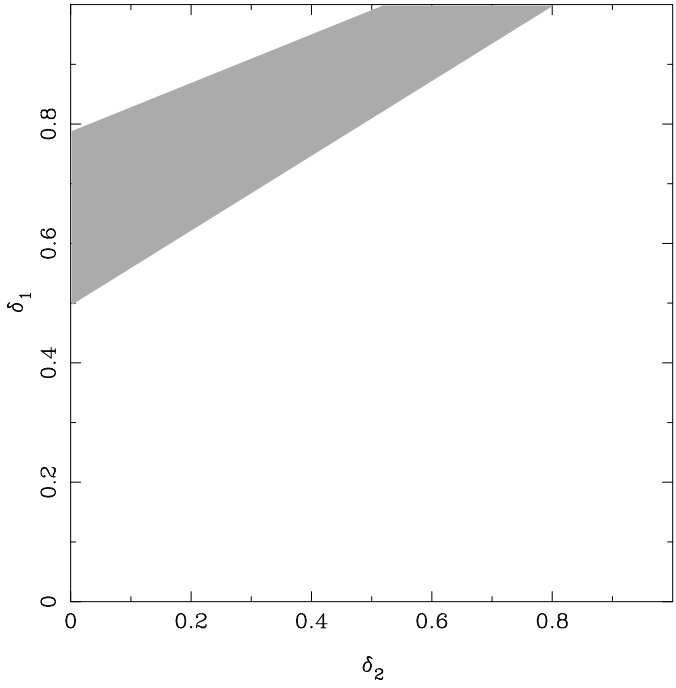}
\caption{Constraints on the values and the variability of the shock parameter $\delta$. The shock parameter $\delta$ relates to the factor $\mu=\delta/(2-\delta)$ by which linear polarization is reduced by shock compression. Our observations provide $\Delta\mu=2(\delta_1-\delta_2)/(2-\delta_1)(2-\delta_2)\approx\sigma_m/\langle m_L\rangle$ for two arbitrary points in time ``1'', ``2''. The gray-shaded area marks the $\delta_1-\delta_2$ plane range consistent with the observed $\sigma_m/\langle m_L\rangle\approx0.33-0.65$ for our sample of sources.}
\label{fig_polshock}
\end{figure}

\noindent
Evidently, $\Delta\mu$ depends on the individual values of $\delta_1$ and $\delta_2$ as well as on their difference $\delta_1-\delta_2$ (see Eq.~\ref{eq_delta-mu}). Accordingly, it is not possible to derive single values for $\delta_{1,2}$ but ranges in a plane spanned by $\delta_1$ and $\delta_2$. In Fig.~\ref{fig_polshock} we present such a plane with the ranges marked that are permitted for $\sigma_m/\langle m_L\rangle\approx0.33-0.65$ as observed for our sources (compare Table~\ref{tab_stat}). Our result points toward the occurrence of fairly violent shocks and/or multiple shocks with varying geometries within our sources. According to Eq.~\ref{eq_delta} and regarding, e.g., the pair $\delta_1\approx1$ and $\delta_2\approx0.8$ in Fig.~\ref{fig_polshock}, our observations indicate the occurrence of $k\lesssim0.9$, or of variations in $\epsilon$ by $\gtrsim10^{\circ}$, or combinations of both.

\subsection{Correlations between flux and polarization}

Immediate correlations between rescaled (Eq.~\ref{eq_rescale}) polarization parameters and source flux (Fig.~\ref{fig_pol-vs-flux},\ref{fig_polflux-vs-flux},\ref{fig_chi-vs-flux}) can occur in several varieties. Because we might observe multiple emission components within the same source, we can write the observed linear polarization vector ${\mathbf P}$ as

\begin{equation}
{\mathbf P} = \sum_i {\mathbf P}_i 
\label{eq_pol-sum}
\end{equation}

\noindent
with two-dimensional vectors ${\mathbf P} = (Q,U)$ with Stokes parameters $Q$ and $U$. To separate out the variability in source flux $S_{\nu}$, we also write a normalized polarization

\begin{equation}
{\mathbf p} \equiv \frac{\mathbf P}{I} = \frac{\sum_i {\mathbf p}_i I_i}{\sum_i I_i} 
\label{eq_pol-vs-flux}
\end{equation}

\noindent
with the Stokes parameter $I$ being identical to $S_{\nu}$; the index $i$ indicates the contribution from the $i$th emission component. Here ${\bf p} = (Q/I,U/I) \equiv (q,u)$, which relates to our observed parameters like $m_L^2 = q^2 + u^2$ and $\chi = 0.5\tan^{-1}(u/q)$. In general, one may assume that all parameters involved can vary independently, leaving the polarization parameters $m_L$, $\chi$ and rescaled fluxes $S^*_{\nu}$ uncorrelated; indeed, this is what our data suggest for most of our targets. Nevertheless, we are able to probe at least two special cases:

\begin{enumerate}

\item  We receive radiation from a single predominant emission component with an arbitrary but approximately fixed degree and angle of polarization. Any change in source flux should be accompanied by a proportional change in polarized flux. In this case we would observe a strong correlation between flux density and polarized flux. The flux would not correlate with degree or angle of polarization.

\item  We receive radiation from two (or more) emission components with each component having an arbitrary but approximately constant $q_i$ and $u_i$. The observed polarization signal is a superposition of the individual components. Any increase (or decrease) in the luminosity of one component increases (decreases) its contribution to the average signal observed. In this scenario, we would observe a tight correlation between flux and degree of polarization as well as between flux and polarization angle\footnote{Strictly speaking, this is only true if we do not fall victim to a ``conspiracy'' where several parameters change simultaneously such that no net effect is left. This would, e.g., be the case if the fluxes from the individual emission components change such that their sum does not.}.

\end{enumerate}

Remarkably, our data indicate the occurrence of those special cases in three objects. In 0415+379, we observe a signature as predicted by scenario 2. While the source flux increases by a factor $\approx$3, the degree of polarization drops by a factor $\approx$4, leading to a strong anticorrelation between the two parameters. Simple arithmetics shows that, when assuming that the degrees of polarization are approximately fixed, we observe increased emission from (at least) one weakly or non-polarized component added to the emission from (at least) one component with a higher level of linear polarization. Our interpretation is supported by recent time-resolved radio interferometric observations. Grossberger et al. (\cite{grossberger2011}) were able to associate the flux peak observed in 2007 with the ejection of a new jet component that adds its emission to various other, separate jet components recognized in radio interferometric maps obtained with the Very Long Baseline Array (VLBA).

For 0507+179, we observe a signature in agreement with scenario 1: while the source flux increases by a factor $\approx$3, the polarized flux increases by a factor $\approx$3, too. There is no significant correlation between flux on the one hand and degree or angle of polarization on the other hand. Radio interferometric VLBA maps obtained by Fey \& Charlot (\cite{fey1997}) at 8.6~GHz show a dominating core and a fainter jet component about 3~mas west of the core. However, as this map was obtained in 1995, any comparison between the data of Fey \& Charlot (\cite{fey1997}) and our results has to be taken with care. We note that our ``one-component scenario'' does not contradict the discussion in Sect.~5.1. The scenarios are reconciled when assuming large-scale regions with coordinated luminosity variations -- like shock fronts -- that have substructure defined by, e.g., characteristic ordering scales of magnetic fields.

In 1418+546, we observe a more complex signature. While the source flux increases by a factor $\approx$3, the polarized flux inceases by a factor $\approx$8, resulting in a significant correlation between the two parameters in agreement with scenario 1. Additionally, there is also a significant, though more scattered, positive correlation between flux and degree of polarization, with $m_L$ rising by a factor $\approx$4. There is no significant correlation between flux and polarization angle. In essence, we seem to observe a single, dominating emission component that tends toward higher polarization levels at higher flux densities. We note the agreement of the behavior of 1418+546 with the signatures expected from recent model calculations based on ``shock in jet'' scenarios (Hughes et al. \cite{hughes2011}). Regarding the lightcurve, our observations indeed appear to cover two outbursts occurring within about two years. Our observations find (a) a correlated variation of flux and polarization, (b) changes of the polarization angle by tens of degrees, and (c) no correlation between flux and polarization angle. We see the best agreement of our data with the scenario of an oblique shock front observed at an angle $>$60$^{\circ}$ to the jet axis (see Fig. 7f of Hughes et al. \cite{hughes2011}). We also note that the calculations of Hughes et al. (\cite{hughes2011}) assume a spectral index $\alpha=0.25$; this may be compared to the observational result $\alpha=0.2\pm0.1$ (see Table~\ref{tab_journal} and Paper~I). Overall, our observations appear to provide new support for the ``oblique shock'' scenario.

\subsection{Rotation measures}

We analyzed the rotation measures of our target AGN according to the procedure outlined in Sect.~3.3. We are able to derive statistically significant rotation measures for 1418+546 and 1637+574 (see Fig.~\ref{fig_chi-vs-lambda} and Sect. 4.5, 4.6). We note that for these two sources we included all data available into our calculation because ``time windowing'' did not increase the significance of the result or reduce the scatter caused by temporal variability. Our calculations assume that any Faraday rotation detected is caused by matter in or within the vicinity of the emitters. A priori, there are two additional sources of Faraday rotation that might distort our analysis: the terrestrial ionosphere and Galactic interstellar matter. For the frequency range of our observations, $\nu>80$~GHz, the ionosphere is able to rotate the polarization angle of electromagnetic radiation by less than 0.01$^{\circ}$ (Thompson et al. \cite{thompson2001}), meaning its influence may be neglected. Galactic interstellar matter introduces rotation measures $|{\rm RM}|\lesssim200$ at radio frequencies (Taylor et al. \cite{taylor2009}). This number is about one order of magnitude smaller than the statistical errors of our rotation measure values, which renders the influence of interstellar matter irrelevant.

We find ${\rm RM}=(-39\pm1_{\rm stat}\pm13_{\rm sys})\times10^3$~rad\,m$^{-2}$ and ${\rm RM=(42\pm1_{\rm stat}\pm11_{\rm sys})\times10^4}$~rad\,m$^{-2}$ for 1418+546 and 1637+574, respectively. For 0415+379, 0507+179, and 0954+658 we derive $3\sigma$ upper limits on $|{\rm RM}|$ of $1.1\times10^4$~rad\,m$^{-2}$, $1.4\times10^4$~rad\,m$^{-2}$, and $1.7\times10^4$~rad\,m$^{-2}$, respectively. Evidently, our observations are subject to strong selection effects. As already noted by Agudo et al. (\cite{agudo2010}), Eq.~\ref{eq_chi} implies that, at mm-wavelengths, changes in polarization angle as small as $\approx5^{\circ}$ require $|{\rm RM}|\approx7\times10^3$~rad\,m$^{-2}$. In addition, our polarization angle data show substantial scatter caused by rapid temporal variability. Therefore rotation measures smaller than a few $10^4$~rad\,m$^{-2}$ probably remain unnoticed simply because they do not imprint a statistically significant signal into a $\chi - \lambda_0^2$ diagram. This is illustrated by the case of 1418+546, where an ${\rm RM}\approx-3.9\times10^4$~rad\,m$^{-2}$ signal still comes with a statistical false-alarm probability of $\approx$0.4\%, barely corresponding to a $3\sigma$ significance threshold. Rotation measure values derived from cm/radio observations are usually located in ranges from a few 10~rad\,m$^{-2}$ to a few 1\,000~rad\,m$^{-2}$ at most (e.g., O'Sullivan \& Gabuzda \cite{osullivan2009}; Taylor et al. \cite{taylor2009}). At mm/radio frequencies, AGN polarization and Faraday rotations are poorly probed, with the most comprehensive study to date probably being the work by Jorstad et al. (\cite{jorstad2007}). Those studies observe values up to $|{\rm RM}|\approx3.4\times10^4$~rad\,m$^{-2}$ (for CTA~102; Jorstad et al. \cite{jorstad2007}). Accordingly, we report in this study the highest rotation measures observed to date in AGN.

From Eq.~\ref{eq_rm} we may estimate the product of magnetic field strength and electron density  averaged along the line of sight $\langle B_{||}n_e\rangle$. For emission region sizes on the order of one parsec (compare, e.g., O'Sullivan \& Gabuzda \cite{osullivan2009} for the case of 1418+546) and RM of order $10^5$~rad\,m$^{-2}$, we find $\langle B_{||}n_e\rangle \sim 0.1$~G\,cm$^{-3}$. For realistic electron densities $n_e\sim10^3$~cm$^{-3}$ (e.g., Koski \cite{koski1978}; Bennert et al. \cite{bennert2009}) this corresponds to average magnetic field strengths $B_{||}\sim10^{-4}$~G.

Observing increasing rotation measures when going from cm/radio to mm/radio frequencies may be expected from Eq.~\ref{eq_rm} for optically thick sources. Higher frequencies probe regions closer to the central engine with higher electron densities and stronger magnetic fields. For realistic emission region geometries, one may assume $n_e\propto l^{-a}$ and $B_{||}\propto l^{-1}$, with $n_e$ being the electron density, $B_{||}$ being the magnetic field parallel to the line of sight, $l$ being the distance from the core where the observed emission predominantly originates, and $a$ being a positive real number (Jorstad et al. \cite{jorstad2007}). For outflows with spherical or conical geometries, $a=2$. From Eq.~\ref{eq_rm} one derives immediately $|{\rm RM}|\propto l^{-a}$. For optically thick sources, distance $l$ and frequency $\nu$ are related like $l\propto\nu^{-1}$, leading to a ``core shift'' effect. For the specific case of 0507+179, Sokolovsky et al. (\cite{sokolovsky2011}) explicitly observed a core shift for the frequency range 1.4--15~GHz. Combining the various relations leads to

\begin{equation}
|{\rm RM}| \propto \nu^a  ,
\label{eq_rm-vs-nu}
\end{equation}

\noindent
When inserting values for $a$ in agreement with observations, i.e. $a\approx1-3$ (O'Sullivan \& Gabuzda \cite{osullivan2009}), one may indeed expect $|{\rm RM}|$ of order $10^5$~rad\,m$^{-2}$. This is additionally supported by observations of Sagittarius~A*, the radio source commonly equated with the supermassive black hole at the center of the Milky Way (e.g., Genzel et al. \cite{genzel2010}, and references therein). Here rotation measures up to $|{\rm RM}|\approx5\times10^5$~rad\,m$^{-2}$ at frequencies around 340~GHz have been reported (Marrone et al. \cite{marrone2006}; Macquart et al. \cite{macquart2006}).

For 1418+546 it is possible to compare our result with RM observations obtained at cm/radio frequencies and from these to derive $a$. O'Sullivan \& Gabuzda (\cite{osullivan2009}) report ${\rm RM}=-501\pm48$~rad\,m$^{-2}$ at observed frequencies around 15~GHz. This can be related to our result, ${\rm RM}=(-39\pm1_{\rm stat}\pm13_{\rm sys})\times10^3$~rad\,m$^{-2}$ at $\nu\approx150$~GHz. From Eq.~\ref{eq_rm-vs-nu} we obtain $a=1.9$ with a formal \emph{statistical} error of only $\pm$0.04. From the systematic error in $|{\rm RM}|$ and the uncertainties in defining the two reference frequencies used in our calculation we expect an additional \emph{systematic} error of approximately 0.3. Eventually, we obtain $a=1.9\pm0.3$ with the error being predominantly of systematic nature. This may be compared to the theoretical value $a=2$ for spherical or conical outflows. Our rotation measure analysis therefore indicates that for the specific case of 1418+546 the assumption of a simple spherical or conical outflow geometry is sufficient.

\begin{table}
\caption{Rotation measures RM for our six target sources. Statistical (``stat.'') errors refer to the formal statistical fit uncertainty. Systematic (``sys.'') errors quantify the additional uncertainty due to temporal variability of the polarization angles. Values with ``$<$'' are upper limits, entries ``---'' denote ``no value available''.}
\label{tab_rm}
\centering
\begin{tabular}{l c c c}
\hline\hline
Source & RM & stat. error & sys. error \\
  & [rad\,m$^{-2}$] & [rad\,m$^{-2}$] & [rad\,m$^{-2}$] \\
\hline
0415+379$^{\mathrm{a}}$ & $<1.1\times10^4$ & --- & --- \\
0507+179$^{\mathrm{a}}$ & $<1.4\times10^4$ & --- & --- \\
0528+134$^{\mathrm{b}}$ & --- & --- & --- \\
0954+658$^{\mathrm{a}}$ & $<1.7\times10^4$ & --- & --- \\
1418+546 & $-3.9\times10^4$ & $1\times10^3$ & $1.3\times10^4$ \\
1637+574 & $4.2\times10^5$ & $1\times10^4$ & $1.1\times10^5$ \\
\hline
\end{tabular}
\begin{list}{}{}
\item[$^{\mathrm{a}}$] Upper limits refer to the absolute value of the rotation measure $|{\rm RM}|$.
\item[$^{\mathrm{b}}$] For this source strong temporal variability of the polarization angle prevents an RM measurement.
\end{list}
\end{table}

\section{Summary and conclusions}

We have analyzed the linear polarization of six active galactic nuclei -- 0415+379 (3C~111), 0507+179, 0528+134 (OG+134), 0954+658, 1418+546 (OQ+530), and 1637+574 (OS+562) -- in the observatory frame frequency range 80--253~GHz, corresponding to a range of rest frame frequencies 88--705~GHz. Our data were obtained in the years 2007--2011 and provide information resolved in time and frequency for all targets. Our sample includes two optically thin, outflow-dominated and four optically thick, core-dominated emitters. This study arrives at the following principal conclusions:

\begin{enumerate}

\item  In agreement with the results of Paper~I, the observed degrees of polarization $m_L$ are lower than the levels expected from the theory of synchrotron emission by up to one order of magnitude. This indicates that the polarization signal is effectively averaged out. Assuming radiation from $N$ emission regions with random orientations and geometries, we expect $m_L \rightarrow m_L'=m_L/\sqrt{N}$. Our results point to values from $N\approx3$ (for two optically thick emitters) to $N\approx1000$ (0415+379, an optically thin emitter).

\item  Degrees of polarization as well as fluxes vary with relative rates of a few per cent per day. We find fluctuation rate ratios $\rho$ (Eq.~\ref{eq_rho}) on the order of unity though typically slightly lower than one. This indicates that the size scales relevant for the polarized emission are compatible with, but probably not smaller than, the size scales relevant for the overall emission.

\item  The degrees of polarization in our sample fluctuate with normalized standard deviations $\sigma_m/\langle m_L\rangle\approx0.33-0.65$. We identify this variability approximately with source-averaged fluctuations expected from shock compression of magnetized plasmas. Our calculations point towards fairly strong shocks and/or complex, variable shock geometries, with compression factors $k\lesssim0.9$ and/or changes in viewing angles $\gtrsim$10$^{\circ}$.

\item  A search for correlations between source fluxes and polarization parameters identifies two special cases. For 0415+379, we observe a strong anticorrelation between flux and degree of polarization. This indicates the presence of (at least) two distinct emission regions with different levels of polarization. For 0507+179 and 1418+546, we find a positive correlation between flux and polarized flux. Here the data point toward emission from a single, predominant component.

\item  For 1418+546 we find (a) a correlated variation of flux and polarization, (b) changes of the polarization angle by tens of degrees, and (c) no correlation between flux and polarization angle. Comparison with the ``oblique shock'' models by Hughes et al. (\cite{hughes2011}) indicates good agreement with the presence of an oblique shock front observed at an angle $>$60$^{\circ}$ to the jet axis.

\item  We derive rotation measures of ${\rm RM}=(-39\pm1_{\rm stat}\pm13_{\rm sys})\times10^3$~rad\,m$^{-2}$ and ${\rm RM=(42\pm1_{\rm stat}\pm11_{\rm sys})\times10^4}$~rad\,m$^{-2}$ for 1418+546 and 1637+574, respectively. These are the highest values reported to date for AGN except for Sagittarius~A*. Our results point to magnetic field strengths on the order of $\sim$10$^{-4}$~G. For 0415+379, 0507+179, and 0954+658 we are able to derive upper limits ${\rm|RM|}<1.7\times10^4$~rad\,m$^{-2}$.

\item  Comparing our rotation measure values with those obtained by other studies at lower frequencies, we are able to constrain the emission region geometry of 1418+546. Exploiting the relation $|{\rm RM}|\propto\nu^a$ we find $a=1.9\pm0.3$, in good agreement with $a=2$ as expected for a spherical or conical outflow.

\item  We do not see a relation between the observed polarization properties of our sources and their physical parameters such as black hole masses or kpc-morphology, as summarized in Table~\ref{tab_journal}.

\end{enumerate}

This paper is the second of a set of two articles. Earth rotation polarimetry of AGN is a by-product of regular PdBI operations; accordingly, the database is going to grow steadily in the future. We therefore expect that a new study -- analogous to the one outlined in this article as well as in Paper~I -- carried out in a few years from now promises new insights already because of substantially improved statistics. In addition, the PdBI crew currently evaluates the implementation of an observing mode that permits observations of all four Stokes parameters. Evidently, such an observing mode would permit additional studies if implemented eventually.

Our work demonstrates the power of mm/radio polarization studies of AGN based on data resolved in time and frequency. Given the strong temporal variability of AGN polarization, it becomes clear that more accurate measurements of differential parameters such as rotation measures require simultaneous multi-frequency observations. New facilities such as the Korean VLBI Network (KVN; e.g., Kim et al. \cite{kim2010}), which permits single-polarization observations simultaneously at
four frequencies -- namely 22, 43, 86, and 129~GHz -- or dual-polarization observations simultaneously at two of the aforementioned frequencies, will be important tools for new discoveries.

\begin{acknowledgements}
We are grateful to the entire IRAM and PdBI staff, whose hard work over many years made this study possible. Our work has made use of the NASA/IPAC Extragalactic Database (NED), which is operated by the Jet Propulsion Laboratory, California Institute of Technology, under contract with the National Aeronautics and Space Administration. We have also made use of data from the MOJAVE database that is maintained by the MOJAVE team (Lister et al. \cite{lister2009}). For data processing and analysis we applied the software package GILDAS, developed and maintained by the GILDAS team (\url{http://www.iram.fr/IRAMFR/GILDAS/}), as well as the software package DPUSER (\url{http://www.mpe.mpg.de/~ott/dpuser/index.html}), developed and maintained by Thomas Ott at MPE Garching. Last but not least, we are grateful to the referee, E. Ros, whose review improved the quality of our paper.
\end{acknowledgements}

\clearpage

\clearpage

\clearpage

\clearpage

\clearpage

\clearpage

\clearpage

\clearpage

\begin{longtable}{l c c c c c c c c c}
\caption{\label{tab_obsjournal} Observations journal of our AGN linear polarization study. For each \emph{source} we quote its name, the number of polarimetric observations $N_{\rm obs}$, and the number of occasions when significant polarization was detected $N_{\rm det}$. For each \emph{observation} we give the observing date, the observing frequency $\nu$, the degree of linear polarization $m_L$, its statistical $1\sigma$ error $\delta m_L$, the polarization angle $\chi$, and its statistical $1\sigma$ error $\delta\chi$. When no polarization was detected, we quote a $3\sigma$ upper limit in the $m_L$ column, denoted by ``$<$''. Entries ``--'' indicate ``no value available''.} \\
\hline\hline
Name &  $N_{\rm obs}$ & $N_{\rm det}$ & observing date & $\nu$ & $m_L$ & $\delta m_L$ & $\chi$ & $\delta\chi$ \\
(B1950) & & & & [GHz] & [\%] & [\%] & $[^{\circ}]$ & $[^{\circ}]$ \\
\hline
\endfirsthead
\caption{continued.} \\
\hline\hline
Name &  $N_{\rm obs}$ & $N_{\rm det}$ & observing date & $\nu$ & $m_L$ & $\delta m_L$ & $\chi$ & $\delta\chi$ \\
(B1950) & & & & [GHz] & [\%] & [\%] & $[^{\circ}]$ & $[^{\circ}]$ \\
\hline
\endhead
\hline
\endfoot
0415+379 & 34 & 17 & 18-MAR-2007 & 89 & 0.7 & 0.1 & 38 & 4 \\
 & ~ & ~ & 25-MAR-2007 & 89 & $<$0.4 & -- & -- & -- \\
 & ~ & ~ & 07-APR-2007 & 93 & $<$1.2 & -- & -- & -- \\
 & ~ & ~ & 11-JUL-2007 & 110 & $<$1.8 & -- & -- & -- \\
 & ~ & ~ & 29-JUL-2007 & 86 & 1.8 & 0.1 & 177 & 2 \\
 & ~ & ~ & 31-JUL-2007 & 86 & 1.4 & 0.1 & 6 & 1 \\
 & ~ & ~ & 03-AUG-2007 & 84 & 1.2 & 0.1 & 6 & 2 \\
 & ~ & ~ & 06-AUG-2007 & 87 & 1.2 & 0.1 & 178 & 1 \\
 & ~ & ~ & 07-AUG-2007 & 84 & 1.5 & 0.2 & 174 & 1 \\
 & ~ & ~ & 08-AUG-2007 & 84 & $<$1.5 & -- & -- & -- \\
 & ~ & ~ & 11-AUG-2007 & 97 & 1.1 & 0.1 & 12 & 2 \\
 & ~ & ~ & 14-AUG-2007 & 84 & $<$1.3 & -- & -- & -- \\
 & ~ & ~ & 15-AUG-2007 & 84 & 1.2 & 0.1 & 175 & 1 \\
 & ~ & ~ & 18-AUG-2007 & 101 & 0.8 & 0.1 & 27 & 4 \\
 & ~ & ~ & 19-AUG-2007 & 113 & $<$0.6 & -- & -- & -- \\
 & ~ & ~ & 20-AUG-2007 & 101 & $<$1.7 & -- & -- & -- \\
 & ~ & ~ & 22-AUG-2007 & 216 & 2 & 0.2 & 106 & 5 \\
 & ~ & ~ & 28-AUG-2007 & 113 & $<$1 & -- & -- & -- \\
 & ~ & ~ & 30-AUG-2007 & 87 & $<$1.2 & -- & -- & -- \\
 & ~ & ~ & 31-AUG-2007 & 85 & $<$1.3 & -- & -- & -- \\
 & ~ & ~ & 29-OCT-2007 & 227 & 4.3 & 0.2 & 117 & 2 \\
 & ~ & ~ & 26-JAN-2008 & 231 & $<$5 & -- & -- & -- \\
 & ~ & ~ & 31-JAN-2008 & 84 & 2.8 & 0.1 & 96 & 1 \\
 & ~ & ~ & 17-FEB-2008 & 231 & $<$1.8 & -- & -- & -- \\
 & ~ & ~ & 22-FEB-2008 & 110 & 4.2 & 0.1 & 110 & 1 \\
 & ~ & ~ & 23-FEB-2008 & 220 & $<$2.5 & -- & -- & -- \\
 & ~ & ~ & 27-FEB-2008 & 89 & 3.9 & 0.1 & 118 & 1 \\
 & ~ & ~ & 18-MAR-2008 & 146 & 5.9 & 0.1 & 106 & 1 \\
 & ~ & ~ & 05-APR-2008 & 93 & 3.5 & 0.1 & 115 & 1 \\
 & ~ & ~ & 19-JUL-2008 & 115 & $<$1.5 & -- & -- & -- \\
 & ~ & ~ & 08-SEP-2008 & 231 & $<$4 & -- & -- & -- \\
 & ~ & ~ & 17-NOV-2008 & 113 & $<$1 & -- & -- & -- \\
 & ~ & ~ & 22-APR-2009 & 153 & 1.3 & 0.2 & 118 & 5 \\
 & ~ & ~ & 06-JAN-2010 & 231 & $<$2.4 & -- & -- & -- \\
0507+179 & 22 & 21 & 06-SEP-2007 & 85 & 4.9 & 0.4 & 106 & 1 \\
 & ~ & ~ & 07-SEP-2007 & 98 & 5.9 & 0.1 & 106 & 1 \\
 & ~ & ~ & 10-SEP-2007 & 86 & 5.9 & 0.4 & 107 & 2 \\
 & ~ & ~ & 01-OCT-2007 & 86 & 6.6 & 0.3 & 106 & 1 \\
 & ~ & ~ & 16-FEB-2008 & 231 & 2 & 0.1 & 47 & 5 \\
 & ~ & ~ & 04-APR-2008 & 110 & 3.1 & 0.1 & 133 & 1 \\
 & ~ & ~ & 05-APR-2008 & 93 & 3.4 & 0.1 & 130 & 1 \\
 & ~ & ~ & 14-NOV-2008 & 146 & 8 & 0.1 & 81 & 1 \\
 & ~ & ~ & 12-DEC-2008 & 231 & 9.2 & 0.1 & 103 & 1 \\
 & ~ & ~ & 28-DEC-2008 & 90 & 9.5 & 0.2 & 88 & 1 \\
 & ~ & ~ & 05-JAN-2009 & 89 & 9.3 & 0.1 & 86 & 1 \\
 & ~ & ~ & 28-JAN-2009 & 231 & 8.7 & 0.4 & 96 & 1 \\
 & ~ & ~ & 29-JAN-2009 & 231 & 11.5 & 0.5 & 92 & 1 \\
 & ~ & ~ & 27-FEB-2009 & 231 & 10.4 & 0.4 & 90 & 1 \\
 & ~ & ~ & 17-MAR-2009 & 90 & 7.6 & 0.1 & 89 & 1 \\
 & ~ & ~ & 06-AUG-2009 & 89 & 7.8 & 0.1 & 107 & 1 \\
 & ~ & ~ & 25-NOV-2009 & 89 & 9.9 & 0.3 & 96 & 1 \\
 & ~ & ~ & 11-JAN-2010 & 90 & 8.1 & 0.1 & 97 & 1 \\
 & ~ & ~ & 03-FEB-2010 & 220 & 9.2 & 1.2 & 102 & 3 \\
 & ~ & ~ & 05-NOV-2010 & 227 & 8.4 & 0.3 & 115 & 2 \\
 & ~ & ~ & 12-NOV-2010 & 89 & 5.6 & 0.1 & 113 & 1 \\
 & ~ & ~ & 22-JAN-2011 & 220 & $<$2.5 & -- & -- & -- \\
0528+134 & 32 & 18 & 03-FEB-2007 & 231 & 1.9 & 0.1 & 107 & 6 \\
 & ~ & ~ & 04-FEB-2007 & 231 & $<$3.4 & -- & -- & -- \\
 & ~ & ~ & 16-FEB-2007 & 231 & 4.1 & 0.4 & 35 & 6 \\
 & ~ & ~ & 11-APR-2007 & 87 & 5.4 & 0.1 & 125 & 1 \\
 & ~ & ~ & 15-APR-2007 & 115 & 5.6 & 0.1 & 126 & 1 \\
 & ~ & ~ & 15-AUG-2007 & 84 & 2.9 & 0.1 & 11 & 1 \\
 & ~ & ~ & 18-AUG-2007 & 101 & 3.4 & 0.1 & 21 & 1 \\
 & ~ & ~ & 19-AUG-2007 & 113 & 2.9 & 0.1 & 30 & 2 \\
 & ~ & ~ & 20-AUG-2007 & 101 & 2.6 & 0.2 & 32 & 3 \\
 & ~ & ~ & 28-AUG-2007 & 113 & 0.8 & 0.1 & 113 & 3 \\
 & ~ & ~ & 30-AUG-2007 & 87 & $<$1.4 & -- & -- & -- \\
 & ~ & ~ & 31-AUG-2007 & 85 & $<$1 & -- & -- & -- \\
 & ~ & ~ & 01-SEP-2007 & 231 & $<$1.9 & -- & -- & -- \\
 & ~ & ~ & 08-SEP-2007 & 216 & $<$2.1 & -- & -- & -- \\
 & ~ & ~ & 22-SEP-2007 & 231 & 3.4 & 0.3 & 102 & 4 \\
 & ~ & ~ & 03-OCT-2007 & 203 & 2.4 & 0.2 & 63 & 4 \\
 & ~ & ~ & 07-OCT-2007 & 226 & $<$5.6 & -- & -- & -- \\
 & ~ & ~ & 12-NOV-2007 & 115 & 2.7 & 0.4 & 2 & 1 \\
 & ~ & ~ & 26-JAN-2008 & 231 & 4.6 & 0.3 & 159 & 2 \\
 & ~ & ~ & 10-FEB-2008 & 231 & $<$2.4 & -- & -- & -- \\
 & ~ & ~ & 14-FEB-2008 & 231 & $<$3.9 & -- & -- & -- \\
 & ~ & ~ & 15-FEB-2008 & 229 & $<$2.1 & -- & -- & -- \\
 & ~ & ~ & 16-FEB-2008 & 231 & $<$2.1 & -- & -- & -- \\
 & ~ & ~ & 22-FEB-2008 & 110 & 3.5 & 0.2 & 3 & 1 \\
 & ~ & ~ & 27-FEB-2008 & 89 & $<$2.8 & -- & -- & -- \\
 & ~ & ~ & 18-MAR-2008 & 146 & 2 & 0.1 & 18 & 2 \\
 & ~ & ~ & 23-SEP-2008 & 113 & 2.9 & 0.1 & 55 & 1 \\
 & ~ & ~ & 27-SEP-2008 & 231 & $<$0.5 & -- & -- & -- \\
 & ~ & ~ & 27-NOV-2008 & 113 & $<$3.6 & -- & -- & -- \\
 & ~ & ~ & 03-DEC-2008 & 113 & 1.9 & 0.1 & 144 & 4 \\
 & ~ & ~ & 07-DEC-2008 & 231 & $<$4 & -- & -- & -- \\
 & ~ & ~ & 24-DEC-2008 & 89 & 4 & 0.1 & 119 & 3 \\
0954+658 & 37 & 35 & 08-MAR-2007 & 235 & 18.2 & 0.4 & 1 & 2 \\
 & ~ & ~ & 25-AUG-2007 & 85 & 10.6 & 0.1 & 175 & 1 \\
 & ~ & ~ & 05-SEP-2007 & 85 & 10.4 & 0.1 & 172 & 1 \\
 & ~ & ~ & 14-SEP-2007 & 84 & 6.5 & 0.1 & 173 & 1 \\
 & ~ & ~ & 16-SEP-2007 & 84 & 6.5 & 0.1 & 171 & 1 \\
 & ~ & ~ & 19-SEP-2007 & 84 & 7.8 & 0.1 & 171 & 1 \\
 & ~ & ~ & 07-FEB-2008 & 208 & 16.2 & 0.1 & 180 & 1 \\
 & ~ & ~ & 13-FEB-2008 & 235 & 12.4 & 0.3 & 177 & 1 \\
 & ~ & ~ & 15-FEB-2008 & 207 & 8.3 & 0.1 & 175 & 1 \\
 & ~ & ~ & 22-FEB-2008 & 208 & 16.5 & 0.4 & 172 & 1 \\
 & ~ & ~ & 07-MAR-2008 & 208 & 9.3 & 0.3 & 176 & 1 \\
 & ~ & ~ & 08-MAR-2008 & 208 & 8.9 & 0.2 & 178 & 1 \\
 & ~ & ~ & 23-JUL-2008 & 94 & 4.1 & 0.1 & 47 & 1 \\
 & ~ & ~ & 14-AUG-2008 & 165 & 1.3 & 0.2 & 15 & 7 \\
 & ~ & ~ & 31-AUG-2008 & 91 & 8 & 0.1 & 148 & 1 \\
 & ~ & ~ & 02-SEP-2008 & 88 & 6.1 & 0.1 & 151 & 1 \\
 & ~ & ~ & 09-SEP-2008 & 100 & 4.1 & 0.1 & 151 & 1 \\
 & ~ & ~ & 10-SEP-2008 & 84 & 4.2 & 0.1 & 153 & 1 \\
 & ~ & ~ & 01-OCT-2008 & 111 & 3.7 & 0.2 & 175 & 1 \\
 & ~ & ~ & 02-OCT-2008 & 92 & 3.6 & 0.2 & 171 & 2 \\
 & ~ & ~ & 07-OCT-2008 & 154 & 5.5 & 0.3 & 169 & 1 \\
 & ~ & ~ & 21-FEB-2009 & 137 & 4.1 & 0.1 & 29 & 1 \\
 & ~ & ~ & 19-JUN-2009 & 101 & 7.1 & 0.2 & 25 & 1 \\
 & ~ & ~ & 25-JUN-2009 & 101 & 7.9 & 0.2 & 14 & 2 \\
 & ~ & ~ & 17-JUL-2009 & 101 & $<$2.3 & -- & -- & -- \\
 & ~ & ~ & 22-JUL-2009 & 101 & $<$2.1 & -- & -- & -- \\
 & ~ & ~ & 24-JUL-2009 & 101 & 1.9 & 0.2 & 129 & 1 \\
 & ~ & ~ & 24-AUG-2009 & 115 & 4.3 & 0.2 & 28 & 1 \\
 & ~ & ~ & 25-AUG-2009 & 115 & 5.8 & 0.2 & 32 & 2 \\
 & ~ & ~ & 26-AUG-2009 & 115 & 4.3 & 0.1 & 27 & 3 \\
 & ~ & ~ & 09-SEP-2009 & 115 & 5 & 0.2 & 16 & 2 \\
 & ~ & ~ & 18-OCT-2009 & 150 & 5.9 & 0.1 & 11 & 2 \\
 & ~ & ~ & 05-APR-2010 & 240 & 7.5 & 0.4 & 179 & 1 \\
 & ~ & ~ & 08-APR-2010 & 82 & 9.8 & 0.2 & 175 & 1 \\
 & ~ & ~ & 07-AUG-2010 & 143 & 7.5 & 0.1 & 166 & 1 \\
 & ~ & ~ & 15-SEP-2010 & 171 & 8.8 & 0.3 & 168 & 1 \\
 & ~ & ~ & 01-JAN-2011 & 234 & 7.8 & 0.3 & 6 & 2 \\
1418+546 & 76 & 52 & 25-JAN-2008 & 94 & 2.8 & 0.1 & 104 & 1 \\
 & ~ & ~ & 26-JAN-2008 & 91 & 2.8 & 0.1 & 105 & 1 \\
 & ~ & ~ & 15-FEB-2008 & 207 & 8 & 0.4 & 116 & 2 \\
 & ~ & ~ & 14-MAR-2008 & 94 & 7.8 & 0.1 & 101 & 1 \\
 & ~ & ~ & 04-MAY-2008 & 96 & 4.9 & 0.2 & 111 & 1 \\
 & ~ & ~ & 05-MAY-2008 & 89 & 5.2 & 0.2 & 112 & 2 \\
 & ~ & ~ & 18-MAY-2008 & 163 & 8.7 & 0.1 & 124 & 1 \\
 & ~ & ~ & 20-MAY-2008 & 157 & 10.8 & 0.6 & 121 & 1 \\
 & ~ & ~ & 22-MAY-2008 & 137 & 7.2 & 0.2 & 119 & 1 \\
 & ~ & ~ & 10-JUN-2008 & 137 & 5.4 & 0.2 & 112 & 3 \\
 & ~ & ~ & 30-JUN-2008 & 93 & 4.1 & 0.2 & 103 & 2 \\
 & ~ & ~ & 05-JUL-2008 & 101 & 3.5 & 0.1 & 103 & 2 \\
 & ~ & ~ & 08-JUL-2008 & 154 & 7.2 & 0.3 & 105 & 2 \\
 & ~ & ~ & 09-JUL-2008 & 154 & 7.1 & 0.6 & 103 & 4 \\
 & ~ & ~ & 18-JUL-2008 & 112 & 4.2 & 0.3 & 114 & 2 \\
 & ~ & ~ & 16-AUG-2008 & 158 & 3.7 & 0.5 & 115 & 2 \\
 & ~ & ~ & 19-SEP-2008 & 108 & 3.1 & 0.2 & 124 & 1 \\
 & ~ & ~ & 24-SEP-2008 & 95 & 2.9 & 0.2 & 127 & 1 \\
 & ~ & ~ & 30-SEP-2008 & 160 & $<$6 & -- & -- & -- \\
 & ~ & ~ & 19-OCT-2008 & 158 & $<$8.6 & -- & -- & -- \\
 & ~ & ~ & 27-NOV-2008 & 158 & 2.9 & 0.1 & 121 & 2 \\
 & ~ & ~ & 03-JAN-2009 & 96 & $<$1.6 & -- & -- & -- \\
 & ~ & ~ & 04-JAN-2009 & 96 & $<$2.1 & -- & -- & -- \\
 & ~ & ~ & 05-JAN-2009 & 91 & $<$3.1 & -- & -- & -- \\
 & ~ & ~ & 14-JAN-2009 & 137 & 2.3 & 0.1 & 112 & 1 \\
 & ~ & ~ & 30-JAN-2009 & 163 & 5.8 & 0.2 & 114 & 1 \\
 & ~ & ~ & 14-FEB-2009 & 163 & 2.3 & 0.1 & 132 & 2 \\
 & ~ & ~ & 12-MAR-2009 & 87 & $<$1.8 & -- & -- & -- \\
 & ~ & ~ & 13-MAR-2009 & 163 & 3.7 & 0.3 & 150 & 4 \\
 & ~ & ~ & 14-MAR-2009 & 160 & $<$2.4 & -- & -- & -- \\
 & ~ & ~ & 01-APR-2009 & 155 & $<$2.9 & -- & -- & -- \\
 & ~ & ~ & 04-APR-2009 & 152 & $<$3 & -- & -- & -- \\
 & ~ & ~ & 05-APR-2009 & 165 & $<$4.1 & -- & -- & -- \\
 & ~ & ~ & 09-APR-2009 & 137 & $<$2.3 & -- & -- & -- \\
 & ~ & ~ & 28-MAY-2009 & 152 & $<$1.1 & -- & -- & -- \\
 & ~ & ~ & 20-JUL-2009 & 109 & 2.5 & 0.2 & 121 & 2 \\
 & ~ & ~ & 10-AUG-2009 & 95 & $<$1.8 & -- & -- & -- \\
 & ~ & ~ & 11-AUG-2009 & 153 & $<$4.7 & -- & -- & -- \\
 & ~ & ~ & 22-AUG-2009 & 95 & $<$1.2 & -- & -- & -- \\
 & ~ & ~ & 29-AUG-2009 & 115 & $<$3.2 & -- & -- & -- \\
 & ~ & ~ & 19-SEP-2009 & 107 & 3.4 & 0.1 & 137 & 2 \\
 & ~ & ~ & 02-OCT-2009 & 115 & 5.3 & 0.1 & 132 & 1 \\
 & ~ & ~ & 15-OCT-2009 & 253 & 5.2 & 0.4 & 127 & 7 \\
 & ~ & ~ & 27-OCT-2009 & 115 & 6.1 & 0.2 & 139 & 2 \\
 & ~ & ~ & 30-OCT-2009 & 253 & 11.1 & 0.8 & 151 & 1 \\
 & ~ & ~ & 01-NOV-2009 & 253 & $<$10.6 & -- & -- & -- \\
 & ~ & ~ & 17-NOV-2009 & 159 & 6.8 & 0.2 & 137 & 1 \\
 & ~ & ~ & 15-DEC-2009 & 115 & 5.4 & 0.4 & 145 & 2 \\
 & ~ & ~ & 17-DEC-2009 & 115 & 4.6 & 0.2 & 143 & 2 \\
 & ~ & ~ & 05-JAN-2010 & 115 & 3.1 & 0.3 & 143 & 2 \\
 & ~ & ~ & 20-JAN-2010 & 155 & 3.5 & 0.2 & 128 & 3 \\
 & ~ & ~ & 21-JAN-2010 & 152 & 3.5 & 0.1 & 129 & 5 \\
 & ~ & ~ & 23-JAN-2010 & 115 & 4.1 & 0.1 & 132 & 1 \\
 & ~ & ~ & 25-JAN-2010 & 115 & 5.6 & 0.1 & 135 & 3 \\
 & ~ & ~ & 29-JAN-2010 & 115 & 4.4 & 0.3 & 130 & 1 \\
 & ~ & ~ & 04-MAR-2010 & 115 & 5.2 & 0.2 & 114 & 2 \\
 & ~ & ~ & 08-MAR-2010 & 115 & 6 & 0.2 & 117 & 1 \\
 & ~ & ~ & 09-MAR-2010 & 115 & 6.5 & 0.2 & 116 & 2 \\
 & ~ & ~ & 11-MAR-2010 & 115 & 6.4 & 0.5 & 113 & 2 \\
 & ~ & ~ & 12-MAR-2010 & 115 & 6.1 & 0.2 & 114 & 1 \\
 & ~ & ~ & 29-MAY-2010 & 82 & 4.7 & 0.1 & 109 & 1 \\
 & ~ & ~ & 03-JUN-2010 & 82 & 4.3 & 0.1 & 113 & 1 \\
 & ~ & ~ & 04-JUN-2010 & 97 & 4.7 & 0.1 & 115 & 1 \\
 & ~ & ~ & 07-JUN-2010 & 82 & 4.6 & 0.1 & 115 & 1 \\
 & ~ & ~ & 18-JUN-2010 & 110 & 6 & 0.4 & 123 & 1 \\
 & ~ & ~ & 31-JUL-2010 & 146 & 4.6 & 0.1 & 114 & 1 \\
 & ~ & ~ & 12-SEP-2010 & 90 & 2.9 & 0.3 & 132 & 2 \\
 & ~ & ~ & 17-SEP-2010 & 145 & $<$5.4 & -- & -- & -- \\
 & ~ & ~ & 19-SEP-2010 & 221 & $<$5.1 & -- & -- & -- \\
 & ~ & ~ & 26-SEP-2010 & 145 & 3.3 & 0.1 & 138 & 1 \\
 & ~ & ~ & 21-OCT-2010 & 243 & $<$2.8 & -- & -- & -- \\
 & ~ & ~ & 27-OCT-2010 & 111 & $<$1.5 & -- & -- & -- \\
 & ~ & ~ & 04-NOV-2010 & 171 & $<$3.6 & -- & -- & -- \\
 & ~ & ~ & 06-NOV-2010 & 221 & $<$2 & -- & -- & -- \\
 & ~ & ~ & 04-JAN-2011 & 111 & $<$2.8 & -- & -- & -- \\
 & ~ & ~ & 27-JAN-2011 & 157 & 2.1 & 0.1 & 125 & 5 \\
1637+574 & 26 & 17 & 13-FEB-2007 & 81 & 2.4 & 0.1 & 156 & 1 \\
 & ~ & ~ & 22-FEB-2007 & 80 & 1.3 & 0.1 & 136 & 3 \\
 & ~ & ~ & 06-MAR-2007 & 80 & $<$0.6 & -- & -- & -- \\
 & ~ & ~ & 12-MAR-2007 & 104 & $<$1.1 & -- & -- & -- \\
 & ~ & ~ & 18-MAR-2007 & 80 & 0.8 & 0.1 & 13 & 2 \\
 & ~ & ~ & 29-AUG-2007 & 87 & 3 & 0.1 & 124 & 1 \\
 & ~ & ~ & 01-SEP-2007 & 89 & 2.9 & 0.1 & 125 & 1 \\
 & ~ & ~ & 06-OCT-2007 & 102 & 2.9 & 0.2 & 148 & 3 \\
 & ~ & ~ & 26-OCT-2007 & 87 & 4.4 & 0.1 & 143 & 1 \\
 & ~ & ~ & 09-APR-2008 & 110 & $<$1.2 & -- & -- & -- \\
 & ~ & ~ & 23-APR-2008 & 110 & 3 & 0.1 & 105 & 1 \\
 & ~ & ~ & 24-APR-2008 & 111 & 2.6 & 0.1 & 110 & 1 \\
 & ~ & ~ & 27-APR-2008 & 100 & 2.6 & 0.1 & 108 & 1 \\
 & ~ & ~ & 04-MAY-2008 & 90 & 3 & 0.1 & 112 & 1 \\
 & ~ & ~ & 06-JUN-2008 & 104 & 2.8 & 0.1 & 96 & 1 \\
 & ~ & ~ & 08-JUN-2008 & 104 & 3.5 & 0.1 & 94 & 1 \\
 & ~ & ~ & 13-JUN-2008 & 108 & 4.1 & 0.1 & 94 & 1 \\
 & ~ & ~ & 19-JUN-2008 & 108 & 4.7 & 0.2 & 97 & 3 \\
 & ~ & ~ & 07-JUL-2008 & 96 & 3.9 & 0.1 & 87 & 1 \\
 & ~ & ~ & 23-JUL-2008 & 115 & 2.9 & 0.1 & 77 & 4 \\
 & ~ & ~ & 24-JUL-2008 & 115 & $<$9.3 & -- & -- & -- \\
 & ~ & ~ & 20-SEP-2008 & 115 & $<$2.2 & -- & -- & -- \\
 & ~ & ~ & 25-APR-2009 & 108 & $<$1.7 & -- & -- & -- \\
 & ~ & ~ & 28-APR-2009 & 105 & $<$1.1 & -- & -- & -- \\
 & ~ & ~ & 13-MAY-2009 & 92 & $<$1.9 & -- & -- & -- \\
 & ~ & ~ & 26-AUG-2009 & 92 & $<$1.6 & -- & -- & -- \\

\end{longtable}

\end{document}